\documentclass[11pt]{article}

\usepackage[margin=1in]{geometry}
\usepackage{amsmath,amssymb,amsthm,bm}
\usepackage{booktabs}
\usepackage{longtable}
\usepackage{array}
\usepackage{graphicx}
\usepackage{subcaption}
\usepackage{float}
\usepackage{tikz}
\usepackage{algorithm}
\usepackage{algpseudocode}
\usetikzlibrary{arrows.meta,positioning,fit,calc}
\usepackage{flafter}
\usepackage[section]{placeins}
\usepackage{microtype}
\usepackage{natbib}
\usepackage{enumitem}
\usepackage{xcolor}
\usepackage{url}
\usepackage[colorlinks=true,linkcolor=black,citecolor=black,urlcolor=black]{hyperref}
\newcommand{\E}{\mathbb E}
\newcommand{\R}{\mathbb R}
\newcommand{\Norm}{\mathcal N}
\newcommand{\Gam}{\operatorname{Gamma}}
\newcommand{\Bern}{\operatorname{Bernoulli}}
\newcommand{\diag}{\operatorname{diag}}
\newcommand{\tr}{\operatorname{tr}}
\newcommand{\ELBO}{\mathcal L}
\newcommand{\method}[1]{\textsc{#1}}

\newcommand{\GBmulti}{\method{gg-BFA} (mf)}
\newcommand{\GBtwo}{\method{gg-BFA} (2f)}
\newcommand{\GBsparse}{\method{gg-BFA} (2f+s)}

\newcommand{\diagop}{\operatorname{diag}}
\newcommand{\Ind}[1]{\mathbf 1\!\left\{#1\right\}}

\newcolumntype{P}[1]{>{\raggedright\arraybackslash}p{#1}}

\newtheorem{theorem}{Theorem}
\newtheorem{proposition}{Proposition}[section]
\newtheorem{lemma}[proposition]{Lemma}
\newtheorem{corollary}[proposition]{Corollary}
\newtheorem{apptheorem}[proposition]{Theorem}
\newtheorem{assumptioninner}{Assumption}
\newenvironment{assumption}[2]{\renewcommand{\theassumptioninner}{#1}\begin{assumptioninner}[#2]}{\end{assumptioninner}}
\theoremstyle{definition}
\newtheorem{definition}[proposition]{Definition}
\newtheorem{example}[proposition]{Example}
\theoremstyle{remark}
\newtheorem{appremark}[proposition]{Remark}
\theoremstyle{plain}

\newcommand{\KL}{\operatorname{KL}}
\newcommand{\IW}{\operatorname{IW}}
\newcommand{\IG}{\operatorname{IG}}
\newcommand{\Wis}{\operatorname{Wishart}}
\newcommand{\op}{\mathrm{op}}
\newcommand{\Fam}{\mathcal F}
\newcommand{\Qfam}{\mathcal Q}
\newcommand{\ld}{\ell^{\mathrm d}_n}
\newcommand{\En}{\mathcal E_n}
\newcommand{\Lt}{L_t}
\newcommand{\Ut}{U_t}
\newcommand{\Kstop}{K_{\mathrm{stop}}}
\newcommand{\Ntot}{N_{\mathrm{tot}}}
\newcommand{\mKL}{m^{\mathrm{KL}}}
\newcommand{\Pst}{P_*}

\title{\bfseries Graph-Guided Bayesian Factor Analysis:\\
A Fast Integrated EFA--CFA Approach\\
for High-Dimensional Data}
\author{Shitao Fan$^{1}$, Lizhen Lin$^{2}$, Zhenyao Ye$^{1}$, and Shuo Chen$^{1,3}$,\\
  \normalsize for the Alzheimer's Disease Neuroimaging Initiative\thanks{Data used in preparation of this article were obtained from the Alzheimer's Disease Neuroimaging Initiative (ADNI) database (\url{adni.loni.usc.edu}). As such, the investigators within the ADNI contributed to the design and implementation of ADNI and/or provided data but did not participate in analysis or writing of this report. A complete listing of ADNI investigators can be found at: \url{https://adni.loni.usc.edu/wp-content/uploads/how_to_apply/ADNI_Acknowledgement_List.pdf}.}\\
  \normalsize and for the Alzheimer's Disease Metabolomics Consortium\thanks{Data used in preparation of this article were generated by the Alzheimer's Disease Metabolomics Consortium (ADMC). As such, the investigators within the ADMC provided data but did not participate in analysis or writing of this report. A complete listing of ADMC investigators can be found at: \url{https://sites.duke.edu/adnimetab/team/}.}\\[1ex]
  \normalsize $^{1}$University of Maryland School of Medicine, Baltimore, MD, USA\\[0.5ex]
  \normalsize $^{2}$Department of Mathematics, University of Maryland, College Park, MD, USA\\[0.5ex]
  \normalsize $^{3}$Division of Biostatistics and Bioinformatics, Department of Epidemiology and Public Health}
\date{November 2026}
\makeatletter\def\@fnsymbol#1{\ensuremath{\ifcase#1\or *\or **\fi}}\makeatother 

\begin{document}
\maketitle

\begin{abstract}
Factor analysis has been central to the analysis of high-dimensional biomedical data. However, existing exploratory and confirmatory factor analysis methods face several limitations: less interpretable factor findings, difficulty in identifying which variables can and cannot be represented by low-dimensional factors, challenges in estimating the covariance among factors, and maximum-likelihood fits that break down when variables outnumber subjects. We propose graph-guided Bayesian factor analysis (gg-BFA) to address these gaps. We first use a Bayesian model to explore and learn the latent graph structure, identifying which variables can and cannot be represented by low-dimensional factors and how many factors are needed. A dense-subgraph search proposes partitions of the variables into communities, each defining one factor, and a forward stepwise search compares their fractional variational posteriors and stops when the evidence lower bound first decreases. We show that this search recovers the true communities even when the number of variables grows much faster than the sample size. We then develop a confirmatory Bayesian model whose variational posterior of the covariance contracts at the parametric rate up to a logarithmic factor, yielding posterior inference for the factors and their covariance, as well as predictions of missing values. Its four members range from a community mask to several cross-loadings per variable. In five biomedical panels, the complete analysis took under ten minutes each on one GPU, and cross-loadings improved held-out covariance and the prediction of masked measurements. In TCGA Kidney, six of the first eight factors were enriched for known pathways, and 190 of the 210 factor correlations had 95\% credible intervals excluding zero. Overall, gg-BFA is computationally fast and provides interpretable factor analysis findings and reliable inference results.
\end{abstract}

\section{Introduction}
\label{sec:introduction}

Biomedical studies such as omics profiling and medical imaging measure hundreds or thousands
of variables on each subject.  The dependence among these variables is often concentrated
within groups of related variables, such as genes in the same pathway or measures from the
same brain region.  The scientific questions behind these studies go beyond individual
variances and pairwise correlations.  We want to know which molecular features or
imaging-derived phenotypes vary together, which latent processes drive their joint variation,
and whether the resulting dependence structure can predict the missing measurements of a new
subject.  Factor analysis is a latent variable approach that explains the correlations among
many observed variables through a small number of unobserved common factors.  The model
implies \(\operatorname{Cov}(x_i)=\Lambda\Phi\Lambda^\top+\Psi\), where \(x_i\) is the vector
of \(p\) variables for subject \(i\), \(\Lambda\) is the \(p\times K\) matrix of loadings with
\(K\) much smaller than \(p\), \(\Phi\) is the factor covariance, and \(\Psi\) is the diagonal
matrix of residual variances.  This representation is useful when each factor corresponds to
one such group, for example a pathway or a brain region, so that \(\Phi\) describes how these
groups co-vary.

In practice, the factor structure is often found in two steps
\citep{gerbing1996efa,fabrigar1999efa}.  Exploratory factor analysis (EFA) chooses the number
of factors and assigns factor memberships to the variables.  Confirmatory factor analysis
(CFA) then estimates the loadings of this structure and computes the factor scores.  Through
these two steps, we represent each subject by \(K\) factor scores instead of \(p\) correlated
variables, and each factor stays interpretable through the variables assigned to it.

Methodological work on factor analysis, both frequentist and Bayesian, has focused mainly on
the EFA step.  Frequentist work has developed estimators of the number of factors, the
loadings, and the covariance matrix in high dimensions
\citep{bai2002determining,bai2012statistical,fan2013large}.  Bayesian work has chosen the
number of factors by model comparison \citep{lopes2004bayesian} and has introduced shrinkage
and sparsity-inducing priors on \(\Lambda\) that determine the number of factors and the
pattern of nonzero loadings jointly
\citep{carvalho2008sparse,bhattacharya2011sparse,rockova2016fast,fruhwirth2025sparse}.  These
priors can improve covariance estimation, but the factors are then defined by the learned
loadings and need not correspond to a recognizable group of variables
\citep{dunson2026tutorial}.  The confirmatory step has received less attention.  Its Bayesian
formulations have been developed largely for structural equation models whose loading pattern
is specified in advance \citep{lee1981bayesiancfa,muthen2012bayesian}, so the posterior
describes uncertainty, but a cross-loading fixed at zero stays zero.  The two-step procedure
itself, in which the output of EFA defines the model fitted by CFA, has been studied even
less.  More importantly, factor memberships are typically obtained by rotating the EFA
loadings and keeping those above a threshold, which may miss the groups of strongly correlated
variables.  Because the CFA step and the factor scores inherit these memberships, any error in
them carries over to the fitted factor covariance and to every analysis based on the factor
scores.

Graph-based methods offer a different starting point.  Viewing the dependence structure as a
weighted graph, with the variables as nodes and their pairwise associations as weighted edges,
adaptive dense-subgraph methods extract communities of strongly associated variables from
large, noisy association matrices \citep{wu2022adsd}.  The interconnected community network
model adds that communities may remain connected through bridge variables, which belong
primarily to one community but also carry information about another \citep{wu2021icn}.
Semi-confirmatory factor analysis (SCFA) assigns one factor to each community and fixes the
loading matrix to the binary community mask \citep{yang2024scfa}.  The mask gives each factor a
direct community interpretation and admits a closed-form estimator.  However, all variables of
a community then share the same primary loading, bridge variables cannot carry signal from a
second community, the number of communities is chosen outside the model, and SCFA provides no
joint posterior for the factor scores, loadings, factor covariance, and residual variation.  A
related nonfactor approach, BCME, uses estimated groups to regularize the within- and
between-block correlation structure directly \citep{lei2026bcme}.

\begin{figure}[!htbp]
\centering
\includegraphics[width=\textwidth]{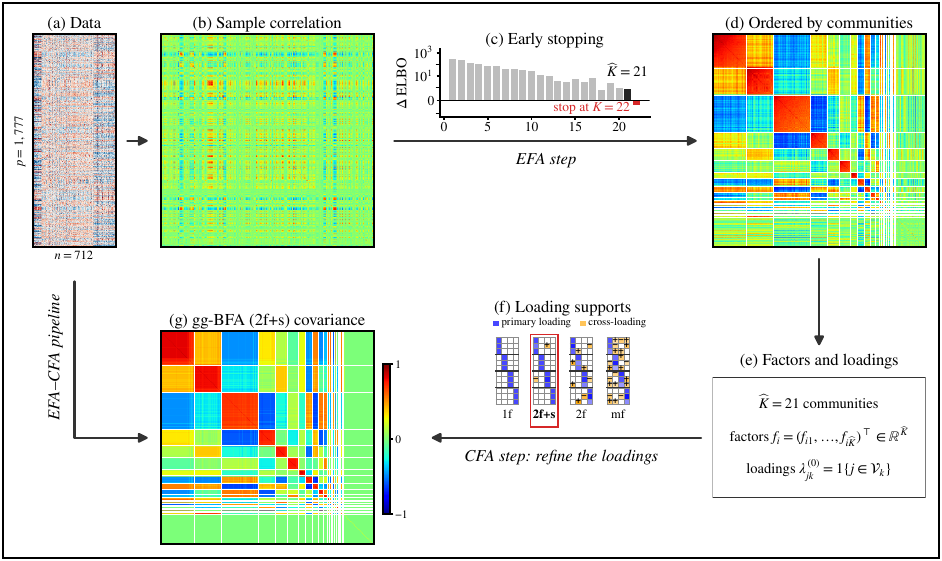}
\caption{Overview of the two steps of gg-BFA. Step 1 explores and learns the latent structure from the input data (a) and the sample correlation matrix (b). The output graph structure reveals (i) which variables can be assigned to the same community (non-zero loadings); (ii) how many communities there are; and (iii) which variables are `independent' and cannot be represented by factors. Step 2 is the confirmatory step, yielding the loadings, the factors, the covariance of the factors, and all corresponding inference (e,f,g). The TCGA Kidney data (712 subjects, 1,777 genes) are used for illustration.  (a)~Data
\(X^\top\), genes in rows.  (b)~Sample correlation in the original gene order.  (c)~Early
stopping of the forward stepwise search (Algorithm~\ref{alg:selection}): the best ELBO first
decreases at \(K=22\), so \(\widehat K=21\).  (d)~Sample correlation ordered by the selected
communities.  (e)~Factors and initial 0/1 loadings.  (f)~Loading supports of the four
confirmatory members (schematic; rows are variables, columns are factors); 2f+s, used in~(g),
is framed.  (g)~Covariance fitted by gg-BFA (2f+s), on the correlation scale and in the order
of~(d).}
\label{fig:method}
\end{figure}

To address these challenges, we propose a \textbf{g}raph-\textbf{g}uided two-step procedure for
\textbf{B}ayesian \textbf{f}actor \textbf{a}nalysis (gg-BFA, see Figure~\ref{fig:method}).  The
graph-based approach treats the sample correlation matrix as a weighted graph and groups
strongly correlated variables into communities \citep{wu2022adsd,yang2024scfa}.  Each community
defines one factor, and the factor covariance \(\Phi\) accounts for the dependence between
communities.  In the confirmatory step, the primary community of each variable defines what its
factor represents, while cross-loadings let a variable carry signal from other communities, as
bridge variables do.  The four members of gg-BFA range from the binary community mask,
gg-BFA (1f), through a sparse set of single cross-loadings, gg-BFA (2f+s), and one
cross-loading per variable, gg-BFA (2f), to several cross-loadings per variable, gg-BFA (mf)
(Figure~\ref{fig:method}f).  The Bayesian framework then gives credible intervals for the factor
covariance (Figure~\ref{fig:results}) and predicts the missing measurements of a new subject
from the same posterior.  Bayesian factor models are usually fitted by Markov chain Monte Carlo
\citep{gelman2013bda}, which becomes costly when high-dimensional models must be refitted
across partitions, loading structures, and cross-validation folds.  We fit each member by
mean-field variational inference \citep{blei2017vi}, whose consistency and contraction rates
are established under suitable conditions
\citep{wang2019frequentistvb,yang2020alphavb,zhang2020variational,ohn2024adaptivevb}.  Because
mean-field approximations can understate uncertainty \citep{giordano2018covariances}, we
evaluate interval calibration separately and use variational bagging
\citep{fan2025variationalbagging} as a supplement, which we call bagged gg-BFA.

Our method makes several novel contributions.  First, gg-BFA integrates the two steps.  A
forward stepwise search chooses the communities and their number, and the confirmatory step
fits the same communities, so each factor keeps its community interpretation
(Algorithm~\ref{alg:selection}).  Second, we show that the search recovers the true
communities and that the variational posterior of the covariance contracts at the parametric
rate up to a logarithmic factor, even when the number of variables grows much faster than the
sample size (Section~\ref{sec:selection-theory}).  The stopping rule comes from the
early-stopped aggregation framework of \citet{ohn2026esa}, which we adapt to this setting with
a guarantee that the search does not stop too late.  When no finite partition is exact but the
approximation error decays geometrically in the number of communities, the search stops after
a number of communities that grows logarithmically with the sample size, and the selected
model attains the parametric rate up to a polylogarithmic factor without knowing the rate of
decay (Appendix~\ref{th:sec:misspec}).  Third, both steps work when the variables outnumber
the subjects, where the maximum-likelihood EFA--CFA pipelines of \texttt{psych} and \texttt{lavaan}
cannot be applied because the sample covariance matrix is singular
(Section~\ref{sec:sim-selection}).  Finally, with mean-field variational inference on one GPU,
the whole pipeline took less than ten minutes on each of five biomedical panels, including two
with more variables than subjects (Section~\ref{sec:applications}).

Our evaluation asks, for the exploratory step, whether the search recovers the true number of
communities and the true partition when they exist, whether its stopping point and covariance
error behave as the theory predicts when no finite partition is exact, and how its estimated
covariance compares with those of exploratory-to-confirmatory pipelines.  For the confirmatory
step, we ask how much of the covariance the community assignment alone explains, whether some
variables must also load on the factors of other communities, whether this flexibility carries
over to new subjects and to the prediction of hidden measurements, what it costs in
computation, and whether the variational intervals are calibrated.  Simulations with known
factors, loadings, and covariances separate point recovery, misspecification, and calibration,
whereas in five real-data panels (ADNI Nightingale metabolomics, TCGA Kidney and TCGA miRNA
molecular data, and ABCD and UK Biobank imaging data) the evaluation focuses on covariance
representation and out-of-sample behavior.  Section~\ref{sec:model} presents the two steps of
gg-BFA and the theory of the exploratory step, Section~\ref{sec:evaluation} defines the two
comparisons, and Sections~\ref{sec:experiments} and~\ref{sec:applications} report the
simulation and real-data results.  Section~\ref{sec:discussion} discusses
the implications and limitations.

\section{Methods}
\label{sec:model}

Let \(x_1,\ldots,x_n\in\R^{p_0}\) be the standardized observations of all \(p_0\) variables on
\(n\) subjects.  A factor model with \(K\) factors assumes
\(x_i\mid f_i\sim\Norm_{p_0}\{\Lambda f_i,\diag(\psi)\}\) and \(f_i\sim\Norm_K(0,\Phi)\), so that the
covariance of \(x_i\) is \(\Omega=\Lambda\Phi\Lambda^\top+\diag(\psi)\).  Here \(f_i\) contains the
factor scores of subject \(i\), \(\Lambda\) is the loading matrix, \(\Phi\) is the factor covariance,
and \(\psi\) contains the residual variances.  The first step, EFA, chooses \(K\) and the pattern of
nonzero loadings.  The second step, CFA, then estimates \(\Lambda\), \(\Phi\), and \(\psi\) with this
pattern fixed.

We view the sample correlation matrix as a weighted graph on the \(p_0\) variables.  Variables that
load on the same factor are strongly correlated, so they form a dense community in this graph, and
correlated factors connect their communities \citep{wu2021icn,yang2024scfa}.  We therefore use the
communities as the loading pattern, with one factor per community.  The exploratory step takes the
place of EFA.  Section~\ref{sec:icn-graph} defines the graph structure,
Section~\ref{sec:icons-estimation} the candidate communities, Section~\ref{sec:selection} the forward
stepwise search that chooses the communities and their number (Algorithm~\ref{alg:selection}), and
Section~\ref{sec:selection-theory} its guarantees.  The confirmatory step takes the place of CFA.
Sections~\ref{sec:bfa-target}--\ref{sec:masked-cell} give the Bayesian factor model on the selected
partition, its four cross-loading members, variational inference, posterior summaries, and
prediction.  A partition may also be supplied in advance.  The confirmatory step treats the
partition as fixed and does not propagate the uncertainty of community discovery into its
posterior.

Table~\ref{tab:option-comparison} compares the four members of the framework.
gg-BFA (1f) fixes the binary loading mask and shares the residual variance
within each community.  The other three members estimate variable-specific
primary loadings and residual variances but impose different restrictions on
the cross-loadings.  Thus, 1f serves as a fixed community reference, whereas
comparisons among 2f+s, 2f, and mf isolate the effect of increasing
cross-loading flexibility.

\begin{table}[!htbp]
\centering
\caption{Loading and residual-variance restrictions for the four gg-BFA
members.  The suffix describes the loading support for a single variable and
does not denote the total number of factors \(K\).  We use \(\kappa=0.10\) in
the experiments unless stated otherwise.}
\label{tab:option-comparison}
\small
\setlength{\tabcolsep}{3pt}
\begin{tabular}{@{}lP{0.28\textwidth}P{0.20\textwidth}P{0.29\textwidth}@{}}
\toprule
Option & Loading structure & Residual variance & Role in the framework\\
\midrule
gg-BFA (1f) & Primary loading fixed at one; all cross-loadings fixed at zero
& Shared within each primary community
& Fixed community summary and internal reference\\
gg-BFA (2f+s) & At most one cross-loading per row and posterior draw; at most a fraction
\(\kappa\) of the rows in community \(h\) have a cross-loading
& Variable-specific
& Sparse cross-community overlap with a community-level budget\\
gg-BFA (2f) & At most one cross-loading per row and posterior draw
& Variable-specific
& Variable-specific overlap with one secondary community\\
gg-BFA (mf) & Any subset of nonprimary cross-loading components may be active
& Variable-specific
& Variable-specific overlap with multiple secondary communities\\
\bottomrule
\end{tabular}
\end{table}

\FloatBarrier

\subsection{Interconnected-community graph structure}
\label{sec:icn-graph}

Let \(\mathcal G=(\mathcal V,\mathcal E,W)\) be a weighted undirected graph for
\(p_0\) observed variables, where \(\mathcal V=\{1,\ldots,p_0\}\) and the
symmetric matrix \(W=(w_{jr})\) records pairwise associations, with
\(w_{jj}=0\).  A community partition consists of \(K\) disjoint, nonempty
subsets \(\mathcal V_1,\ldots,\mathcal V_K\).  Define
\begin{equation}
  \mathcal V_s=\biguplus_{k=1}^K\mathcal V_k,
  \qquad
  \mathcal V_0=\mathcal V\setminus\mathcal V_s,
  \label{eq:icn-partition}
\end{equation}
Variables in \(\mathcal V_s\) are retained for community-oriented factor
modeling, whereas \(\mathcal V_0\) contains variables that were not assigned
to any retained community and is not included in the current confirmatory fit.
These variables are not converted into singleton factors.  Each induced
subgraph \(\mathcal G_k=(\mathcal V_k,\mathcal E_k)\) forms a dense
within-community backbone.  The communities need not be independent.  For
\(k\ne\ell\), define the between-community edge set
\begin{equation}
  \mathcal E_{k\ell}^{*}
  =\{(j,r)\in\mathcal E:j\in\mathcal V_k,
                         r\in\mathcal V_\ell\}.
  \label{eq:icn-cross-edges}
\end{equation}
Communities \(k\) and \(\ell\) are interconnected when
\(\mathcal E_{k\ell}^{*}\ne\varnothing\) \citep{wu2021icn}.  Each variable
retains one primary community, although some variables may connect that
community to others through between-community edges.  This structure differs
from a block-diagonal graph, which excludes such edges, and from an
overlapping-community representation, in which a variable has multiple
primary memberships.

The graph partition determines the primary factor labels but does not fix the
complete loading matrix.  Community \(\mathcal V_k\) defines factor \(k\).
The factor covariance \(\Phi\) represents dependence between the
community-level factors, whereas cross-loadings allow individual variables to
carry signal associated with a nonprimary community.  Consequently, a
posterior cross-loading is not a one-to-one estimate of a particular edge in
\(\mathcal E_{k\ell}^{*}\).

\subsection{Estimating the primary communities by dense-subgraph search}
\label{sec:icons-estimation}

We estimate the primary communities by the greedy dense-subgraph search of SCFA \citep{wu2022adsd,yang2024scfa}.  Let
\(X^{\mathrm{fit}}\in\mathbb R^{n_{\mathrm{fit}}\times p_0}\) denote the
centered and scaled observations available when the partition is estimated.
This is the full analysis sample for descriptive fitting and the training
sample for held-out evaluation.  We form the sample correlation matrix
\(\widehat W=(\widehat w_{jr})\) and set its diagonal to zero.  For a candidate
percentile \(q\in(0,100)\), let
\begin{equation}
  c_q=\operatorname{Quantile}_{q/100}
  \{\widehat w_{jr}:1\le j<r\le p_0\}.
  \label{eq:icons-percentile-cutoff}
\end{equation}
The thresholded adjacency matrix retains positive correlations that reach this
cutoff:
\begin{equation}
  \widehat w_{jr}^{(q)}
  =\widehat w_{jr}\Ind{\widehat w_{jr}\ge c_q,\ \widehat w_{jr}>0},
  \qquad j\ne r,
  \label{eq:icons-threshold}
\end{equation}
For a candidate vertex set \(C\subseteq\mathcal V\), define its generalized
weighted density by
\begin{equation}
  D_{q,\lambda_{\mathrm{den}}}(C)
  =\frac{\displaystyle
     \sum_{j<r;\,j,r\in C}\widehat w_{jr}^{(q)}}
    {|C|^{2\lambda_{\mathrm{den}}}},
  \qquad \lambda_{\mathrm{den}}\ge0.
  \label{eq:icons-density}
\end{equation}
The parameter \(\lambda_{\mathrm{den}}\) controls the penalty on community
size.  For a fixed pair \((q,\lambda_{\mathrm{den}})\), the algorithm applies
minimum-weighted-degree peeling within every thresholded connected component
containing at least \(m_{\min}\) variables.  Along each peeling path, it retains
the intermediate vertex set with the largest generalized density.  The
highest-scoring candidate is recorded as the next community, and its vertices
are removed before the procedure is repeated.

Discovery stops when no eligible positive-edge component remains or when the
number of detected communities reaches a prespecified upper bound
\(K_{\max}\).  Thus, \(K_{\max}\) is a cap, and the realized number of
communities \(K\) may be smaller.  Variables that are not selected remain in
\(\mathcal V_0\).  The resulting communities define
\begin{equation}
  h_j=k\quad\Longleftrightarrow\quad
  j\in\widehat{\mathcal V}_k,
  \qquad
  A_{jk}=\Ind{h_j=k}.
  \label{eq:icons-primary-mask}
\end{equation}
The minimum block size \(m_{\min}\), the optional cap \(K_{\max}\), and the
candidate grids are specified separately for each experiment.

Running the dense-subgraph search for every tuning pair \((q,\lambda_{\mathrm{den}})\) in a
prespecified grid \(\mathcal H\) gives a finite collection of nested community
paths: along each path, the first \(K\) communities form a candidate partition
with \(K\) communities.  For the exploratory step the paths are run to
their natural end without a cap.  Section~\ref{sec:selection} describes how
one path and the number of communities are chosen from this collection.

The dense-subgraph search and the confirmatory step play different roles.  The search groups the variables and assigns
each variable to one primary community.  The confirmatory step uses this community assignment
to define the variable's primary factor.  Individual graph edges are not
converted directly into cross-loadings.  Instead, after the primary factors
have been defined, the confirmatory step uses the observed data to determine whether a
variable also loads on factors associated with other communities.  The same
procedure can begin with a community partition supplied by scientific
knowledge rather than estimated by the dense-subgraph search.

\subsection{Exploratory step: early-stopped community selection}
\label{sec:selection}

The forward stepwise search (Algorithm~\ref{alg:selection}) chooses one candidate path and the number of
communities by comparing one-factor-per-community models fitted to all \(p_0\)
variables.

\paragraph{Selection model.}
For a candidate partition with communities \(\mathcal V_1,\ldots,\mathcal V_K\)
and remaining variables \(\mathcal V_0\), the selection model assigns one factor
with unit loadings to each community and treats \(\mathcal V_0\) as a background
block without a factor:
\begin{equation}
  x_i\sim\Norm_{p_0}\bigl\{0,\Omega_K(\Phi,v)\bigr\},
  \qquad
  \Omega_K(\Phi,v)=A_K\Phi A_K^\top+\diagop\bigl(v_{h_1},\ldots,v_{h_{p_0}}\bigr),
  \label{eq:selection-model}
\end{equation}
where \(A_K\) is the \(p_0\times K\) membership matrix, \(h_j\in\{0,\ldots,K\}\)
is the community of variable \(j\), and \(v_h\) is the residual variance shared
by the variables in \(\mathcal V_h\).  The priors are
\(\Phi\sim\IW_K(K+2,I_K/2)\) and, independently, \(v_h\sim\IG(2,1/2)\).
Unlike the confirmatory step, which is fitted to the community-assigned variables only, the
selection model keeps every variable, so that models with different \(K\)
describe the same data.

\paragraph{Variational posterior and ELBO.}
For a learning rate \(\alpha\in(0,1)\), the \(\alpha\)-fractional posterior of
the selection model raises the likelihood to the power \(\alpha\),
\(\Pi_{K,\alpha}(d\Phi,dv\mid X)\propto p(X\mid\Phi,v)^\alpha\,\Pi_K(d\Phi,dv)\),
where the factor scores are integrated out of the likelihood and \(\Pi_K\) is
the prior above.  The search approximates it by the \(\alpha\)-fractional
variational posterior \(\widehat Q\), the maximizer of the \(\alpha\)-fractional
evidence lower bound over a tractable family \(\mathcal Q_K\), and compares
candidate partitions by the attained maximum,
\begin{equation}
  \mathrm{ELBO}=\max_{Q\in\mathcal Q_K}
  \Bigl[\alpha\,\E_Q\bigl\{\log p(X\mid\Phi,v)\bigr\}-\KL\bigl(Q\,\Vert\,\Pi_K\bigr)\Bigr],
  \label{eq:selection-elbo}
\end{equation}
where \(\mathcal Q_K\) contains the products of an inverse-Wishart
distribution for \(\Phi\) and inverse-gamma distributions for the \(v_h\).
Maximizing the same objective over all distributions would return the
fractional posterior itself, with maximum
\(\log\int p(X\mid\Phi,v)^\alpha\,\Pi_K(d\Phi,dv)\), the logarithm of the
fractional marginal likelihood; restricting the maximization to
\(\mathcal Q_K\) gives its variational approximation \(\widehat Q\), and the
ELBO is a lower bound on this logarithm.  Larger values are better.  The
covariance estimate of a fitted model is computed from \(\widehat Q\).  The
learning rate balances the likelihood against the prior.  We set \(\alpha=\alpha_0\), the learning rate of
\citet{lyddon2019llb} evaluated at the null model in which all variables share
one variance; it is approximately the effective number of variables divided by
\(p_0\) and is computed before any community is fitted
(Appendix~\ref{app:selection-implementation}).

\paragraph{Search.}
Every distinct partition formed by the first \(K\) communities of a candidate path is a candidate with \(K\) communities.  At each \(K\), the search fits every
candidate with \(K\) communities and keeps the largest ELBO; the partition
that attains it may change from one \(K\) to the next.  The search stops at the
first \(K\) at which this largest ELBO decreases and returns the best visited
model (Algorithm~\ref{alg:selection}).  Models with more communities than the
stopping point are never fitted.  The selected partition defines the primary
communities used by the confirmatory step below.

\paragraph{Two-stage acceleration.}
Fitting every candidate at every \(K\) can be costly for large panels.  The
two-stage version (Algorithm~\ref{alg:two-stage}) first compares the
candidates at a common reference size and keeps one path, a single empirical-Bayes choice of the tuning pair; it then moves along the kept path and
stops at the first decrease of the ELBO.  On our panels it fits about an order
of magnitude fewer models, and the real-data analyses use it.  It inherits the
guarantees of Section~\ref{sec:selection-theory} only when the kept path
contains the true partition and the bias decreases sufficiently along it
(Appendix~\ref{th:sec:icons}).  Appendix~\ref{app:selection-implementation}
gives the implementation details of both versions, including the Monte Carlo
evaluation of the ELBO.

\begin{algorithm}[!htbp]
\caption{Exploratory step of gg-BFA: forward stepwise search with early stopping}
\label{alg:selection}
\begin{algorithmic}[1]
\Require standardized data \(X\in\R^{n\times p_0}\); grid \(\mathcal H\) of tuning pairs
\State For every tuning pair \((q,\lambda_{\mathrm{den}})\in\mathcal H\), run the dense-subgraph search on \(X\) to obtain a nested
  path of communities.
\State Compute the learning rate \(\alpha_0\) from the null model.
\For{\(K=0,1,2,\ldots\)}
  \State Fit every distinct partition formed by the first \(K\) communities
    of the paths, and let \(\mathrm{ELBO}_K\) be the largest of their
    ELBOs~\eqref{eq:selection-elbo}.
  \State Stop if \(K\ge1\) and \(\mathrm{ELBO}_K<\mathrm{ELBO}_{K-1}\), or if
    no path has more than \(K\) communities.
\EndFor
\State \Return \(\widehat K=\arg\max\mathrm{ELBO}_K\) over the visited \(K\)
  (ties go to the larger \(K\)), its partition, and its fractional
  variational posterior \(\widehat Q\).
\end{algorithmic}
\end{algorithm}

\begin{algorithm}[!htbp]
\caption{Two-stage acceleration of the exploratory step}
\label{alg:two-stage}
\begin{algorithmic}[1]
\Require standardized data \(X\in\R^{n\times p_0}\); grid \(\mathcal H\) of tuning pairs
\State Run the dense-subgraph search and compute \(\alpha_0\) as in Algorithm~\ref{alg:selection}.
\State \textbf{Stage 1.} Let \(K_{\mathrm{ref}}\) be the length of the shortest
  path.  Among the distinct partitions formed by the first \(K_{\mathrm{ref}}\)
  communities of the paths, keep the path whose partition has the largest
  ELBO.
\State \textbf{Stage 2.} Along the kept path, compute the ELBOs
  \(\mathrm{ELBO}_0,\mathrm{ELBO}_1,\ldots\) of its first \(K\) communities
  and stop at the first \(K\) with \(\mathrm{ELBO}_K<\mathrm{ELBO}_{K-1}\), or
  at the end of the path.
\State \Return the \(K\) with the largest ELBO among the layers visited in
  stage~2 (ties go to the larger \(K\)), its partition, and its fractional
  variational posterior \(\widehat Q\).
\end{algorithmic}
\end{algorithm}

\subsection{Theoretical guarantees}
\label{sec:selection-theory}

Two results support the exploratory step, and both extend, under the conditions of
Appendix~\ref{th:app:highdim}, to a number of variables that grows much faster than the sample size.  Theorem~\ref{thm:selection} shows that, with probability tending
to one, the search stops at most one step after the true number of communities and returns the true
partition.  Theorem~\ref{thm:covariance} shows that the variational posterior of the covariance then
contracts around the truth at the parametric rate up to a logarithmic factor.

We analyze the exploratory step for Gaussian data.  The analysis adapts the early-stopping
theory of \citet{ohn2026esa}; complete statements and proofs are given in
Appendix~\ref{app:selection-theory}.  The theorems concern independent
Gaussian observations from the selection model~\eqref{eq:selection-model},
a fixed learning rate, fixed candidate paths, and exact maximization of the
ELBO.  Appendix~\ref{th:sec:icons} discusses what lies outside this setting:
standardization, the data-dependent paths and \(\alpha_0\), numerical
approximation, and the two-stage version.  Write \(\Omega_*\) for the true
covariance, \(\widehat Q\) for the variational posterior of the selected
model, and \(d_*=(K_*+1)(K_*+2)/2\) for the number of parameters of the true
model (one fewer when its background is empty).

\begin{theorem}[Selection of the community structure]\label{thm:selection}
Suppose that \(\Omega_*\) has the form~\eqref{eq:selection-model} with
\(K_*\ge1\) communities, each containing at least two variables, and a
positive-definite factor covariance; that the first \(K_*\) communities of
some candidate path form the true partition; and that the bias of the best
candidate decreases sufficiently with each added community, in the sense of
Assumption~(W) in Appendix~\ref{th:sec:exact}.  If \(p_0\) and the paths do
not change with \(n\), then, with probability tending to one, the search stops at
\(K_*+1\), or at \(K_*\) if no path has more communities, and returns
\(\widehat K=K_*\) together with the true partition.
\end{theorem}

\begin{theorem}[Covariance estimation]\label{thm:covariance}
Under the conditions of Theorem~\ref{thm:selection}, let \(A_n\to\infty\)
and \(r_n=\{d_*\log n/(n\alpha(1-\alpha))\}^{1/2}\).  Then
\[
  \widehat Q\Bigl(\bigl\Vert\Omega_*^{-1/2}\,\Omega\,\Omega_*^{-1/2}-I\bigr\Vert_F
  >A_nr_n\Bigr)\to0
\]
in probability, where \(\Omega=\Omega_{\widehat K}(\Phi,v)\) is the covariance
of the selected model under \(\widehat Q\).  Moreover, the posterior mean
\(\widehat\Omega=\E_{\widehat Q}\Omega\) satisfies
\(\Vert\Omega_*^{-1/2}\widehat\Omega\,\Omega_*^{-1/2}-I\Vert_F\le A_nr_n\)
with probability tending to one.
\end{theorem}

Both results rest on two deterministic curves that, with high probability,
bound the gap between \(\alpha\) times the log-likelihood of the truth and the
ELBO of every model.  The lower curve is driven by the R\'enyi divergence
between the truth and the model, so the gap of a model that misses a
community is at least of order \(n\).  The upper curve is driven by the
Kullback--Leibler divergence plus a parameter cost of order \(d_K\log n\), so
the gap of the true model is at most of order \(\log n\).  A model with more
than \(K_*\) communities can reproduce \(\Omega_*\) only with a singular
factor covariance, where the inverse-Wishart prior has little mass; this
gives a penalty at least of order \(n^{1/3}\), and the search stops at
\(K_*+1\).  The first part of Theorem~\ref{thm:covariance} then follows from a
PAC-Bayes bound for the selected model.  For the posterior mean, contraction
alone is not enough, because a little mass far from the truth can move the
mean; we use in addition that every member of the variational family
concentrates around its own mean.  For a fixed learning rate, the rate is the
parametric rate \(\sqrt{d_*/n}\) up to a factor of order \(\sqrt{\log n}\).
No estimator converges faster than the parametric rate uniformly near the
truth, even one that knows the true partition (Appendix~\ref{th:sec:exact}),
so the selected model attains the optimal rate up to this factor without
knowing \(K_*\).  Assumption~(W) is substantive: it compares a R\'enyi bias
with a Kullback--Leibler bias, and it can fail when the communities are large
relative to \(1/\alpha\), even though the true partition is a candidate.

Appendix~\ref{th:app:highdim} shows that, when the number of communities is
fixed and the candidate models satisfy a uniform separation condition, the
selection guarantee and the contraction of the variational posterior extend
to \(p_0\) that grows with \(n\) and exceeds it, with \(d_*\log n\) replaced by
\(d_*\log n+(K_*+1)\log p_0\); the guarantee for the posterior mean is proved
only in fixed dimension.  Appendix~\ref{th:app:highdim} also gives a sequence
with fixed communities and a growing background in which the condition
holds.
Appendix~\ref{th:sec:misspec} states, under an assumption on the two curves
that describes a bias decaying geometrically in \(K\), that the search stops
after about \((\log n)/\gamma\) communities and that, without knowing the decay
rate \(\gamma\), the selected model attains the parametric rate up to a
polylogarithmic factor.  For the two-stage version, the conclusions of
Theorems~\ref{thm:selection} and~\ref{thm:covariance} hold when, with
probability tending to one, the kept path contains the true partition and
Assumption~(W) holds along it (Appendix~\ref{th:sec:icons}).  Compared with
\citet{ohn2026esa}, who show that early stopping does not stop too early for
variational empirical Bayes, the finite candidate paths allow a weaker
separation condition and an additional guarantee that the search does not
stop too late.

\subsection{Probability model of the confirmatory step and identification}
\label{sec:bfa-target}

After community discovery, we retain the \(p=|\mathcal V_s|\) assigned
variables and relabel them as \(1,\ldots,p\).  Let \(x_i\in\R^p\) denote the
centered and scaled measurements for subject \(i\).  Conditional on the primary
community labels \(h=(h_1,\ldots,h_p)\), the confirmatory step assumes
\begin{equation}
  x_i\mid f_i,\Lambda,\psi
  \sim \Norm(\Lambda f_i,\diagop(\psi)),
  \qquad
  f_i\sim\Norm(0,\Phi),
  \label{eq:bfa-target}
\end{equation}
The vector \(f_i\in\R^K\) contains the subject's scores on the \(K\)
community-oriented factors.  The loading matrix
\(\Lambda=(\lambda_1,\ldots,\lambda_p)^\top\in\R^{p\times K}\) connects these
factors to the observed variables, and \(\Phi\in\mathbb S_{++}^K\) describes
dependence among the factors.  The residual variances in
\(\psi=(\psi_1,\ldots,\psi_p)^\top\in\R_+^p\) are variable-specific for the
learned-loading members and shared within communities for 1f.

Integrating out the factor scores gives the observation covariance
\begin{equation}
  \Omega:=\operatorname{Cov}(x_i\mid\Lambda,\Phi,\psi)
  =\Lambda\Phi\Lambda^\top+\diagop(\psi).
  \label{eq:bfa-covariance-target}
\end{equation}
This is the covariance matrix estimated and evaluated in the real-data
comparisons.

For a variable \(j\) assigned to primary factor \(h_j\), we treat the
nonprimary loadings \(\eta_{jk}\), \(k\ne h_j\), as the free loading
parameters.  The primary loading is then determined by
\begin{equation}
  \lambda_j
  =e_{h_j}+\sum_{k\ne h_j}\eta_{jk}(e_k-e_{h_j}),
  \qquad
  \lambda_{j,h_j}=1-\sum_{k\ne h_j}\eta_{jk},
  \label{eq:unit-row-sum}
\end{equation}
where \(e_k\) is the \(k\)th coordinate vector.  Consequently, every loading
row satisfies
\begin{equation}
  \mathbf 1_K^\top\lambda_j=1.
  \label{eq:unit-row-sum-constraint}
\end{equation}
When all cross-loadings are zero, \(\lambda_j=e_{h_j}\), which recovers the
binary community mask.  If a single cross-loading equals \(0.2\), for example,
the corresponding primary loading becomes \(0.8\).  The loadings are signed,
so this constraint does not require them to be nonnegative.

This parameterization places the loading rows on a common scale and keeps the
no-cross-loading model aligned with the community mask.  It does not, by
itself, remove every possible factor rotation.  The community labels and
shrinkage toward the binary mask provide the additional orientation used by
the model.  A variable's primary factor therefore records its assigned
community, but it need not have the largest loading in absolute value.

The three learned-loading members use the same likelihood, factor-score model,
residual-variance prior, and factor-covariance prior.  They differ only in the
support and prior assigned to the cross-loadings.  Their common priors are
\begin{align}
  \psi_j&\sim\operatorname{IG}(a_0,b_0),
  &
  \Phi&\sim\operatorname{IW}_K
  \left(\nu_\Phi^{(0)},\Psi_\Phi^{(0)}\right),
  \label{eq:gb-model}\\
  \nu_\Phi^{(0)}&=K+1+\nu_0,
  &
  \Psi_\Phi^{(0)}&=\nu_0\tau_f I_K.
  \label{eq:full-phi-prior}
\end{align}
Under this parameterization, \(\E(\Phi)=\tau_fI_K\).  Thus, the prior centers
the factor covariance on independent factors with common variance \(\tau_f\),
while the posterior remains free to estimate dependence between factors.

\paragraph{The fixed-loading member: gg-BFA (1f).}
The fixed-loading member, gg-BFA (1f), imposes
\begin{equation}
  \eta_{jk}\equiv0\quad(k\ne h_j),\qquad
  \Lambda=A,\qquad \psi_j=v_{h_j},\qquad
  v_h\sim\operatorname{IG}(a_0,b_0),
  \label{eq:vb-scfa-restriction}
\end{equation}
Each variable therefore loads one on its primary factor and zero on all other
factors, while variables in the same community share a residual variance.  The
factor scores, community-specific residual variances, and the full factor
covariance \(\Phi\) remain unknown and are estimated from the data.  Unlike
frequentist CFA estimators, 1f provides variational posterior distributions for
these quantities.  Because it also ties residual variances within communities,
1f is not obtained simply by turning off the cross-loadings in one of the
learned-loading members.  We retain SCFA and ML-CFA as separate frequentist
comparison methods.

\subsection{Cross-loading specifications}
\label{sec:cross-loading-options}

For each nonprimary factor \(k\ne h_j\), let \(\gamma_{jk}=1\) indicate that
variable \(j\) has an active cross-loading on factor \(k\).  The three
learned-loading members use the same signed row-sum parameterization but impose
different restrictions on these indicators.  The mf member permits multiple
active cross-loadings in one row, 2f permits at most one, and 2f+s adds a
community-level limit so that only a sparse subset of variables within each
primary community may cross-load.

\subsubsection{gg-BFA (mf)}

gg-BFA (mf) allows any combination of the \(K-1\) nonprimary loading
components to be active.  Each component has a continuous spike-and-slab
prior:
\begin{align}
  \gamma_{jk}&\sim\Bern(\pi),\qquad k\neq h_j,\\
  \eta_{jk}\mid\gamma_{jk}
  &\sim
  \begin{cases}
    \Norm(0,\sigma_{\mathrm{slab}}^2), & \gamma_{jk}=1,\\
    \Norm(0,\sigma_{\mathrm{spike}}^2), & \gamma_{jk}=0,
  \end{cases}
  \label{eq:spike-slab}
\end{align}
Here, \(\pi\) is the prior probability that a nonprimary loading is active, and
\(\sigma_{\mathrm{spike}}^2\ll\sigma_{\mathrm{slab}}^2\).  Because the spike
variance is small but nonzero, an inactive component is strongly shrunk toward
zero rather than fixed exactly at zero.  The posterior probability
\(\Pr(\gamma_{jk}=1\mid X)\) measures the evidence for that cross-loading.
The primary loading is then obtained from the signed row-sum constraint.

\subsubsection{gg-BFA (2f)}

gg-BFA (2f) allows a variable to load on its primary factor and at most one
additional factor.  For variable \(j\), define
\(s_j\in\{0\}\cup\bigl(\{1,\ldots,K\}\setminus\{h_j\}\bigr)\), where \(s_j=0\)
denotes no cross-loading and \(s_j=k\) selects factor \(k\) as the single
cross-loading target.  For variables whose primary factor is \(h\), we use
\begin{align}
  \theta_h&\sim\operatorname{Beta}(a_\theta,b_\theta),
  &
  \omega_{h,-h}&\sim\operatorname{Dirichlet}(\alpha_\omega\mathbf 1),\\
  \Pr(s_j=0\mid\theta_h)&=1-\theta_h,
  &
  \Pr(s_j=k\mid\theta_h,\omega_h)&=\theta_h\omega_{hk},\\
  \eta_j\mid s_j=k&\sim\Norm(0,\sigma_a^2),
  &
  \lambda_j&=
  \begin{cases}
    e_{h_j}, & s_j=0,\\
    e_{h_j}+\eta_j(e_k-e_{h_j}), & s_j=k.
  \end{cases}
  \label{eq:two-loading-family}
\end{align}
The parameter \(\theta_h\) controls how often variables in community \(h\)
cross-load, while \(\omega_h\) distributes those cross-loadings across the
other communities.  The Gaussian variable \(\eta_j\) determines the magnitude
and sign of the selected cross-loading.  Equivalently,
\(\gamma_{jk}=\Ind{s_j=k}\).  By construction, every posterior draw contains
at most one cross-loading in each row.  A posterior mean row may contain small
entries in several nonprimary columns because it averages over mutually
exclusive target states.

\subsubsection{gg-BFA (2f+s)}

gg-BFA (2f+s) retains the single-target loading structure of 2f and adds a
community-level sparsity constraint.  Let
\(\mathcal B_h=\{j:h_j=h\}\), \(g_j=\Ind{s_j\ne0}\), and define
\begin{equation}
  L_h=\left\lfloor\kappa|\mathcal B_h|\right\rfloor,
  \qquad
  \sum_{j\in\mathcal B_h}g_j\le L_h,
  \qquad 0\le\kappa\le1.
  \label{eq:two-sparse-budget}
\end{equation}
Thus, \(L_h\) is the maximum number of variables in community \(h\) that may
cross-load in one posterior draw.  The gate vector has the truncated Bernoulli
prior
\begin{equation}
  p(g_{\mathcal B_h})\propto
  \Ind{\sum_{j\in\mathcal B_h}g_j\le L_h}
  \prod_{j\in\mathcal B_h}\pi_g^{g_j}(1-\pi_g)^{1-g_j}.
  \label{eq:two-sparse-gate-prior}
\end{equation}
The parameter \(\pi_g\) controls the prior preference for cross-loading before
the community-level constraint is applied.  Conditional on \(g_j=1\), the
target community and cross-loading magnitude follow the same categorical and
Gaussian distributions as in 2f.  The parameter \(\kappa\) limits the number
of cross-loading rows rather than the magnitude of their loadings.  We use
\(\kappa=0.10\) unless stated otherwise; the dataset-specific settings are
reported in Section~\ref{sec:experiments}.

At the level of individual loading rows, the four structures form the nested
sequence
\begin{equation}
  \mathcal L_{\mathrm{1f}}
  \subset \mathcal L_{\mathrm{2f+s}}
  \subset \mathcal L_{\mathrm{2f}}
  \subset \mathcal L_{\mathrm{mf}}.
  \label{eq:loading-support-nesting}
\end{equation}
For 2f and 2f+s, setting every row to the state \(s_j=0\) recovers the binary
loading mask.  For mf, setting \(\gamma_{jk}=0\) is not sufficient because its
continuous spike has nonzero variance; the cross-loading amplitudes must also
be fixed exactly at zero.  This nesting describes the loading structures only.
The complete 1f model additionally differs by sharing residual variances within
each community.  Moving from 1f to mf therefore increases the range of
cross-community covariance patterns that the loading matrix can represent.  Figure~\ref{fig:method}f shows the four
loading supports schematically.

\FloatBarrier

\subsection{Variational inference for the confirmatory step}
\label{sec:mf-vb}

Posterior inference is based on a variational approximation.  Let \(\theta\)
collect the latent variables and unknown parameters of a loading
specification.  We choose \(q\) from a tractable family \(\mathcal Q\) to
minimize the reverse Kullback--Leibler divergence from the exact posterior,
\begin{equation}
  q^*
  =\arg\min_{q\in\mathcal Q}\KL\bigl\{q(\theta)\,\Vert\,p(\theta\mid X)\bigr\}
  =\arg\max_{q\in\mathcal Q}\ELBO(q),
  \quad
  \ELBO(q)=\E_q\{\log p(X,\theta)\}-\E_q\{\log q(\theta)\},
  \label{eq:vi-objective}
\end{equation}
where the evidence lower bound \(\ELBO(q)\) satisfies
\(\log p(X)=\ELBO(q)+\KL\{q(\theta)\,\Vert\,p(\theta\mid X)\}\).  The family
factorizes into blocks,
\begin{equation}
  q(\theta)=\prod_{g=1}^G q_g(\theta_g),
  \label{eq:mean-field}
\end{equation}
for the factor covariance, the subject-specific factor scores, the residual
variances, and the loading parameters; a block keeps the internal dependence
required by its loading specification.  We optimize \(q\) by coordinate
ascent: holding the other blocks fixed, the update for block \(g\) is
\(\log q_g^*(\theta_g)=\E_{q_{-g}}\{\log p(X,\theta)\}+\text{constant}\),
and iterating these updates increases \(\ELBO(q)\) until the convergence
criterion is met.

\subsubsection{Posterior factorization}

For gg-BFA (1f), the mean-field variational family in
Eq.~\eqref{eq:mean-field} separates the posterior blocks for the factor
covariance, subject-specific factor scores, and community-specific residual
variances.  The loading matrix is fixed at the binary community mask \(A\),
and every nonprimary loading indicator is fixed at zero.  Its variational
family is therefore
\begin{equation}
  q_{\mathrm{1f}}(F,\Phi,v,\Lambda,\gamma)
  =q(\Phi)\prod_{i=1}^n q(f_i)\prod_{h=1}^K q(v_h)
    \delta_A(\Lambda)\delta_0(\gamma),
  \label{eq:fixed-mf-family}
\end{equation}
Here, \(\delta_A(\Lambda)\) and \(\delta_0(\gamma)\) are point masses at the
fixed loading matrix and inactive cross-loading states.  The score factor
\(q(f_i)\) is multivariate Gaussian, \(q(v_h)\) is inverse-Gamma, and
\(q(\Phi)\) is inverse-Wishart.  Thus, fixing the loadings does not remove
posterior inference for the scores, factor covariance, or residual variances.
All variables assigned to community \(h\) share \(v_h\), so their residual
variances are tied rather than estimated separately.

The three learned-loading members use the same mean-field construction but
replace the fixed loading matrix with an option-specific loading block
\(\xi_o\):
\begin{equation}
  q_o(F,\Phi,\psi,\xi_o)
  =
  q(\Phi)
  \prod_{i=1}^n q(f_i)
  \prod_{j=1}^p q(\psi_j)
  q_o(\xi_o),
  \qquad o\in\{\mathrm{2f+s},\mathrm{2f},\mathrm{mf}\},
  \label{eq:common-mf-family}
\end{equation}
The factor covariance, subject-specific scores, and residual variances form
separate mean-field blocks. Here, \(q(f_i)=\Norm(m_i,V_f)\),
\(q(\psi_j)=\operatorname{IG}(a_j,b_j)\), and \(q(\Phi)\) is inverse-Wishart.
Unlike 1f, these models estimate a separate residual variance for every
variable. The block \(q_o(\xi_o)\) contains the loading variables and retains
the internal dependence required by the corresponding cross-loading
restriction.

For gg-BFA (mf), each possible nonprimary loading has its own inclusion
indicator and loading amplitude. The corresponding variational block
factorizes across variables and nonprimary factors:
\begin{equation}
  q_{\mathrm{mf}}(\xi_{\mathrm{mf}})
  =
  \prod_{j=1}^p\prod_{k\neq h_j}
  q(\gamma_{jk})q(\eta_{jk}\mid\gamma_{jk}).
  \label{eq:mf-family}
\end{equation}
For each candidate cross-loading,
\(q(\gamma_{jk})=\Bern(\rho_{jk})\), where \(\rho_{jk}\) is its posterior
inclusion probability. Conditional on \(\gamma_{jk}=c\),
\(q(\eta_{jk}\mid\gamma_{jk}=c)\) is Gaussian for \(c\in\{0,1\}\). This
construction preserves the dependence between the inclusion indicator and
loading amplitude within each pair, while treating different pairs as
separate mean-field blocks. The primary loading is recovered through the
signed unit-row-sum constraint in Eq.~\eqref{eq:unit-row-sum}; consequently,
the elements of a complete loading row are not variationally independent.

gg-BFA (2f) represents the cross-loading decision for each variable by a
single categorical state. For a variable whose primary community is \(h\),
the state \(s_j=0\) denotes no cross-loading, whereas \(s_j=k\) selects one
target factor \(k\ne h\). Its structured variational family is
\begin{equation}
  q_{\mathrm{2f}}(\xi_{\mathrm{2f}})
  =
  \prod_{h=1}^K\left\{
    q(\theta_h)q(\omega_h)
    \prod_{j:h_j=h}q(s_j)q(\eta_j\mid s_j)
  \right\},
  \label{eq:two-mf-family}
\end{equation}
Here, \(q(\theta_h)\) is Beta and describes the community-level propensity to
cross-load. The Dirichlet factor \(q(\omega_h)\) distributes this probability
among the \(K-1\) allowable target factors. Each \(q(s_j)\) is categorical over
the no-cross-loading state and the \(K-1\) possible targets, while
\(q(\eta_j\mid s_j)\) is Gaussian for an active target. Because the target is
represented by one categorical state, every posterior draw contains at most
one cross-loading in each row. The row therefore has \(K\) possible states
rather than the \(2^{K-1}\) subsets allowed by mf.

gg-BFA (2f+s) retains the single-target structure of 2f but adds a
community-level sparsity constraint. Let \(g_j\) indicate whether variable
\(j\) is allowed to cross-load. For the variables \(\mathcal B_h\) assigned to
primary community \(h\), the gate vector \(g_{\mathcal B_h}\) must satisfy
\[
  \sum_{j\in\mathcal B_h}g_j\le L_h,
  \qquad
  L_h=\left\lfloor\kappa|\mathcal B_h|\right\rfloor.
\]
Because this constraint couples the gates within a community, they are
retained as one structured variational block:
\begin{align}
  q_{\mathrm{2f+s}}(\xi_{\mathrm{2f+s}})
  &=\prod_{h=1}^K\left\{
    q(g_{\mathcal B_h})q(\omega_h)
    \prod_{j\in\mathcal B_h}
    q(t_j\mid g_j=1)q(\eta_j\mid t_j,g_j=1)
  \right\},
    \label{eq:two-sparse-mf-family}\\
  q(g_{\mathcal B_h})
  &\propto
    \Ind{\sum_{j\in\mathcal B_h}g_j\le L_h}
    \exp\!\left\{\sum_{j\in\mathcal B_h}\ell_jg_j\right\}.
    \label{eq:truncated-bernoulli-q}
\end{align}
Conditional on \(g_j=1\), \(t_j\) selects one of the \(K-1\) nonprimary
factors and \(\eta_j\) gives the corresponding loading amplitude. The quantity
\(\ell_j\) is the current variational log-odds that variable \(j\) cross-loads,
after averaging over its possible target factors. The overall approximation
remains mean-field across the major blocks in
Eq.~\eqref{eq:common-mf-family}, but \(q(g_{\mathcal B_h})\) does not factorize
across variables. Retaining this joint truncated-Bernoulli factor ensures that
every posterior draw satisfies the community-level budget. Its normalizing
constant and marginal gate probabilities are computed by a forward--reverse
dynamic program in \(O(p_hL_h)\) time, where \(p_h=|\mathcal B_h|\).

Across the four gg-BFA members, the score, factor-covariance, and
residual-variance updates depend on the loading distribution through its
posterior first and second moments. For 1f, these moments are fixed by the
binary community mask; for the three learned-loading members, they are
obtained from the corresponding variational block. The same update formulas
can therefore be used after substituting the appropriate loading moments.
Because factor \(k\) remains associated with the same primary community in
every model, factor-specific posterior summaries can be compared across
loading specifications. Their numerical values may nevertheless differ
because each loading structure allocates the observed covariance differently.

The resulting factor-score distributions are Gaussian, and the
factor-covariance variational factor is inverse-Wishart. Complete coordinate
updates for the factor scores, factor covariance, loadings, and residual
variances, together with computational and convergence details, are given in
Appendix~\ref{app:implementation}.

\subsection{Posterior quantities and covariance estimation}
\label{sec:posterior-outputs}

All four gg-BFA members are summarized through five posterior objects: the
subject-specific factor scores \(q(F)\), loading matrix \(q_o(\Lambda)\),
cross-loading states \(q_o(\gamma)\), residual variances \(q_o(\psi)\), and
factor covariance \(q(\Phi)\). Some of these distributions are degenerate when
the corresponding quantities are fixed by the model.

For 1f, \(q(\Lambda)=\delta_A\) and \(q(\gamma)=\delta_0\), while its residual
distribution is induced by the \(K\) community-level factors \(q(v_h)\).
Posterior intervals for the factor scores, residual variances, and factor
covariance remain available, but loading and cross-loading uncertainty is not
applicable because those quantities are fixed. The three learned-loading
members instead estimate one residual variance for each variable. For 2f,
\(\gamma_{jk}=\Ind{s_j=k}\); for 2f+s,
\(\gamma_{jk}=g_j\Ind{t_j=k}\), which also preserves the dependence induced by
the community-level sparsity budget. Primary community membership is supplied
by the graph partition and is therefore not treated as a posterior estimand.

Posterior means and marginal intervals are available in closed form for most
parameters. We use draws from the fitted inverse-Wishart distribution
\(q(\Phi)\) to obtain intervals for the factor covariances and correlations.
Loading rows are sampled jointly from their fitted variational distributions,
preserving the signed row sum and the restrictions imposed by 2f and 2f+s.
The posterior mean loading matrix can be less sparse than an individual draw
because it averages over alternative loading states.

To return from the latent representation to the covariance among the observed
variables, each posterior draw combines the loading matrix, factor covariance,
and residual variances. The loading matrix connects each variable to the
community factors, \(\Phi\) describes how those factors vary together, and
\(\psi\) accounts for the variation that remains specific to each variable.
After integrating out the subject-specific factor scores, the induced
observation covariance is
\begin{equation}
  \Omega
  =\operatorname{Cov}(x_i\mid\Lambda,\Phi,\psi)
  =\Lambda\Phi\Lambda^\top+\diagop(\psi).
  \label{eq:gb-covariance}
\end{equation}
We estimate the observation covariance by averaging these induced covariance
matrices under the fitted variational posterior:
\begin{equation}
  \widehat\Omega_o
  =\E_{q_o}\!\left\{\Lambda\Phi\Lambda^\top
    +\diagop(\psi)\right\}.
  \label{eq:gb-posterior-mean-covariance}
\end{equation}
This posterior mean covariance is used for covariance reconstruction and
Gaussian prediction.

For two distinct variables \(j\) and \(l\), the residual variances drop out
and
\[
  \Omega_{jl}=\lambda_j^\top\Phi\lambda_l.
\]
If both variables load only on their primary factors, their covariance is
determined by the corresponding entry of \(\Phi\). For \(k\ne h_j\), a
nonzero loading \(\lambda_{jk}\) allows factor \(k\) to contribute to variable
\(j\) in addition to its primary factor \(h_j\). Through
\(\lambda_j^\top\Phi\lambda_l\), this additional loading changes the covariance
between variable \(j\) and other variables according to their loadings and the
\(k\)th row of \(\Phi\). The posterior inclusion probability
\(\Pr_q(\gamma_{jk}=1)\) is the fitted probability that this additional loading
is active. The primary label \(h_j\) remains fixed at the community assigned by
the graph.

\subsection{Prediction from observed measurements}
\label{sec:masked-cell}

Let \(x^{\mathrm{new}}\in\mathbb R^p\) denote a test subject who was not used
for model fitting. Within this subject, we mask a set of variable indices
\(H\) and retain the remaining indices
\(O=\{1,\ldots,p\}\setminus H\) as observed. Thus,
\[
  x^{\mathrm{new}}
  =
  \begin{pmatrix}
    x_O\\
    x_H
  \end{pmatrix},
\]
where \(x_O\) is available to the prediction procedure and \(x_H\) is used
only to evaluate its predictions.

Let \(\mu\) and \(C\) denote the mean and observation covariance estimated
from the training subjects. For index sets \(A\) and \(B\), \(C_{AB}\) denotes
the submatrix of \(C\) with rows in \(A\) and columns in \(B\). Under the
Gaussian model,
\[
  x_H\mid x_O
  \sim
  \Norm\!\left(
    \mu_H+C_{HO}C_{OO}^{-1}(x_O-\mu_O),
    C_{HH}-C_{HO}C_{OO}^{-1}C_{OH}
  \right).
\]
We use the conditional mean as the point prediction and the conditional
covariance to describe its uncertainty:
\begin{align}
  \widehat x_H^{\mathrm G}
  &=\mu_H+C_{HO}C_{OO}^{-1}(x_O-\mu_O),\\
  \widehat V_H^{\mathrm G}
  &=C_{HH}-C_{HO}C_{OO}^{-1}C_{OH}.
  \label{eq:gaussian-predict}
\end{align}
This Gaussian prediction rule provides a common comparison across methods.
Bayesian methods use the posterior mean covariance in
Eq.~\eqref{eq:gb-posterior-mean-covariance}; non-Bayesian methods use their
corresponding covariance estimates. We compute the conditional moments by
linear solves. If \(C_{OO}\) cannot be factorized, the prediction is recorded
as a failure without adding a ridge term, jitter, or an eigenvalue floor.

The four gg-BFA members also provide a model-specific posterior prediction.
After fitting the global model to the training subjects, we use only
\(x_O\) to infer the new subject's factor scores.  The posterior mean of those
scores gives the point prediction for \(x_H\), while posterior draws propagate
uncertainty in the local scores, loadings, and residual variation.  The masked
values \(x_H\) are used only to evaluate the resulting predictions.  This
model-specific analysis supplements the common Gaussian comparison above;
its full formulas and sampling construction are given in
Appendix~\ref{app:local-vb-prediction}.

\section{Empirical evaluation design}
\label{sec:evaluation}

The evaluation has two parts, one for each component of the method.
Comparison~1 concerns the choice of the communities: it compares
the exploratory step of gg-BFA with three pipelines that discover a factor
structure by exploratory factor analysis and then confirm it.  Comparison~2
concerns the confirmatory fit: it holds one partition fixed and compares the
gg-BFA members with other confirmatory factor methods.  Comparison~1 uses simulations, in which the truth is known; Comparison~2 uses simulations and
five real-data panels.
Section~\ref{sec:comparison-selection} describes Comparison~1, Section~\ref{sec:comparison-arms}
Comparison~2, and Section~\ref{sec:metric-definitions} the summaries used to report them.

The real-data analyses address three questions: how well a model represents
the covariance in the complete sample, whether that covariance generalizes to
new subjects, and whether observed measurements can predict measurements
hidden for evaluation.  The first analysis is descriptive.  The latter two use
the same train--test splits, with standardization, structure discovery, and
model fitting based only on the training subjects.

\subsection{Comparison 1: choosing the structure}
\label{sec:comparison-selection}

In the simulations we run the forward stepwise search of Algorithm~\ref{alg:selection}, the version
analyzed by the theory; in the five panels we use its two-stage version
(Algorithm~\ref{alg:two-stage}), as described in Section~\ref{sec:selection}
and Appendix~\ref{app:selection-implementation}.  The output is the selected
partition, the number of communities
\(\widehat K\), and the fitted selection model, whose covariance estimate
is~\eqref{eq:selection-model} evaluated at the posterior means of \(\Phi\)
and \(v\) and covers all \(p_0\) variables.  Two of the comparison pipelines
share their first step: the number of factors is chosen by parallel analysis
\citep{horn1965parallel}, as implemented in \texttt{psych}
\citep{revelle2026psych}, and is capped at 20.  These two pipelines then fit a
maximum-likelihood exploratory factor model with that number of factors in
\texttt{psych} or \texttt{lavaan} \citep{rosseel2012lavaan}, each with its
default rotation, keep the loadings with absolute value at least \(0.30\) as
the loading pattern, and fit that pattern by maximum-likelihood confirmatory
factor analysis.  The third pipeline is the exploratory-to-confirmatory
procedure of \texttt{semopy} \citep{igolkina2020semopy}, which chooses its own
number of factors and loading pattern.  Variables without a loading above the
threshold are treated as independent of all other variables.  All settings
were fixed before any results were examined, and a pipeline that fails, for
example because the sample covariance matrix is singular or the
confirmatory model is not identified, is reported as a failure rather than
repaired.

The simulations for Comparison~1 follow two designs.  In S1, the data
follow the selection model~\eqref{eq:selection-model} with five communities
and a background block, with either \(p_0=200\) or \(p_0=2n\), so that the
dimension can exceed the sample size.  In S2, forty communities have factor
variances that decrease geometrically.  The model with the forty true
communities is exact, but at our sample sizes the search stops well before
forty, so S2 shows how many of the weakening communities are retained as
\(n\) grows; it illustrates the regime of Appendix~\ref{th:sec:misspec}
without verifying its assumption.  For each design and sample size we generate 50
data sets (Appendix~\ref{app:sim-selection-design}), and all procedures are
run on the same data sets.  As in the
real-data analyses, the variables are standardized with their sample standard
deviations and \(\alpha_0\) is computed from the data.  The theory assumes
neither, so the design with \(p_0=2n\), in which \(\alpha_0n\) stays bounded,
checks the selection in high dimension rather than the rate of
Appendix~\ref{th:app:highdim}, which we check separately under its own
assumptions (Appendix~\ref{th:sec:highdim-sim}).  We report how often
\(\widehat K\) and the partition are correct, the adjusted Rand index
\citep{hubert1985ari}, and the relative error
\(\Vert\Omega_*^{-1/2}\widehat\Omega\,\Omega_*^{-1/2}-I\Vert_F\) of the
posterior mean \(\widehat\Omega\), computed against the population
correlation matrix; Theorem~\ref{thm:covariance} bounds this error for the
exact model in fixed dimension.

We do not repeat this comparison on the five panels.  In two of them, TCGA Kidney and TCGA
miRNA, the variables outnumber the subjects (Table~\ref{tab:realdata-design-current}), so the
sample covariance matrix is singular and the maximum-likelihood pipelines of \texttt{psych} and
\texttt{lavaan} cannot be applied.  On the real data, the exploratory step chooses the partition of
the complete pipeline (Section~\ref{sec:raw-to-fit}).

\subsection{Comparison 2: confirmatory fitting on a common partition}
\label{sec:comparison-arms}

The simulations for Comparison~2 examine loading structure in settings where
the truth is known.  The first asks whether extra flexibility is useful when
the binary community mask is correct.  The second introduces cross-community
signal and compares loading restrictions of different complexity.  A
supplementary stress test removes the generating community structure
altogether.  These settings separate covariance recovery from parameter
recovery and interval calibration.

Comparison~2 asks whether heterogeneous primary loadings or cross-loadings
improve on a fixed community representation. We compare the
four gg-BFA members with SCFA \citep{yang2024scfa}, ML-CFA
\citep{joreskog1969cfa}, MFVB-CFA \citep{dang2022sem}, and PCFA-VA
\citep{jin2025pcfa}. Together, they provide a progression from fixed community
loadings, through learned primary strengths, to models that permit
cross-loadings. Within each analysis, every CFA method receives the same
primary community assignment.

Raw loading values are not directly comparable across all methods because
their identification constraints and loading scales differ. We instead
describe loading complexity by the expected number of active nonprimary
components per selected variable. This value is computed from posterior
inclusion probabilities for gg-BFA and PCFA-VA and is zero by construction
for the four models without cross-loadings. Elementwise loading recovery is
used only in simulations where the fitted and generating coordinates share
the same scale.

Held-out comparisons with maximum-likelihood exploratory factor analysis,
FactorVerse \citep{dunson2026tutorial}, and BCME \citep{lei2026bcme} at the same
number of factors are reported in Appendix~\ref{app:current-real-data}.  Unlike
the pipelines of Comparison~1, these estimators stop at an exploratory or
block-covariance fit and do not produce a confirmatory loading pattern.

\paragraph{Common partition for comparing CFA methods.}
The comparison of confirmatory fits holds the partition fixed, so that the
methods differ only in how they estimate loadings and covariance given the same
primary communities.  For this comparison the partition is obtained from the dense-subgraph search with the fixed number of communities \(K\) of Section~\ref{sec:realdata-design},
and the tuning pair \((q,\lambda_{\mathrm{den}})\) is chosen from a prespecified
grid by the following criterion.  For each candidate partition, let \(\widehat S\) denote the sample
covariance matrix, let
\(\widehat\Sigma_{\mathrm{shared}}
=A\widehat\Phi_{\mathrm{SCFA}}A^\top\) denote the covariance represented by its
fixed-loading community summaries, and let \(\widehat\Sigma_u\) denote the
corresponding residual covariance.  We evaluate
\begin{equation}
  J(q,\lambda_{\mathrm{den}})
  =\left\|\operatorname{offdiag}(\widehat\Sigma_u)\right\|_F
   +\left\|\operatorname{offdiag}
      \left(\widehat S-\widehat\Sigma_{\mathrm{shared}}\right)\right\|_F,
  \label{eq:icons-tuning}
\end{equation}
where
\(\operatorname{offdiag}(M)=M-\diagop\{\diag(M)\}\).  The first term measures
residual dependence after constructing the community summaries, whereas the
second measures covariance not represented by their shared block structure.

We select the candidate with the smallest value of \(J\).  Candidates that
produce a nonpositive residual-variance estimate are treated as invalid rather
than repaired.  Deterministic ties are resolved by preferring fewer selected
variables, fewer communities, a higher correlation percentile, and then a
smaller value of \(\lambda_{\mathrm{den}}\).  For held-out evaluation, the
correlation matrix, candidate grid, tuning criterion, and final partition are
computed from the training subjects only and then held fixed for the test
subjects.  Descriptive full-data analyses use the complete analysis sample.
This criterion only fixes a common partition for comparing CFA methods; it is
not part of gg-BFA.

\subsection{Primary and supporting evaluation summaries}
\label{sec:metric-definitions}

Each analysis is organized around one primary evaluation measure. Additional
measures describe parameter recovery, uncertainty calibration, loading
complexity, and computational cost. Table~\ref{tab:evaluation-map} summarizes
the role of each measure, and complete definitions are given in
Appendix~\ref{app:evaluation-metrics}.

\begin{table}[!htbp]
\centering
\caption{Evaluation map for the simulation and real-data analyses. Complete
definitions are given in Appendix~\ref{app:evaluation-metrics}.}
\label{tab:evaluation-map}
\small
\renewcommand{\arraystretch}{1.20}
\begin{tabular}{@{}P{0.27\textwidth}P{0.25\textwidth}P{0.40\textwidth}@{}}
\toprule
Analysis & Primary measure & Additional measures \\
\midrule
\multicolumn{3}{@{}l}{\emph{Comparison 1: choosing the structure}}\\
Structure simulations (S1, S2)
  & Correct \(\widehat K\) and partition (S1); covariance error
    \(\Vert\Omega_*^{-1/2}\widehat\Omega\,\Omega_*^{-1/2}-I\Vert_F\) (S1, S2)
  & ARI; stopping point; relative Frobenius error; runtime \\
\midrule
\multicolumn{3}{@{}l}{\emph{Comparison 2: confirmatory fitting on a common partition}}\\
Fixed-community simulation
  & Covariance \(d_F\)
  & Parameter RMSE; coverage; complexity; runtime \\
\midrule
Cross-loading simulation
  & Loading RMSE
  & Covariance \(d_F\); TPR and FDR; coverage; runtime \\
\midrule
No-community stress test
  & Off-diagonal correlation RMSE
  & Covariance \(d_F\); test NLL \\
\midrule
\multicolumn{3}{@{}l}{\emph{Real data}}\\
Complete-data covariance representation
  & Descriptive covariance \(d_F\)
  & Selected \(K\) of the exploratory step; loading complexity; runtime \\
\midrule
Held-out covariance generalization
  & Gaussian NLL per variable
  & Covariance \(d_F\); SRMR; runtime \\
\midrule
Masked-measurement prediction
  & Masked-cell RMSE
  & MAE; NLPD; coverage; interval width; runtime \\
\bottomrule
\end{tabular}
\end{table}

Within each comparison, all methods are evaluated on the same subjects,
variables, and train--test splits.  In Comparison~1, every covariance estimate
covers all \(p_0\) variables.  In Comparison~2, the primary analysis uses the
variables assigned to communities by the training-data partition, and results
extended to all retained variables, including variables not assigned to a
community, are reported separately in Appendix~\ref{app:current-real-data}.

\section{Simulation studies}
\label{sec:experiments}

Section~\ref{sec:sim-selection} reports the structure simulations of
Comparison~1, in which the communities and their number are unknown.  The two main
simulations of Comparison~2 treat the primary community labels as known and fit the
competing methods to the same generated data within each replicate, which separates the
effect of the loading restriction from errors in community discovery.  They are reported in
Appendix~\ref{sec:experiment1-current-results} and~\ref{sec:experiment2-current-results}, with
complete generating mechanisms, setting-level results, and secondary endpoints in the rest of
Appendix~\ref{app:simulation-details}; Section~\ref{sec:sim-summary} summarizes what the
simulations establish.

\subsection{Choosing the communities}
\label{sec:sim-selection}

Design S1 is an exact selection model with \(K_*=5\) communities.  The five
factors have unit variances and pairwise correlations \(0.3\), and the
residual variances make the correlation between two variables of the same
community equal to \(0.60\), \(0.50\), \(0.45\), \(0.40\), and \(0.35\) in the
five communities; the remaining variables form a background of independent
variables with unit variances.  In the fixed-dimension version, \(p_0=200\)
and the communities contain 30, 26, 22, 18, and 14 variables.  In the
high-dimensional version, \(p_0=2n\) and the communities keep the same shares
of the variables, so that they grow with \(n\).  Both versions use
\(n\in\{200,500,1000,2000\}\).  Design S2 has 40 communities of 10 variables
and 100 background variables, so that \(p_0=500\), and uses
\(n\in\{200,500,1000,2000,5000\}\).  Its factor variances decrease
geometrically, \(1.5\times0.8^{h-1}\) for community \(h\), its factor
correlations are \(0.2\), and all residual variances are one.  The model with
the 40 communities is exact, but the weakest communities cannot be detected
at these sample sizes, so S2 shows how the number of selected communities
grows with \(n\).

For each design and sample size we generate 50 data sets and run the exploratory step and its
two-stage version on the same candidate paths with the same \(\alpha_0\).  On the same
data sets we also run the three exploratory-to-confirmatory pipelines of
Section~\ref{sec:comparison-selection}.  In S1
we also fit the selection model at the true partition; the error of this
oracle is what the exploratory step would attain if the partition were known.  For S1,
Theorems~\ref{thm:selection} and~\ref{thm:covariance} predict that the search
stops at \(K=6\), returns the true partition, and that the variational posterior of the covariance
contracts at the rate \(r_n\).  Tables~\ref{tab:sim-selection} and~\ref{tab:sim-pipelines} summarize the results, and
Appendix~\ref{app:sim-selection-design} gives further numerical checks.

\begin{table}[!htbp]
\centering
\caption{Simulation results for gg-BFA in designs S1 and S2, with 50 data sets per row: the
structure selected by the exploratory step and its two-stage version.  Factor membership recovery
accuracy: proportion of data sets in which the true communities were recovered.  Covariance
difference (\(\mathrm{Cov}_F^d\)): \(\Vert\Omega_*^{-1/2}\widehat\Omega\,\Omega_*^{-1/2}-I\Vert_F\),
averaged over the data sets, for the posterior mean \(\widehat\Omega\) and the population
correlation matrix \(\Omega_*\).  \(\widehat K\): mean number of selected communities.  Oracle: the
selection model fitted at the true partition.  In S2 the 40 true communities are never all selected, so
Accuracy is not shown.  For S2 with \(n\ge2000\), the iteration limit of the optimizer of the
exploratory step was raised from 600 to 3000 because the fits did not converge within 600
iterations.}
\label{tab:sim-selection}
\small
\begin{tabular}{@{}lrrrrrrrrr@{}}
\toprule
 & & & \multicolumn{4}{c}{gg-BFA} & \multicolumn{3}{c}{Two-stage version}\\
\cmidrule(lr){4-7}\cmidrule(l){8-10}
Design & \(n\) & \(p_0\) & \(\widehat K\) & Accuracy & \(\mathrm{Cov}_F^d\) & Oracle & \(\widehat K\) & Accuracy & \(\mathrm{Cov}_F^d\)\\
\midrule
S1, \(p_0=200\) & 200 & 200 & 5.00 & 0.98 & 0.64 & 0.61 & 4.70 & 0.74 & 1.06\\
 & 500 & 200 & 5.00 & 1.00 & 0.36 & 0.36 & 4.58 & 0.62 & 1.17\\
 & 1000 & 200 & 5.00 & 1.00 & 0.28 & 0.28 & 4.64 & 0.64 & 1.10\\
 & 2000 & 200 & 5.00 & 1.00 & 0.20 & 0.21 & 4.56 & 0.56 & 1.25\\
\addlinespace
S1, \(p_0=2n\) & 200 & 400 & 5.00 & 0.98 & 0.78 & 0.77 & 4.80 & 0.84 & 1.14\\
 & 500 & 1000 & 5.00 & 1.00 & 0.74 & 0.74 & 4.86 & 0.86 & 1.29\\
 & 1000 & 2000 & 5.00 & 1.00 & 0.77 & 0.76 & 4.84 & 0.78 & 2.50\\
 & 2000 & 4000 & 5.00 & 1.00 & 0.73 & 0.73 & 4.96 & 0.68 & 5.59\\
\addlinespace
S2 & 200 & 500 & 11.70 & -- & 2.66 & -- & 7.96 & -- & 4.12\\
 & 500 & 500 & 13.98 & -- & 1.59 & -- & 13.36 & -- & 1.91\\
 & 1000 & 500 & 15.52 & -- & 1.21 & -- & 15.08 & -- & 1.30\\
 & 2000 & 500 & 16.60 & -- & 0.95 & -- & 16.62 & -- & 0.96\\
 & 5000 & 500 & 18.38 & -- & 0.75 & -- & 18.36 & -- & 0.75\\
\bottomrule
\end{tabular}

\end{table}

\begin{table}[!htbp]
\centering
\caption{Simulation results for three comparable EFA--CFA pipelines
(Section~\ref{sec:comparison-selection}) on the data sets of Table~\ref{tab:sim-selection}.
Accuracy and \(\mathrm{Cov}_F^d\) as in Table~\ref{tab:sim-selection}, for the estimated covariance
of the confirmatory model, averaged over the completed runs; a run that did not complete counts as
incorrect.  \(\widehat K\): mean number of factors.  Singular: no results, as the sample covariance
matrix is singular (\(p_0\ge n\)), so the program was not run; not estimable: a factor had fewer
than two indicators; NA: numerical failure in every run.  In S2 Accuracy is not shown, as in Table~\ref{tab:sim-selection}.  \texttt{psych} and
\texttt{lavaan} gave identical results and share one set of columns.}
\label{tab:sim-pipelines}
\small
\begin{tabular}{@{}lrrrrrrrr@{}}
\toprule
 & & & \multicolumn{3}{c}{\texttt{psych}, \texttt{lavaan}} & \multicolumn{3}{c}{\texttt{semopy}}\\
\cmidrule(lr){4-6}\cmidrule(l){7-9}
Design & \(n\) & \(p_0\) & \(\widehat K\) & Accuracy & \(\mathrm{Cov}_F^d\) & \(\widehat K\) & Accuracy & \(\mathrm{Cov}_F^d\)\\
\midrule
S1, \(p_0=200\) & 200 & 200 & \multicolumn{3}{c}{singular} & 7.38 & 0.00 & 1.87\\
 & 500 & 200 & 5.00 & 1.00 & 0.79 & 5.46 & 0.34 & 0.90\\
 & 1000 & 200 & 5.00 & 1.00 & 0.56 & 5.08 & 0.92 & 0.57\\
 & 2000 & 200 & 5.00 & 1.00 & 0.40 & 5.00 & 1.00 & 0.40\\
\addlinespace
S1, \(p_0=2n\) & 200 & 400 & \multicolumn{3}{c}{singular} & 9.18 & 0.00 & 2.91\\
 & 500 & 1000 & \multicolumn{3}{c}{singular} & 6.15 & 0.18 & 2.03\\
 & 1000 & 2000 & \multicolumn{3}{c}{singular} & \multicolumn{3}{c}{NA}\\
 & 2000 & 4000 & \multicolumn{3}{c}{singular} & \multicolumn{3}{c}{NA}\\
\addlinespace
S2 & 200 & 500 & \multicolumn{3}{c}{singular} & 20.74 & -- & 6.68\\
 & 500 & 500 & \multicolumn{3}{c}{singular} & 10.80 & -- & 2.41\\
 & 1000 & 500 & \multicolumn{3}{c}{not estimable} & 9.66 & -- & 1.95\\
 & 2000 & 500 & \multicolumn{3}{c}{not estimable} & 10.04 & -- & 1.70\\
 & 5000 & 500 & \multicolumn{3}{c}{not estimable} & 10.80 & -- & 1.48\\
\bottomrule
\end{tabular}

\end{table}

In S1, the exploratory step behaved as the theory predicts.  It returned the true partition
whenever that partition was among the candidates, which failed only in
one data set at \(n=200\) in each version, as can happen in small samples; it
always stopped at \(K=6\); and its error was essentially that of the oracle.
With \(p_0=200\), the error decreased at the parametric rate up to a logarithmic factor, consistent with
Theorem~\ref{thm:covariance}.  With \(p_0=2n\), the selection was equally
accurate, but the error did not decrease.  This comes from the sample
standardization, which the theory does not include: the variables of a
community share the sampling error of their factor variance, and dividing by
the sample standard deviations adds to the squared error a term proportional
to the community sizes divided by \(n\), which stays constant when the
communities grow with \(n\).  This version therefore checks the selection
when \(p_0>n\).  The rate in growing dimension is checked under the
assumptions of Appendix~\ref{th:app:highdim}, on the original scale and with
a fixed \(\alpha\), in Appendix~\ref{th:sec:highdim-sim}, where the
contraction decreases in proportion to the rate of
Corollary~\ref{th:cor:rate-p}.

In S2, the number of selected communities increased steadily with \(n\),
roughly linearly in \(\log n\), the search always stopped one community after
the selected number, and the error decreased with \(n\).  This is the
behavior that Corollary~\ref{th:cor:geom} describes for a bias that decays
geometrically in \(K\), and the growth is close to the rate implied by the
stopping point in its proof.

The two-stage version returned the true partition only when the path kept by
its first stage contained it.  In S1 this happened in only part of the data
sets, and not more often for larger \(n\); otherwise the two-stage version
usually returned too few communities, with a much larger error.  In S2 it
selected fewer communities than the search of Algorithm~\ref{alg:selection} at the smallest \(n\) and almost the
same number at the largest.  Its advantage is computation: it fitted far
fewer candidate models and took seconds rather than minutes on one GPU.

We compare gg-BFA with the three EFA--CFA pipelines of
Section~\ref{sec:comparison-selection}, run on the same data sets (Table~\ref{tab:sim-pipelines}).  For \texttt{psych} and \texttt{lavaan}, the number of factors is selected by
parallel analysis, rotated loadings with absolute value of at least \(0.30\) are retained, and the
resulting structure is fitted by maximum-likelihood CFA; \texttt{psych} and
\texttt{lavaan} gave identical results in every data set of S1 and S2.  For \texttt{semopy},
the number of factors and the loading structure are both determined by its
built-in exploratory procedure.

When \(n>p_0\) in S1, \texttt{psych} and \texttt{lavaan} recovered the true
structure in every data set, but their covariance error was about twice that
of gg-BFA.  This is partly expected, since the data in S1 are generated from the
1f model that the exploratory step fits, whereas the pipelines estimate one loading per
variable.  \texttt{semopy} recovered the true structure more often as \(n\)
increased.  When \(p_0\ge n\), including \(n=p_0=200\), the sample covariance
matrix is singular, so the maximum-likelihood pipelines of \texttt{psych} and \texttt{lavaan}
cannot be applied.
In these settings of S1, \texttt{semopy} either selected too many factors or failed.  In S2,
the true model has more communities than can be detected, so the model fitted
by the exploratory step is misspecified.  Even in this case, \texttt{semopy}, the only pipeline
that could be run, had an error 1.5 to 2.5 times larger than that of gg-BFA.

\subsection{What the simulations establish}
\label{sec:sim-summary}

For the choice of the communities, the exploratory step did what the theory predicts when the
truth is an exact selection model: it returned the true partition whenever
that partition was a candidate, stopped one community later, and estimated
the covariance as well as a fit at the known partition.  When the communities
weaken gradually, the number it selects grows with \(\log n\) at about the
rate implied by the proof of Corollary~\ref{th:cor:geom}.  The two-stage
version saves most of the computation but can lose the true partition in its
first stage.  Among the EFA--CFA pipelines, \texttt{psych} and \texttt{lavaan} cannot
be applied when \(p_0\ge n\), and \texttt{semopy} rarely recovered the structure of S1
in that case.  When \(n>p_0\), \texttt{psych} and \texttt{lavaan} recovered the structure
of S1 with about twice the covariance error of gg-BFA, and in S2 \texttt{semopy} had 1.5 to
2.5 times the error of gg-BFA.

For the confirmatory fit, the simulations of Comparison~2
(Appendix~\ref{sec:experiment1-current-results} and~\ref{sec:experiment2-current-results}) show
that the
appropriate loading restriction depends on how variables connect across
communities. When the binary community
mask is correct, additional loading flexibility adds complexity without
improving covariance recovery. When variables carry cross-community signal, a
restriction that matches the form of overlap produces a simpler loading
representation with little loss in covariance accuracy, whereas an overly
restrictive model omits genuine signal. The supplementary stress test further
shows that unrestricted factor methods can be preferable when the data have no
meaningful primary-community structure.

These conclusions about loading structure are separate from uncertainty
calibration. Mean-field inference can recover factor scores and covariance
patterns while producing intervals that are too narrow for active
cross-loadings. Variational bagging improves these intervals but does not
restore nominal coverage uniformly. Covariance recovery and predictive
likelihood may also favor different estimators, motivating the separate
real-data analyses of covariance representation, generalization, and
conditional prediction.

\FloatBarrier

\section{Real-data applications: from covariance maps to out-of-sample behavior}
\label{sec:applications}

In real data, the latent factors and population covariance are unknown, so we
evaluate the consequences of each loading structure at the level of the
observed variables. The complete-sample analysis describes covariance
representation, the held-out analysis evaluates the joint distribution in new
subjects, and the masked-measurement analysis evaluates conditional
prediction. A method can perform well for one of these targets without being
preferred for the others.  Section~\ref{sec:realdata-design} describes the panels and the validation
design.  Sections~\ref{sec:r1-covariance-representation}--\ref{sec:r3-conditional-prediction} report
the three targets in turn, Section~\ref{sec:realdata-computation-boundary} the computation, and
Section~\ref{sec:raw-to-fit} the complete pipeline on the full data.

\subsection{Data panels and validation design}
\label{sec:realdata-design}

The applications span one metabolomics panel, two molecular panels, and two
imaging panels (Table~\ref{tab:realdata-design-current}). Within each panel,
the four gg-BFA members, SCFA, ML-CFA, MFVB-CFA, and PCFA-VA use the same
primary community labels. The partition of each panel has a fixed number of communities \(K\), obtained by the
dense-subgraph search.  We chose \(K\) small enough for most methods to run, so it is not an estimate
of the true latent dimension.

Data for the ADNI Nightingale panel were obtained from the Alzheimer's Disease Neuroimaging
Initiative (ADNI) database (\url{adni.loni.usc.edu}).  The ADNI was launched in 2003 as a
public-private partnership, led by Principal Investigator Michael W. Weiner, MD.  The primary goal
of ADNI has been to test whether serial magnetic resonance imaging (MRI), positron emission
tomography (PET), other biological markers, and clinical and neuropsychological assessment can be
combined to measure the progression of mild cognitive impairment (MCI) and early Alzheimer's
disease (AD).
The ABCD panel uses NBDC Release 7.0 of the Adolescent Brain Cognitive Development Study (DOI
\url{https://doi.org/10.82525/8f3w-5260}).

\begin{table}[!htbp]
\centering
\caption{Primary real-data CFA design. The last two columns report the number \(p_s\) of
variables assigned to the \(K\) communities: on the full data, and the range across the five
training folds.}
\label{tab:realdata-design-current}
\small
\begin{tabular}{@{}lrrrrr@{}}
\toprule
Panel & Subjects & Variables & \(K\) & \(p_s\), full data & \(p_s\), training folds\\
\midrule
ADNI Nightingale & 1,646 & 245 & 8 & 172 & 166--172\\
\midrule
TCGA Kidney & 712 & 1,777 & 6 & 1,363 & 1,355--1,479\\
\midrule
TCGA miRNA & 680 & 1,340 & 8 & 889 & 847--875\\
\midrule
ABCD ses-00A & 6,577 & 3,905 & 20 & 3,114 & 3,107--3,168\\
\midrule
UKB baseline IDP & 49,479 & 2,142 & 8 & 1,478 & 1,479--1,562\\
\bottomrule
\end{tabular}
\par\smallskip
\begin{minipage}{\linewidth}
\footnotesize
The eight methods are gg-BFA (1f), gg-BFA (2f+s), gg-BFA (2f),
gg-BFA (mf), SCFA, ML-CFA, MFVB-CFA, and PCFA-VA.
All eight enter the full-data analysis and the first three held-out panels;
PCFA-VA is omitted from ABCD and UKB held-out analyses because of its
computational cost.
\end{minipage}
\end{table}

The complete-sample analysis uses all available subjects. The covariance-
generalization and masked-measurement analyses share five frozen subject folds;
centering, scaling, community discovery, and model fitting use the training
subjects only. ABCD folds are family-disjoint and site-stratified. Within each
test subject, the prediction analysis hides the same rotating 20\% of variables
for every method.

The primary comparison uses the same community-selected variables within each
analysis. Unassigned variables are not converted into singleton factors;
extensions to all retained variables are reported separately in the Appendix.
All estimators use P100 GPUs, and \(K\) is fixed without using held-out outcomes.  Even at
these values, PCFA-VA, whose cost grows with the number of factors, could not be run on the held-out
folds of ABCD and UKB.

\subsection{Covariance representation and loading complexity}
\label{sec:r1-covariance-representation}

The complete-sample comparison asks how closely each loading structure can
represent the observed covariance. Its primary measure, \(d_F\), is descriptive:
it compares a fitted covariance with the sample covariance from the same
subjects and does not measure population recovery. Table~\ref{tab:cfa-r1-kidney}
uses Kidney to place this reconstruction error beside loading complexity and
GPU computation time; complete results for all panels are reported in the
Appendix.

\begin{table}[!htbp]
\centering
\caption{TCGA Kidney full-data covariance representation, expected active
cross-loadings per selected variable, and GPU estimator time. All methods
use the same six communities and selected variables. $d_F$ is distance to the
sample covariance; $\dagger$ marks an unconverged last iterate.}
\label{tab:cfa-r1-kidney}
\small
\begin{tabular}{@{}lrrr@{}}
\toprule
Method & $d_F$ & Cross/row & GPU seconds \\
\midrule
SCFA & 0.366 & 0.000 & 0.196 \\
ML-CFA & 0.362 & 0.000 & 0.601 \\
MFVB-CFA$^{\dagger}$ & 0.674 & 0.000 & 5.032 \\
PCFA-VA & 0.066 & 2.519 & 20.742 \\
gg-BFA (1f) & 0.366 & 0.000 & 1.018 \\
gg-BFA (2f+s) & 0.346 & 0.098 & 2.467 \\
gg-BFA (2f) & 0.258 & 1.000 & 1.635 \\
gg-BFA (mf) & 0.087 & 3.065 & 2.662 \\
\bottomrule
\end{tabular}

\end{table}

Kidney illustrates the progression across loading structures. SCFA and
gg-BFA (1f) give the same fixed-mask representation. Allowing heterogeneous
primary strengths changes the reconstruction only modestly, whereas models
with cross-loadings can represent substantially more of the sample covariance.
Within gg-BFA, mf gives the closest reconstruction and also uses the most
nonprimary loading components. PCFA-VA gives an even smaller descriptive error
in this panel, accompanied by greater computation than the gg-BFA fits. These
results show the expected accuracy--complexity trade-off; they do not establish
that the most flexible model is closest to the unknown population covariance.

Appendix~\ref{app:current-real-data} gives the complete CFA table and the comparison on all
retained variables.

\subsection{Generalizing covariance to new subjects}
\label{sec:r2-heldout-covariance}

Covariance generalization is assessed primarily by Gaussian negative log
likelihood per variable. Held-out \(d_F\) and SRMR describe discrepancies in
covariance scale and correlation structure.

The CFA comparison separates two sources of improvement: heterogeneous loading
strengths within a community and variable-specific signal across communities.
SCFA and gg-BFA (1f) have nearly identical held-out covariance errors.
Estimating the primary loading strengths generally improves the held-out fit,
although the improvement is modest and is not uniform across every covariance
measure. Allowing cross-loadings produces a clearer change: 2f lies between the
no-cross-loading methods and mf, while mf gives the strongest covariance
generalization among the gg-BFA members in every panel. The common primary
labels therefore retain the community interpretation, but primary loading
strengths alone do not capture all of the covariance that persists in new
subjects.

\begin{figure}[!htbp]
\centering
\includegraphics[width=\textwidth]{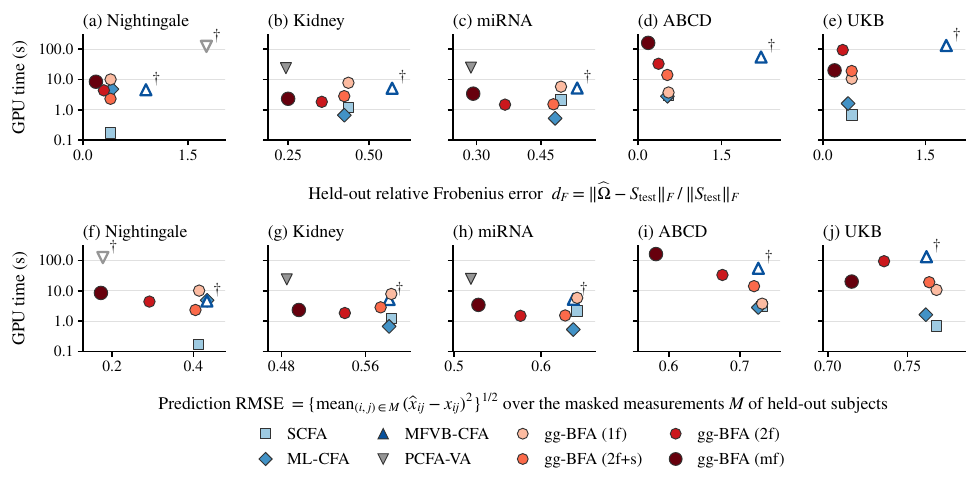}
\caption{Factor analysis results. Held-out errors against GPU time, averaged over five folds:
covariance distance \(d_F\) (top) and prediction RMSE of masked measurements (bottom), where 20\%
of each held-out subject's measurements are treated as missing and predicted from the other 80\%.
gg-BFA in red, mf darkest; hollow and \(\dagger\): not converged. gg-BFA (mf) has the smallest
errors in three of the five data sets; in Kidney and miRNA, the converged PCFA-VA fits are slightly
more accurate but take 7--10 times longer.  All eight estimators use the common factor budget and
are evaluated on the variables assigned to communities on the training subjects; PCFA-VA was not
run on the held-out folds of ABCD and UKB.  Time includes fitting and covariance construction on
P100 GPUs, excluding the shared community discovery and scoring.}
\label{fig:cfa-error-time}
\end{figure}

The converged PCFA-VA fits for Kidney and miRNA give slightly smaller held-out
errors than mf. This similarity concerns covariance fidelity rather than the
meaning of the loading matrix: PCFA-VA learns unspecified loadings under its
own prior, whereas gg-BFA retains community-oriented loading constraints. The
nonconverged PCFA-VA and MFVB-CFA fits remain visible as last-iterate
sensitivities and are not used to support a performance ranking.

The primary Gaussian NLL and the supporting covariance distances describe
different aspects of generalization. NLL evaluates the complete joint density
and is sensitive to poorly estimated low-variance directions, whereas \(d_F\)
and SRMR summarize covariance and correlation discrepancies. Table~\ref{tab:r2-common-current}
and Figure~\ref{fig:cfa-error-time} report these views separately rather than
collapsing them into one score.

\begin{table}[!htbp]
\centering
\caption{Mean held-out Gaussian NLL per selected variable at the common factor budget.
Lower is better; negative values are possible for continuous densities. All
methods in a panel are scored on the same selected variables and test subjects.
\(\dagger\) marks one or more nonconverged folds; \(\ddagger\) marks SCFA's
Nightingale latent-covariance caveat. NR: PCFA-VA was not
run on the held-out folds of ABCD and UKB because of its computational cost. No ranking is assigned to last iterates.}
\label{tab:r2-common-current}
\small
\setlength{\tabcolsep}{4pt}
\begin{tabular}{@{}lrrrrr@{}}
\toprule
Method & Nightingale & Kidney & miRNA & ABCD & UKB \\
\midrule
SCFA & 0.569$^{\ddagger}$ & 0.886 & 0.988 & 1.107 & 1.155 \\
ML-CFA & 0.380 & 0.845 & 0.929 & 1.067 & 1.115 \\
MFVB-CFA & 0.385$^{\dagger}$ & 0.845$^{\dagger}$ & 0.929$^{\dagger}$ & 1.067$^{\dagger}$ & 1.116$^{\dagger}$ \\
PCFA-VA & -0.531$^{\dagger}$ & 0.660 & 0.718 & NR & NR \\
gg-BFA (1f) & 0.570 & 0.886 & 0.988 & 1.107 & 1.155 \\
gg-BFA (2f+s) & 0.351 & 0.835 & 0.912 & 1.051 & 1.119 \\
gg-BFA (2f) & -0.009 & 0.765 & 0.825 & 0.980 & 1.082 \\
gg-BFA (mf) & -0.864 & 0.676 & 0.733 & 0.796 & 1.052 \\
\bottomrule
\end{tabular}

\end{table}

\subsection{Predicting masked measurements}
\label{sec:r3-conditional-prediction}

Point prediction is compared primarily by masked-cell RMSE. NLPD, coverage,
and interval width are reported together to evaluate predictive uncertainty;
MAE provides a complementary measure of absolute prediction error.

The masked-measurement analysis uses the fitted covariance to predict a
subject's hidden measurements from those that remain visible. The pattern
across loading structures is consistent across the five panels
(Figure~\ref{fig:cfa-error-time}, lower row). SCFA and gg-BFA (1f) give almost
the same predictions, and estimating only the primary loading strengths yields
limited improvement. Allowing one cross-loading reduces prediction error more
clearly, while mf has the lowest RMSE among the gg-BFA members in every panel.
The predictive gains therefore arise mainly from dependence that is not
captured by heterogeneous primary strengths alone.

The converged PCFA-VA fits for Kidney and miRNA give slightly smaller prediction
errors than mf, but similar prediction accuracy does not imply the same loading
interpretation. Table~\ref{tab:r3-fixed-selected} therefore reports NLPD,
coverage, and interval width together with RMSE. Coverage cannot be interpreted
without interval width: nominal coverage may result from intervals that are
too wide to be informative.

For the four gg-BFA members, integrating the new subject's local variational
factor posterior gives essentially the same predictions as the common
conditional-Gaussian rule. The local-posterior summaries are reported
separately in Table~\ref{tab:r3-local-variational}; they do not indicate an
additional accuracy or coverage gain over the covariance-based predictor.
Predictive coverage of a noisy future observation also does not resolve the
loading-parameter undercoverage seen in simulation, because residual noise
contributes to the former but not the latter.

\subsection{Computation and stopping status}
\label{sec:realdata-computation-boundary}

All primary CFA fits use the same P100 GPU class.
Figure~\ref{fig:cfa-error-time} places held-out errors beside estimator wall
time, and Table~\ref{tab:runtime-status} gives the detailed times and stopping
status. Estimator time includes fitting and
covariance construction; shared community discovery and scoring are excluded.
The common hardware removes the CPU--GPU distinction from the comparison, but
the algorithms still differ in iteration counts, posterior summaries, data
transfers, and stopping rules.

The fastest method is not necessarily the most accurate, and greater loading
flexibility does not produce a uniform runtime ordering. The converged PCFA-VA
fits for Kidney and miRNA obtain a small improvement over mf in held-out error
but require substantially more computation. PCFA-VA is not run in the ABCD and
UKB held-out analyses because of its full-data computational burden.

Finite output and convergence are reported separately. MFVB-CFA and the
nonconverged PCFA-VA fits are retained as marked last-iterate sensitivities and
do not support performance rankings. The Nightingale and imaging PCFA-VA last
iterates are also sensitive to the numerical implementation, whereas the
converged Kidney and miRNA fits agree with their CPU references. The absence of
ABCD and UKB held-out PCFA-VA results is therefore a computational exclusion,
not a prediction failure.

SCFA has a different limitation on Nightingale: its factor moment can be
indefinite even when the observed-variable covariance is positive definite.
The latter supports covariance scoring and Gaussian prediction, while the
former limits a latent Gaussian interpretation. No covariance repair is used
to obtain the reported scores.

\subsection{The complete pipeline on the full data}
\label{sec:raw-to-fit}

\begin{figure}[!htbp]
\centering
\includegraphics[width=\textwidth]{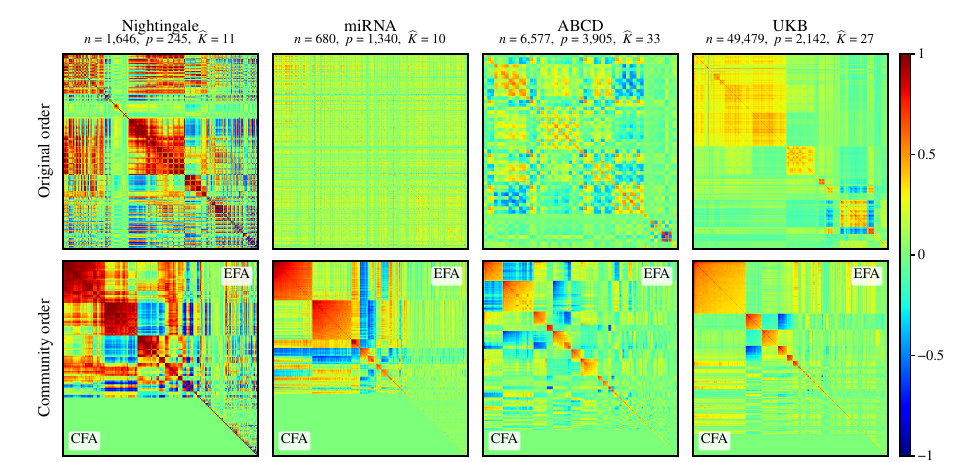}
\caption{Example data sample correlation matrices (top) and gg-BFA estimates (bottom) for the
four panels other than Kidney.  Top: original order.  Bottom: order of the communities selected by
the two-stage version of the exploratory step (Algorithm~\ref{alg:two-stage}), with the sample
correlation above the diagonal (EFA) and the gg-BFA (mf) fit below it (CFA).}
\label{fig:raw-to-fit}
\end{figure}

Figure~\ref{fig:raw-to-fit} shows the complete analysis on the full data for the four panels
other than Kidney, which is shown in Figure~\ref{fig:method}.  In the original order of the
variables, block structure is visible only in Nightingale, ABCD, and UKB, where variables of
the same measurement type are already adjacent.  After the variables are reordered by the
communities selected by the two-stage version of the exploratory step
(Algorithm~\ref{alg:two-stage}), all five panels show dense diagonal blocks and clear
correlations between communities.  This also holds for miRNA and Kidney
(Figure~\ref{fig:method}b and~d), whose original order shows almost no structure.  Among the variables assigned to communities, the correlations fitted by gg-BFA (mf), below the
diagonal, are close to the sample correlations above it, with relative errors between 0.03 and 0.10
over the off-diagonal entries of the five panels; Figure~\ref{fig:method}g shows the 2f+s fit for
Kidney.  For each panel, the community selection and the model fit together took less than ten
minutes.  In the fitted models, the correlations between communities are carried by many small
cross-loadings rather than by a few variables that link two communities.

\begin{figure}[!htbp]
\centering
\includegraphics[width=\textwidth]{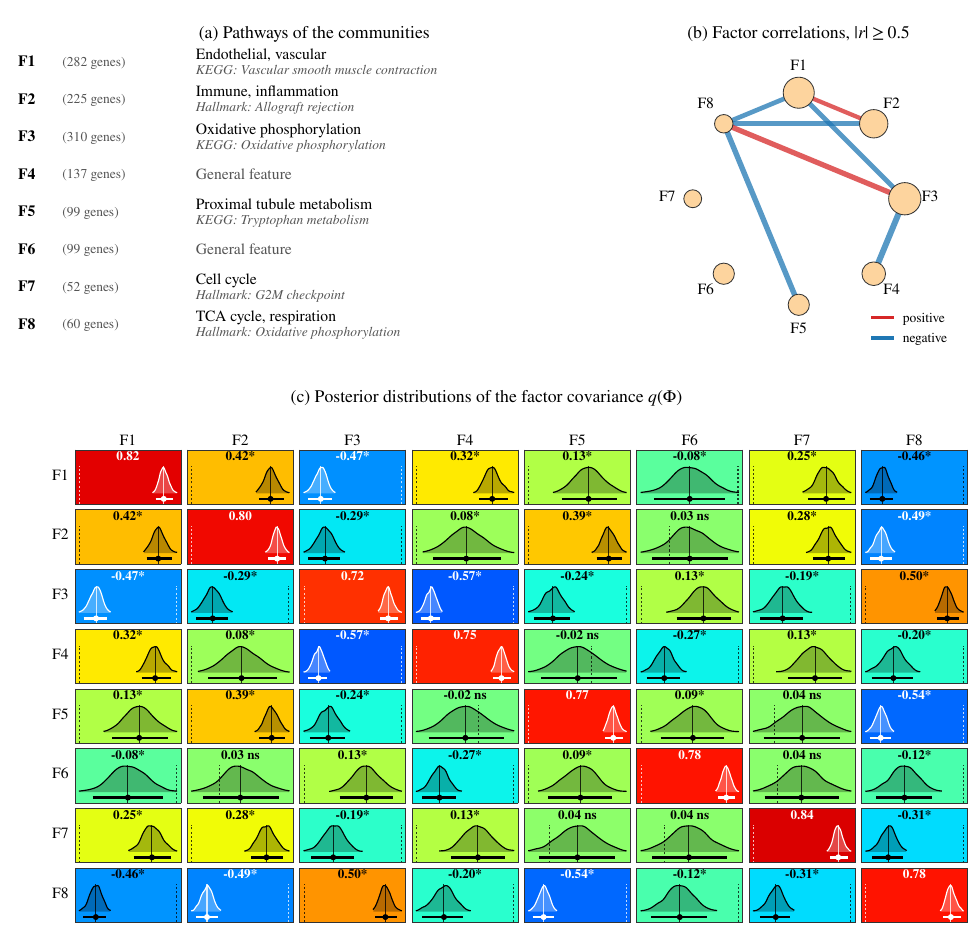}
\caption{The first eight factors (F1--F8) of the TCGA Kidney data under gg-BFA. In (a), we
identify the matched gene pathways for these factors (a factor is labeled as a general feature when
no single dominant pathway is matched). In (b), we use a graph to denote the suprathreshold
interactions among these pathways. In (c), we illustrate the posterior distributions of the
parameters of the covariance matrix of the factors (* indicates that the credible interval excludes
zero; ns otherwise). The member shown is gg-BFA (2f+s), whose \(\Phi\) keeps the correlations
between communities (Section~\ref{sec:discussion}).  (a)~Theme and most significant pathway of each
community (Hallmark, KEGG or Reactome, \(q<0.05\)); a general feature has no pathway with
\(q<0.05\).  (b)~Factor correlations with \(|r|\ge0.5\) whose 95\% credible interval excludes zero.
(c)~Marginal posteriors of the factor covariance \(\Phi\): posterior mean (number and dot), 95\%
credible interval (bar), and zero (dashed).}
\label{fig:results}
\end{figure}

Figure~\ref{fig:results} interprets the first eight of the 21 Kidney communities with
gg-BFA (2f+s), whose factor covariance keeps the correlations between communities
(Section~\ref{sec:discussion}).  Six of the eight communities are enriched for a pathway, among
them endothelial, immune, oxidative phosphorylation, and cell-cycle programs
(Figure~\ref{fig:results}a).  The factor correlations show how these programs vary together
(Figure~\ref{fig:results}b), and the credible intervals give a test for each factor correlation
at no extra cost (Figure~\ref{fig:results}c).  Of the 210 factor correlations, 190 have 95\%
credible intervals that exclude zero.

\FloatBarrier

\subsection{What the real-data analyses establish}

Across the five panels, a fixed community mask provides an economical
covariance representation, while learning only the primary loading strengths
usually gives a modest improvement. The larger gains arise when variables are
allowed to carry signal from outside their assigned communities. Within
gg-BFA, mf gives the strongest covariance and prediction results, while 2f
offers a more restricted loading representation and 2f+s is appropriate when
the number of bridge variables must also remain small.

The comparison does not identify a universal winner. PCFA-VA can produce
slightly smaller held-out errors in the panels where it converges, but it has a
different loading interpretation and a larger computational cost. Covariance
representation, joint density, conditional prediction, loading complexity,
uncertainty, and computation describe different consequences of the same fit.
Model dimension and convergence status must therefore be considered alongside
empirical error rather than treated as secondary implementation details.

\FloatBarrier

\section{Discussion}
\label{sec:discussion}

We proposed gg-BFA, which carries out both steps of factor analysis on the same communities of the
correlation graph.  The exploratory step selects the communities and their number by a forward
stepwise search that stops at the first decrease of the ELBO.  We showed that, with probability
tending to one, it recovers the true communities and its variational posterior of the covariance
contracts at the parametric rate up to a logarithmic factor, even when the number of variables grows
much faster than the sample size.  The confirmatory step fits four Bayesian factor models on these
communities, from no cross-loadings (1f) to many (mf), by mean-field variational inference.
No single loading support was best everywhere: fixed communities suffice when the block summary
is adequate, controlled cross-loadings recover important heterogeneity, and unrestricted loadings are
preferable when the community structure is absent.  gg-BFA therefore does not prescribe one constraint
for every data set; it makes these choices comparable in one framework and measures their costs in
covariance, prediction, runtime, and convergence.

The approach has three advantages.  First, it completes the EFA--CFA pipeline when the variables
outnumber the subjects, where the maximum-likelihood pipelines of \texttt{psych} and \texttt{lavaan}
cannot be applied and \texttt{semopy} selected too many factors or returned no solution.  Second, it
is fast: the whole pipeline took less than ten minutes per panel, whereas PCFA-VA, which also
estimates cross-loadings, took seven to ten times longer than mf and was too slow to run on ABCD and
UKB.  Third, each factor remains interpretable through its community and its matched biological
pathway, and the posterior gives credible intervals for the factor correlations and predicts missing
measurements.  The two steps can also be used separately: a partition chosen by the exploratory step
can be fitted by any confirmatory method, and the confirmatory step can be fitted to a partition
supplied by prior knowledge.

The exploratory step turns the choice of the partition and its size into a comparison of models
on a common scale: every candidate describes all variables, and the search fits models only up to one
step beyond the selected size.  The theory explains why it stops at the right place.  While true
communities are missing, adding the next one removes a bias of order \(n\), which outweighs the cost of
the new parameters.  Once all true communities are present, a further community either splits a true
community, whose two parts then have factors with correlation close to one, or is formed from
background variables, which share no factor, so that its factor variance is close to zero.  Either way
the factor covariance is nearly singular; the inverse-Wishart prior penalizes such covariances, and the
ELBO decreases.  The first argument needs the bias to decrease enough from one community to the next
(Assumption~(W) in Appendix~\ref{th:sec:exact}), which can fail for very large communities, so the
theory gives sufficient conditions rather than the full range in which the search works.

Two practical choices go beyond the theory.  The learning rate balances the likelihood against
the prior; our choice \(\alpha_0\) is approximately the effective number of variables divided by
\(p_0\), so it scales the likelihood down in panels with strong correlation.  The two-stage version
keeps one path after comparing candidates at a common reference size, which reduces the number of
fitted models by about an order of magnitude on our panels and is used in the real-data analyses; its
guarantees additionally require the kept path to contain the true partition.  The theory treats the
learning rate as fixed and analyzes the full search, which the simulations therefore run.

The exploratory-to-confirmatory pipelines choose the number of factors with a retention rule,
derive the loading pattern by rotation and thresholding, and fit the confirmatory model to the same
data.  The exploratory step instead searches a small family of graph-derived partitions and judges every
candidate by the criterion that decides when to stop, which makes its stopping point amenable to
analysis.  In the simulations, where the maximum-likelihood pipelines could be applied, their
covariance error was about twice that of gg-BFA in S1, and in S2 \texttt{semopy} had 1.5 to 2.5 times
the error of gg-BFA (Section~\ref{sec:sim-selection}).  On the real data, \texttt{psych} and
\texttt{lavaan} cannot be applied to TCGA Kidney and TCGA miRNA, whose variables outnumber the
subjects.

Given a partition, the confirmatory step uses it as a coordinate system: the primary community
tells what a factor represents, and the loading option sets how far individual variables may depart
from it.  Fixed masks are easy to name but can miss heterogeneous covariance, while unrestricted
loadings are expressive but can cut the link between a factor and its community; the four members span
this range.  In the simulations, 1f is a concise baseline when a fixed community summary is credible,
2f captures variables that bridge one other community, 2f+s keeps the number of such bridges small,
and mf is most useful when cross-community heterogeneity is widespread and \(K\) is moderate.  The real
data add a second criterion, the interpretation of \(\Phi\).  mf gave the most accurate covariance, but
in Kidney it used about eight cross-loadings per variable, and its factor correlations agreed only
moderately with the correlations between communities in the sample correlation matrix (0.76); those of
2f+s, with 0.09 cross-loadings per variable, agreed almost exactly (0.998).  We therefore recommend mf
for estimating and predicting the covariance, and 2f+s for interpreting \(\Phi\), as in
Figures~\ref{fig:method} and~\ref{fig:results}.  Comparisons with methods that do not use the
communities are given in Appendix~\ref{app:real-data}.

The results also have limits of interpretation and computation.  The factor labels are anchors
rather than proofs of a biological mechanism: some ADNI factor scores change substantially as the
loading support changes, and the signed row-sum constraint and the mask-centered prior do not establish
full rotational identifiability, so factor stability should be checked in each application.  Mean-field
VB recovers factor scores with nearly nominal coverage in the simulations but under-covers active
loadings, and variational bagging improves these intervals without fully calibrating them; near-nominal
predictive coverage on real data concerns future observations, not latent parameters.  On ABCD, with
about 300 discovered factors and roughly one million candidate cross-loadings, mf kept improving its
ELBO but did not reach the common convergence criterion, whereas the two structured learned-loading
members did.

Several choices in this work point to the next steps.  We fit the two steps separately: the
exploratory step returns one partition, and the confirmatory step treats it as known, although the
sensitivity analysis of the dense-subgraph search shows that the selected variables can change widely.
Joining the two in one posterior, for example by weighting the visited models as in the aggregation
step of \citet{ohn2026esa}, would let this uncertainty reach the factors and their credible intervals.
Our theory covers the forward stepwise search on Gaussian data in their original scale with a fixed
learning rate; the two-stage search, the estimated \(\alpha_0\), and the standardization used in
practice still need their own guarantees.  For large \(K\), hierarchical sparsity, stronger
regularization of the loadings, or a reparameterization of \(\Lambda\) and \(\Phi\) could help mf
converge.  The confirmatory models assume Gaussian measurements, but many biomedical panels record
counts, such as sequencing reads, so extending them to non-Gaussian data would widen their use.
Finally, the factors we found in Kidney matched known pathways; whether they reappear in external
cohorts and relate to clinical outcomes would show how far this interpretation carries.

\section*{Data, Code, and Declarations}

The gg-BFA software is available as the Python package \texttt{ggbfa} at
\url{https://github.com/waterism211/ggbfa}.  Public SCFA materials,
including the TCGA matrices, are available from
\url{https://github.com/yiorfun/SCFA}; individual-level ADNI data remain
available to qualified investigators through
\url{https://adni.loni.usc.edu/data-samples/adni-data/} under the ADNI Data Use
Agreement.  Appendix~\ref{app:provenance} identifies the configurations and
saved outputs used here.  The study uses existing data and involves no new
participant recruitment.  Computational tools assisted code, experiment orchestration, and
manuscript editing; the authors remain responsible for the analyses and text.

\section*{Acknowledgments}

Data collection and sharing for this project was funded by the Alzheimer's Disease Neuroimaging
Initiative (ADNI) (National Institutes of Health Grant U01 AG024904), DOD ADNI (Department of Defense
award number W81XWH-12-2-0012), and the Alzheimer's Disease Metabolomics Consortium (National
Institute on Aging R01AG046171, RF1AG051550 and 3U01AG024904-09S4).  ADNI is funded by the National
Institute on Aging, the National Institute of Biomedical Imaging and Bioengineering, and generous
contributions from the organizations listed at
\url{https://adni.loni.usc.edu/wp-content/uploads/how_to_apply/ADNI_Manuscript_Citations.pdf}.

The results shown here are in whole or part based upon data generated by the TCGA Research Network.

Data used in the preparation of this article were obtained from the Adolescent Brain Cognitive
Development\texttrademark{} (ABCD) Study (\url{https://abcdstudy.org/}), NBDC Release 7.0 (DOI
\url{https://doi.org/10.82525/8f3w-5260}), held in the NIH Brain Development Cohorts Data Sharing
Platform (\url{https://www.nbdc-datahub.org/}).  The ABCD Study\textregistered{} is supported by the
National Institutes of Health and additional federal partners under award numbers U01DA041022, 041025,
041028, 041048, 041089, 041093, 041106, 041117, 041120, 041134, 041148, 041156, 041174, 050987,
050988, 050989, 051016, 051018, 051037, 051038, 051039 and U24DA041123, 041147; see
\url{https://abcdstudy.org/federal-partners.html} for a full list of supporters.  ABCD Consortium
investigators designed and implemented the study and/or provided data but did not necessarily
participate in the analysis or writing of this report.  This manuscript reflects the views of the
authors and may not reflect the opinions or views of the NIH or ABCD Consortium investigators.

This research has been conducted using the UK Biobank Resource under Application Number 107106.  We
thank the UK Biobank group that collected the data made available for this work.

\clearpage

\appendix

\paragraph{Appendix guide.}
The appendix follows the order of the main text.
Appendix~\ref{app:selection-implementation} gives the implementation of the exploratory step,
Appendix~\ref{app:selection-theory} states its theory, Appendix~\ref{th:app:proofs} contains the
proofs, and Appendix~\ref{th:app:highdim} extends the theory to growing dimension.
Appendix~\ref{app:implementation} gives the implementation of the confirmatory step, and
Appendix~\ref{app:evaluation-metrics} defines the evaluation summaries.
Appendix~\ref{app:simulation-details} contains the simulation designs and the simulation results
that are not in the main text.  Appendix~\ref{app:real-data} supplements the real-data analyses with the implementations of the
comparison methods and the complete results at the common factor budget.
Appendix~\ref{app:reproducibility} lists the computations behind the reported results.

\section{Implementation of the exploratory step}
\label{app:selection-implementation}

This appendix gives the settings and numerical details of the exploratory step (Algorithm~\ref{alg:selection}) and of its two-stage version (Algorithm~\ref{alg:two-stage}).

\paragraph{Candidate paths.}
The data are centered and scaled to unit variance with divisor $n$. The grid $\mathcal H$ of the dense-subgraph search contains the $50$ tuning pairs with $q\in\{90,91,\ldots,99\}$ and $\lambda_{\mathrm{den}}\in\{0.4,0.5,\ldots,0.8\}$. We use $m_{\min}=2$ and no cap $K_{\max}$, so each path runs until no eligible component remains. The common partition of the CFA comparison (Section~\ref{sec:comparison-arms}) comes from the same dense-subgraph search with the fixed number of communities $K$ of Table~\ref{tab:realdata-design-current}; it is not chosen by the exploratory step.

\paragraph{Learning rate.}
The null model places all $p_0$ variables in the background block with a common variance $v_0$. Its maximum-likelihood estimate is $\widehat v_0=(np_0)^{-1}\sum_{i=1}^n\Vert x_i\Vert^2$, which equals one for standardized data. For a model with parameter $\theta$, \citet{lyddon2019llb} set the learning rate to $\tr(J\Lambda^{-1}J)/\tr(J)$, where $J$ is the Fisher information of one observation and $\Lambda$ is the variance of its score. For the one-parameter null model, with $J$ and $\Lambda$ estimated by plug-in, this gives
\begin{equation}
  \alpha_0=\frac{2p_0\widehat v_0^2}{n^{-1}\sum_{i=1}^n\bigl(\Vert x_i\Vert^2-p_0\widehat v_0\bigr)^2}.
  \label{eq:alpha0}
\end{equation}
If the data are Gaussian with correlation matrix $R$, the denominator estimates $2\Vert R\Vert_F^2$, so that $\alpha_0\approx p_0/\Vert R\Vert_F^2$. The quantity $p_0^2/\Vert R\Vert_F^2$ is an effective number of variables: it equals $p_0$ for independent variables and one when all variables are perfectly correlated. Thus $\alpha_0$ is approximately the effective number of variables divided by $p_0$. The learning rate is computed once, before any community is fitted, and is used for every candidate. The algorithm stops with an error if $\alpha_0\notin(0,1)$; this happens, for example, when the variables are nearly independent, a case for which the selection is not designed.

\paragraph{Computing the ELBO of a partition.}
For a partition with $K$ communities, the likelihood of the selection model~\eqref{eq:selection-model} depends on the data only through three statistics: the $K\times K$ sample covariance of the scaled community sums $y_{ih}=|\mathcal V_h|^{-1/2}\sum_{j\in\mathcal V_h}x_{ij}$, the within-community sums of squares that remain after the community sums are removed, and the sum of squares of the background variables. The variational family fixes the degrees of freedom and the shapes at their conjugate values, $\Phi\sim\IW_K(K+2+\alpha n,\tilde\Psi)$ and $v_h\sim\IG(2+\alpha n|\mathcal V_h|/2,\tilde\beta_h)$, with free $\tilde\Psi$ and $\tilde\beta_h$. For the background block, the optimal variational distribution is conjugate, and its contribution to the ELBO is computed exactly. The Kullback--Leibler terms and the within-community terms also have closed forms. The remaining term, the expected log-likelihood of the community sums, is a $K$-dimensional Gaussian expectation, which we estimate by Monte Carlo. We write $\Phi$ and $v$ as transformations of fixed draws from the standard Bartlett and gamma distributions, so that the estimated ELBO is a smooth function of $(\tilde\Psi,\tilde\beta)$, and maximize it by L-BFGS-B with $256$ draws and at most $600$ iterations ($3000$ in design S2 with $n\ge2000$, where $600$ were not enough, and in the simulation check of Appendix~\ref{th:sec:highdim-sim}), starting from a moment-based value and from a shrunken version of it. The better converged solution is kept, and the fit is reported as a failure if neither start converges. The ELBO is then evaluated at the solution with $2048$ new draws, which also gives its Monte Carlo standard error.

\paragraph{Comparisons under Monte Carlo error.}
Both versions compare ELBOs that carry Monte Carlo standard errors, with the following rules. When the ELBOs are exact, they reduce to those of Algorithms~\ref{alg:selection} and~\ref{alg:two-stage}; the two-standard-error rule guards against Monte Carlo error and is not part of the theory (Appendix~\ref{th:sec:icons}).
\begin{enumerate}
\item \emph{Comparing two ELBOs.} One ELBO is smaller than another when the difference exceeds twice its standard error, which is computed from the two standard errors as if they were independent.
\item \emph{Stopping.} The search stops at the first $K$ whose ELBO is smaller than that of $K-1$ in this sense. When the difference is within two standard errors, the ELBO at $K$ is evaluated again with $16384$ draws, while earlier ELBOs are kept fixed; if the difference is still within two standard errors, the step counts as no decrease and the search continues. In Algorithm~\ref{alg:selection}, $\mathrm{ELBO}_K$ is the largest estimated ELBO among the candidates with $K$ communities, and a re-evaluation applies to all of them; in the two-stage version, stage~2 compares the models along the kept path.
\item \emph{Keeping a path.} Stage~1 of the two-stage version keeps the path whose first $K_{\mathrm{ref}}$ communities have the largest estimated ELBO, with exact ties going to the first pair in the grid order.
\item \emph{Choosing $\widehat K$.} The selected $\widehat K$ is the largest visited $K$ whose ELBO is within two standard errors of the largest visited ELBO, excluding the layer that triggered a significant decrease, if any; when the paths end without such a decrease, the last layer remains eligible.
\end{enumerate}

\section{Theory for the exploratory step}
\label{app:selection-theory}

This appendix gives the complete statements behind Section~\ref{sec:selection-theory}. We study the search on the selection model~\eqref{eq:selection-model} for Gaussian data and ask three questions: whether the search stops too early, before it reaches a good model; whether it stops too late, after fitting models that are larger than necessary; and how close the selected model is to the truth. Our analysis follows the strategy that \citet{ohn2026esa}, hereafter ESA, use for variational empirical Bayes; Appendix~\ref{th:sec:esa} explains how it differs.

We analyze the forward stepwise search of Algorithm~\ref{alg:selection}, which chooses the partition anew at every $K$: the ELBO of model $K$ is maximized over all candidate partitions with $K$ communities, and the search stops at the first decrease. Its two-stage version (Algorithm~\ref{alg:two-stage}) keeps one path, which reduces the computation; Appendix~\ref{th:sec:icons} gives the conditions under which the results carry over to it.

The appendix is organized as follows. Appendix~\ref{th:sec:picture} gives an overview of the argument, and Appendix~\ref{th:sec:setting} introduces the notation. Appendix~\ref{th:sec:curves} constructs a lower and an upper curve that bound the ELBO deficit of every model with high probability, and Appendix~\ref{th:sec:theorems} derives the not-too-early, not-too-late and risk guarantees from them. Appendices~\ref{th:sec:exact} and~\ref{th:sec:misspec} apply these guarantees to an exact 1f model and to a misspecified model with geometrically decaying bias. Appendix~\ref{th:sec:esa} compares the analysis with ESA, and Appendix~\ref{th:sec:icons} discusses the choices that the exploratory step makes from the data. The proofs are in Appendix~\ref{th:app:proofs}, and Appendix~\ref{th:app:highdim} extends the results of Appendix~\ref{th:sec:exact} to dimensions $p_0$ that grow with $n$, including $p_0\gg n$.

\subsection{Overview of the argument}\label{th:sec:picture}

\begin{figure}[!htbp]
\centering
\begin{tikzpicture}[x=1.15cm,y=2.3cm]
  \fill[blue!10] (0,2.000) -- (1,1.778) -- (2,1.531) -- (3,1.255) -- (4,0.954) -- (5,0.477) -- (6,0.875) -- (7,0.929) -- (8,0.978)
       -- (8,0.875) -- (7,0.813) -- (6,0.740) -- (5,0.176) -- (4,0.845) -- (3,1.176) -- (2,1.462) -- (1,1.716) -- (0,1.944) -- cycle;
  \draw[->] (-0.4,0) -- (8.5,0) node[right] {$K$};
  \draw[->] (-0.4,0) -- (-0.4,2.3);
  \node[rotate=90,anchor=south] at (-0.55,1.15) {\small ELBO deficit (log scale, schematic)};
  \foreach \k in {0,...,8} \draw (\k,0) -- (\k,-0.04) node[below] {\small $\k$};
  \draw[thick,blue] (0,2.000) -- (1,1.778) -- (2,1.531) -- (3,1.255) -- (4,0.954) -- (5,0.477) -- (6,0.875) -- (7,0.929) -- (8,0.978);
  \draw[thick,red,dashed] (0,1.944) -- (1,1.716) -- (2,1.462) -- (3,1.176) -- (4,0.845) -- (5,0.176) -- (6,0.740) -- (7,0.813) -- (8,0.875);
  \draw[thin] (0,1.972) -- (1,1.747) -- (2,1.497) -- (3,1.215) -- (4,0.900) -- (5,0.327) -- (6,0.808);
  \foreach \x/\y in {0/1.972,1/1.747,2/1.497,3/1.215,4/0.900,5/0.327,6/0.808} \fill (\x,\y) circle (1.6pt);
  \draw[dotted] (5,0) -- (5,2.15) node[above] {$K_*$};
  \draw[densely dotted] (4,0.845) -- (5,0.845);
  \draw[->,thin] (2.7,0.40) node[left,align=right] {\scriptsize $\Ut(K_*)<\Lt(K_*-1)$:\\[-1pt]\scriptsize ELBO still increases} -- (4.9,0.47);
  \draw[densely dotted] (5,0.477) -- (6,0.477);
  \draw[->,thin] (6.7,0.30) node[right,align=left] {\scriptsize $\Lt(K_*+1)>\Ut(K_*)$:\\[-1pt]\scriptsize ELBO decreases; stop} -- (6.05,0.72);
  \node[right] at (8.05,1.04) {\small upper curve $\Ut(K)$};
  \node[right] at (8.05,0.80) {\small lower curve $\Lt(K)$};
\end{tikzpicture}
\caption{Schematic of the two curves under the exact model, on a log scale. With high probability the ELBO deficit $\Gamma(K)$ (dots) lies in the shaded band between the lower curve $\Lt(K)$ (dashed) and the upper curve $\Ut(K)$ (solid). Because $\Ut(K)<\Lt(K-1)$ for $K\le K_*$, the ELBO increases up to $K_*$. Because $\Lt(K_*+1)>\Ut(K_*)$, it decreases at $K_*+1$, where the search stops. Models beyond $K_*+1$ are never visited.}
\label{th:fig:curves}
\end{figure}
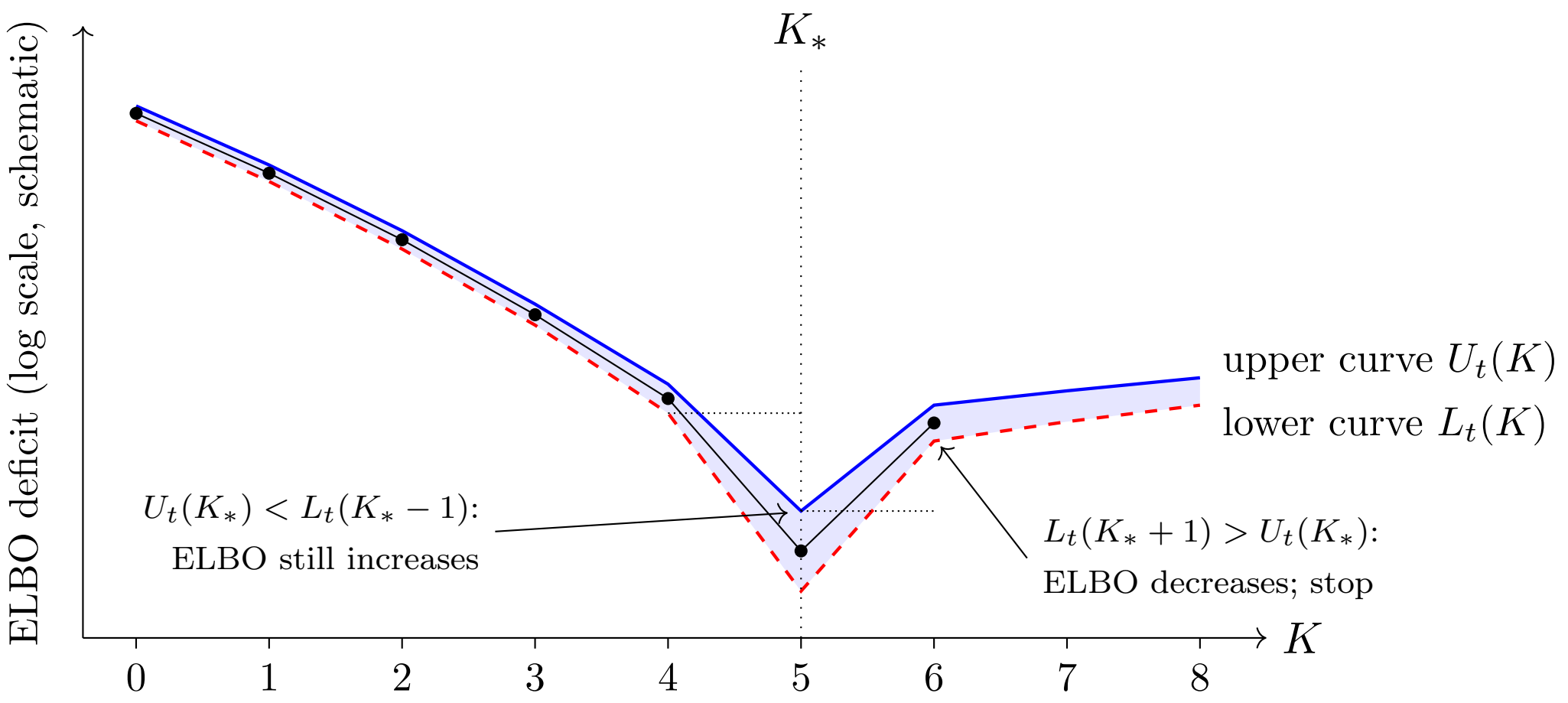

In this section, we describe the idea of the analysis before giving the formal statements. Write $\mathrm{ELBO}_K$ for the ELBO of model $K$. Since $\alpha\ell_n^*$, which is $\alpha$ times the negative log-likelihood of the truth, does not depend on $K$, comparing the ELBOs is the same as comparing the \emph{ELBO deficits} $\Gamma(K)=-\alpha\ell_n^*-\mathrm{ELBO}_K$, the amounts by which the ELBOs fall short of $\alpha$ times the log-likelihood of the truth; the search stops at the first increase of the deficit. The analysis rests on two deterministic curves, a \emph{lower curve} $\Lt(K)$ and an \emph{upper curve} $\Ut(K)$, such that
\[
\Pst\bigl(\Lt(K)<\Gamma(K)\le\Ut(K)\bigr)\ge1-2e^{-t}
\]
for every $K$ and $t>0$. That is, with high probability, the deficit of each model lies in the band between the two curves (Figure~\ref{th:fig:curves}).

The lower curve expresses that a wrong model cannot have a small deficit. It is approximately $n\alpha$ times the smallest R\'enyi divergence from the truth to model $K$, so if every candidate with $K$ communities is far from the truth, the deficit of model $K$ is large whatever the data (Lemma~\ref{th:lem:lower}). The upper curve expresses that the maximization in the ELBO can only help. It is the deficit of one fixed, well-chosen variational distribution, which the maximization over the variational family and the candidates can only improve. It is approximately $n\alpha$ times the smallest KL divergence from the truth to model $K$, plus the parameter cost $\frac{d_K}2\log(\alpha n)$, plus a random fluctuation (Lemma~\ref{th:lem:upper}).

These two curves yield the three guarantees of Appendix~\ref{th:sec:theorems} as follows. First, if $\Ut(K)<\Lt(K-1)$, then $\Gamma(K)<\Gamma(K-1)$ with high probability, so the ELBO still increases at $K$ and the search does not stop there (Theorem~\ref{th:thm:early}). Second, if $\Lt(K)\ge\Ut(k^\circ)$ for some $k^\circ<K$, then $\Gamma(K)>\Gamma(k^\circ)$ with high probability. Since the search can go beyond $K$ only if the ELBO increases at every step from $k^\circ$ to $K$, it stops at or before $K$ (Theorem~\ref{th:thm:late}). Third, the risk of the selected model is at most its deficit plus the logarithm of the number of candidates, and its deficit is at most that of any visited model; in particular, it is at most the upper curve at a good visited model (Theorem~\ref{th:thm:risk}).

Under the exact 1f model with $K_*$ communities, the two curves have the shape shown in Figure~\ref{th:fig:curves} (Proposition~\ref{th:prop:curves}). For $K<K_*$, both curves are of order $n$, because model $K$ misses at least one community. At $K=K_*$, the upper curve drops to $\frac{d_*}2\log n$. For $K>K_*$, the lower curve rises again to at least order $n^{1/3}$, because the extra communities can fit the truth only with a singular factor covariance, near which the inverse-Wishart prior has little mass. Hence the first argument applies to every $K\le K_*$ under condition (W), and the second applies at $K=K_*+1$ without further conditions.

\subsection{Setting and notation}\label{th:sec:setting}

In this section, we introduce the notation and describe the search that we analyze.

\paragraph{Data.}
$x_1,\ldots,x_n$ are independent $\Norm_{p_0}(0,\Omega_*)$ vectors with $\Omega_*\succ0$, and $S=n^{-1}\sum_ix_ix_i^\top$. We write $\Pst=\Norm_{p_0}(0,\Omega_*)$ for the law of one observation, and also use $\Pst$ and $\E_*$ for probabilities and expectations under the joint law of $x_1,\ldots,x_n$. For $\Omega\succ0$, the negative log-likelihood is $\ell_n(\Omega)=\frac n2\{p_0\log2\pi+\log|\Omega|+\tr(\Omega^{-1}S)\}$, and $\ell_n^*=\ell_n(\Omega_*)$ is its value at the truth.

\paragraph{Models.}
A partition $\rho=(\mathcal V_1,\ldots,\mathcal V_K;\mathcal V_0)$ consists of $K\ge0$ disjoint communities, each with at least two variables, and the background $\mathcal V_0$ of the remaining variables. The parameter is $\theta=(\Phi,\tau)$, where $\Phi\succ0$ is the $K\times K$ factor covariance and $\tau_h=1/v_h>0$ is the residual precision of $\mathcal V_h$, $h=0,\ldots,K$. The model covariance is
\[
\Omega_\rho(\theta)=A_\rho\Phi A_\rho^\top+D_\rho(\tau),
\]
where $A_\rho$ is the $p_0\times K$ membership matrix and $D_\rho(\tau)$ is diagonal with entry $v_h$ for the variables of $\mathcal V_h$. This is the selection model~\eqref{eq:selection-model}, indexed by the partition and with the residual variances parametrized by their precisions. Restricted to the community-assigned variables, its covariance is that of gg-BFA (1f), and we call it the 1f model with $K$ communities. Model $K$ has $d_K=(K+1)(K+2)/2$ free parameters when its background is non-empty and $d_K-1$ otherwise. Along a path, each model embeds into the next: setting the row and column of $\Phi$ for the added community to zero and its residual variance equal to the background variance reproduces the smaller model. The models along a path are thus nested in the closure of their parameter spaces, where $\Phi$ is positive semidefinite; the priors put mass only on $\Phi\succ0$. We write $P^\rho_\theta=\Norm_{p_0}(0,\Omega_\rho(\theta))$, $\ld(\rho,\theta)=\ell_n(\Omega_\rho(\theta))-\ell_n^*$ for the log-likelihood ratio, and $\KL(\rho,\theta)=\KL(\Pst\Vert P^\rho_\theta)$.

\paragraph{Candidates.}
The dense-subgraph search is run once for each tuning pair in a finite grid $\mathcal H$ (Section~\ref{sec:icons-estimation}), and each run returns a nested path of communities. $\Fam_K$ is the set of distinct partitions formed by the first $K$ communities of the paths that have at least $K$ communities; $\Fam_0$ contains only the partition with every variable in the background. Let $N_K=|\Fam_K|\le|\mathcal H|$, $\Ntot=\sum_KN_K$, and let $M$ be the largest $K$ with $\Fam_K\neq\emptyset$. The families are treated as non-random, that is, they do not depend on the data; Appendix~\ref{th:sec:icons} discusses the choices that the exploratory step makes from the data.

\paragraph{Prior and variational family.}
For partition $\rho$, the prior is $\Pi_K(\cdot\mid\rho)=\IW_K(\nu_K,cI_K)\otimes\Gam(a_\tau,b_\tau)^{\otimes(K+1)}$ on $(\Phi,\tau)$, with Gamma in shape--rate form and $a_\tau>1$; the implementation uses $\nu_K=K+2$, $c=1/2$, $a_\tau=2$ and $b_\tau=1/2$, which gives the prior of Section~\ref{sec:selection}, since $\tau_h\sim\Gam(2,1/2)$ exactly when $v_h\sim\IG(2,1/2)$. For a learning rate $\alpha\in(0,1)$, the variational family $\Qfam_K(\rho)$ consists of the distributions $\IW_K(\nu_K+\alpha n,\tilde\Psi)\otimes\bigotimes_{h=0}^K\Gam(a_\tau+\alpha n|\mathcal V_h|/2,\tilde\beta_h)$ with free $\tilde\Psi\succ0$ and $\tilde\beta_h>0$.

\paragraph{ELBO and algorithm.}
For $Q\in\Qfam_K(\rho)$, the objective of~\eqref{eq:selection-elbo} at $Q$ is $-\alpha\ell_n^*-F(K,\rho,Q)$, where the \emph{deficit}
\[
F(K,\rho,Q)=\alpha\int\ld(\rho,\theta)\,Q(d\theta)+\KL\bigl(Q,\Pi_K(\cdot\mid\rho)\bigr),
\]
which balances the fit to the data, measured by the expected log-likelihood ratio, against the complexity of the model, measured by the KL divergence from the prior. The deficit of partition $\rho$ is $\Gamma(K;\rho)=\min_{Q\in\Qfam_K(\rho)}F(K,\rho,Q)$, attained at its variational posterior, and that of model $K$ is $\Gamma(K)=\min_{\rho\in\Fam_K}\Gamma(K;\rho)$; minima are assumed to be attained. The minimization over $\rho\in\Fam_K$ treats the partition as a hyperparameter of the prior that is chosen from the data, which is an empirical Bayes choice. The ELBO of model $K$, maximized over $\Fam_K$, is $\mathrm{ELBO}_K=-\alpha\ell_n^*-\Gamma(K)$. The search computes $\mathrm{ELBO}_0,\mathrm{ELBO}_1,\ldots$, and stops at
\[
\Kstop=\min\{K\ge1:\mathrm{ELBO}_K<\mathrm{ELBO}_{K-1}\},
\]
or $\Kstop=M$ if the ELBO never decreases.
In other words, the search stops when the ELBO decreases for the first time. It selects $\widehat K=\arg\max_{0\le K\le\Kstop}\mathrm{ELBO}_K$ (ties go to the larger $K$), with partition $\widehat\rho$ and variational distribution $\widehat Q$. Since the search fits only the models with $K\le\Kstop$, an upper bound on $\Kstop$ also bounds its computational cost. This is the forward stepwise search of Algorithm~\ref{alg:selection}; the partition is chosen anew at every $K$. Its two-stage version (Algorithm~\ref{alg:two-stage}) keeps one path; Appendix~\ref{th:sec:icons} gives the conditions under which the results carry over. The theory concerns exact ELBOs and exact maximization; Appendix~\ref{th:sec:icons} also gives a sufficient condition under which numerical approximation does not change the decisions.

\subsection{The two curves}\label{th:sec:curves}

In this section, we construct the two curves described in Appendix~\ref{th:sec:picture}. The proofs of the results in this section and in Appendix~\ref{th:sec:theorems} are given in Appendix~\ref{th:app:proofs-general}.

\subsubsection{The lower curve}

We first define the population quantity that drives the lower curve.

\begin{definition}[R\'enyi risk and lower curve]\label{th:def:lower}
The \emph{R\'enyi risk} of $(\rho,\theta)$ is
\[
R_n(\rho,\theta)=-\log\E_*\exp\{-\alpha\,\ld(\rho,\theta)\}.
\]
For the Gaussian model, $R_n=n\alpha\,D_{1-\alpha}(\Pst\Vert P^\rho_\theta)$, where $D_{1-\alpha}$ is the R\'enyi divergence of order $1-\alpha$ (Lemma~\ref{th:lem:closed}); it is zero only when $\Omega_\rho(\theta)=\Omega_*$. Averaging over the prior and minimizing over the candidates gives
\[
\En(K,\rho)=-\log\int e^{-R_n(\rho,\theta)}\,\Pi_K(d\theta\mid\rho),\qquad
\En(K)=\min_{\rho\in\Fam_K}\En(K,\rho).
\]
The lower curve is $\Lt(K)=\En(K)-\log N_K-t$.
\end{definition}

The quantity $\En(K,\rho)$ is the log-exp-sum of the R\'enyi risk over the prior on model $K$ with partition $\rho$. By Lemma~\ref{th:lem:dv}, it can be written as
\[
\En(K,\rho)=\min_Q\Bigl\{\int R_n(\rho,\theta)\,Q(d\theta)+\KL\bigl(Q,\Pi_K(\cdot\mid\rho)\bigr)\Bigr\},
\]
where the minimum is over all probability measures $Q$ and is attained by the distribution proportional to $e^{-R_n(\rho,\cdot)}\,\Pi_K(\cdot\mid\rho)$. This relationship implies that $\En(K,\rho)$ is the optimal trade-off over model $K$ between the approximation error $\int R_n\,dQ$ and the estimation error $\KL(Q,\Pi_K(\cdot\mid\rho))$. It is also the population counterpart of the deficit $\Gamma(K;\rho)$, in which the log-likelihood ratio $\alpha\ld$ is replaced by the R\'enyi risk $R_n$ and the variational family by all distributions. Moreover, since $\int e^{-R_n}\,d\Pi\le e^{-\inf R_n}$, we have $\En(K)\ge n\alpha\,m_K$, where
\[
m_K=\min_{\rho\in\Fam_K}\inf_\theta D_{1-\alpha}(\Pst\Vert P^\rho_\theta)
\]
is the smallest R\'enyi divergence from the truth to model $K$. Thus $\En(K)$ is of order $n$ whenever model $K$ cannot represent the truth.

The next lemma shows that the deficit of model $K$ falls below the lower curve only with small probability.

\begin{lemma}[A wrong model cannot have a large ELBO]\label{th:lem:lower}
For every $K$ and $t>0$, $\Pst\bigl(\Gamma(K)\le\Lt(K)\bigr)\le e^{-t}$. Moreover, for every fixed $\rho\in\Fam_K$, $\Pst\bigl(\Gamma(K;\rho)\le\En(K,\rho)-t\bigr)\le e^{-t}$.
\end{lemma}

The term $\log N_K$ in the lower curve is the price of choosing the partition from $N_K$ candidates: the proof bounds the deficit of each candidate separately and then takes a union bound. Because the candidate families are finite, this price is additive and only logarithmic in the number of candidates.

\subsubsection{The upper curve}

We next construct the upper curve. For each $K$, fix an \emph{oracle}: a partition $\rho^\circ_K\in\Fam_K$ and a distribution $Q^\circ_K\in\Qfam_K(\rho^\circ_K)$, both non-random. Since $\Gamma(K)$ is a minimum over partitions and variational distributions, $\Gamma(K)\le F(K,\rho^\circ_K,Q^\circ_K)$, and the next lemma computes the right-hand side explicitly.

\begin{lemma}[The deficit at a fixed $Q$]\label{th:lem:decomp}
For fixed $\rho$ and $Q\in\Qfam_K(\rho)$, let $z_i=\Omega_*^{-1/2}x_i$ and $B_Q=\Omega_*^{1/2}\,\E_Q[\Omega_\rho(\theta)^{-1}]\,\Omega_*^{1/2}-I$. Then
\[
F(K,\rho,Q)=\underbrace{\alpha n\,\E_Q\KL(\rho,\theta)}_{\text{bias}}+\underbrace{\KL\bigl(Q,\Pi_K(\cdot\mid\rho)\bigr)}_{\text{parameter cost}}+\underbrace{\frac\alpha2\sum_{i=1}^n\bigl(z_i^\top B_Qz_i-\tr B_Q\bigr)}_{\text{fluctuation (mean zero)}} .
\]
\end{lemma}

The first term of the decomposition is the approximation error of model $K$, the second is the estimation error, and the third is a centered Gaussian quadratic form with standard deviation $\alpha\sqrt{n/2}\,\Vert B_Q\Vert_F$. Bounding the third term by a tail inequality for Gaussian quadratic forms (Lemma~\ref{th:lem:chaos}) leads to the following definition.

\begin{definition}[Upper curve]
With the oracle $(\rho^\circ_K,Q^\circ_K)$ and $B=B_{Q^\circ_K}$,
\[
\Ut(K)=\alpha n\,\E_{Q^\circ_K}\KL(\rho^\circ_K,\theta)+\KL\bigl(Q^\circ_K,\Pi_K(\cdot\mid\rho^\circ_K)\bigr)+\alpha\Bigl(\sqrt{2n(t+1)}\,\Vert B\Vert_F+(t+1)\Vert B\Vert_{\op}\Bigr).
\]
\end{definition}

The following lemma shows that the deficit of model $K$ exceeds the upper curve only with small probability.

\begin{lemma}[The maximization can only help]\label{th:lem:upper}
For every $K$ and $t>0$, $\Pst\bigl(\Gamma(K)>\Ut(K)\bigr)\le e^{-t}$, and the same holds for $\Gamma(K;\rho^\circ_K)$.
\end{lemma}

The upper curve depends on the choice of the oracle. In Appendices~\ref{th:sec:exact} and \ref{th:sec:misspec}, we use conjugate oracles centered at a fixed parameter $\theta_c$ (defined before Lemma~\ref{th:lem:oracle}), for which Lemma~\ref{th:lem:oracle} gives
\[
\Ut(K)\approx\alpha n\KL(\rho^\circ_K,\theta_c)+\frac{d_K}2\log(\alpha n)+O\bigl(\sqrt{nt}\bigr).
\]
That is, the upper curve is $\alpha n$ times the KL bias of the centre, plus the parameter cost $\frac{d_K}2\log(\alpha n)$, plus a fluctuation of order $\sqrt{nt}$, whose leading term vanishes when the centre is the truth.

\begin{appremark}[Two divergences]\label{th:rem:divergences}
The lower curve involves the R\'enyi divergence $D_{1-\alpha}$, whereas the upper curve involves the KL divergence. Since $D_{1-\alpha}\le\KL$ (Lemma~\ref{th:lem:closed}), the two curves differ even when the parameter cost and the fluctuation are ignored. This gap is the reason why Assumption~(W) in Appendix~\ref{th:sec:exact} is stronger than a strict decrease of the bias. The two divergences are close when $\alpha$ is small relative to the strength of the communities; see Example~\ref{th:ex:onemissing}.
\end{appremark}

\subsection{Stopping and risk guarantees}\label{th:sec:theorems}

In this section, we show that the not-too-early, not-too-late and risk properties of the forward stepwise search follow from the two curves. The results hold for any true covariance $\Omega_*$, and they use only Lemmas~\ref{th:lem:lower} and \ref{th:lem:upper}, together with Lemma~\ref{th:lem:pacbayes} below for the risk bound.

\subsubsection{Not-too-early stopping}

The following theorem shows that the search does not stop prematurely, provided that, at each $K$ up to $K^\dagger$, the upper curve of model $K$ lies below the lower curve of model $K-1$.

\begin{apptheorem}[Not too early]\label{th:thm:early}
Let $1\le K^\dagger\le M$ and $t>0$. If
\[
\Ut(K)<\Lt(K-1)\qquad\text{for every }1\le K\le K^\dagger,
\]
then, with probability at least $1-2K^\dagger e^{-t}$, $\mathrm{ELBO}_K>\mathrm{ELBO}_{K-1}$ for all $K\le K^\dagger$, and hence $\Kstop\ge\min\{K^\dagger+1,M\}$.
\end{apptheorem}

The condition $\Ut(K)<\Lt(K-1)$ compares the best case of model $K$ with the worst case of model $K-1$. It requires that the bias removed by the $K$-th community outweighs the parameter cost of model $K$, the price $\log N_{K-1}$ of the search over the candidates, and the fluctuations. With high probability, the deficit of model $K$ is then at most $\Ut(K)$ and that of model $K-1$ is more than $\Lt(K-1)$, so the ELBO increases at $K$. The failure probability $2K^\dagger e^{-t}$ comes from applying Lemmas~\ref{th:lem:lower} and \ref{th:lem:upper} once at each step; with $t=\log n$, it is $2K^\dagger/n$.

\subsubsection{Not-too-late stopping}

We next show that the search excludes models that are too large, that is, stopping occurs not too late. This supports the computational efficiency of the search, since the models beyond the stopping index are never fitted.

\begin{apptheorem}[Not too late]\label{th:thm:late}
Let $0\le k^\circ<K\le M$ and $t>0$. If $\Lt(K)\ge\Ut(k^\circ)$, then $\Pst(\mathrm{ELBO}_K\ge\mathrm{ELBO}_{k^\circ})\le2e^{-t}$; in particular $\Pst(\Kstop\ge K+1)\le2e^{-t}$.
\end{apptheorem}

The condition $\Lt(K)\ge\Ut(k^\circ)$ requires that the worst case of model $K$ is still worse than the best case of a smaller model $k^\circ$. With high probability, model $K$ then has a smaller ELBO than model $k^\circ$. Since the search can go beyond $K$ only if the ELBO increases at every step from $k^\circ$ to $K$, which would give $\mathrm{ELBO}_K\ge\mathrm{ELBO}_{k^\circ}$, the search stops at or before $K$. Unlike Theorem~\ref{th:thm:early}, the condition involves a single pair $(k^\circ,K)$, so the failure probability does not grow with $K$.

\subsubsection{Risk of the selected model}

Finally, we bound the risk of the selected model. Since the selected model is random, we need a bound that holds simultaneously for all candidates and all variational distributions. The following lemma, a PAC-Bayes bound, provides it.

\begin{lemma}[Risk is controlled by the deficit]\label{th:lem:pacbayes}
For $t>0$, with probability at least $1-e^{-t}$, simultaneously for all $K$, $\rho\in\Fam_K$ and $Q\in\Qfam_K(\rho)$,
\[
\int R_n(\rho,\theta)\,Q(d\theta)\le F(K,\rho,Q)+\log\Ntot+t .
\]
\end{lemma}

Combining Lemma~\ref{th:lem:pacbayes} with the maximality of the selected ELBO, we obtain the following risk bound.

\begin{apptheorem}[Risk of the selected model]\label{th:thm:risk}
Let $k^\circ\le M$ and $t>0$. With probability at least $1-\Pst(\Kstop<k^\circ)-2e^{-t}$,
\[
\int R_n(\widehat\rho,\theta)\,\widehat Q(d\theta)\le\Ut(k^\circ)+\log\Ntot+t ,
\]
and, for $0<\varepsilon\le\frac12$,
\[
\widehat Q\Bigl(\bigl\Vert\Omega_*^{-1/2}\Omega_{\widehat\rho}(\theta)\Omega_*^{-1/2}-I\bigr\Vert_F>\varepsilon\Bigr)\le\frac{9\{\Ut(k^\circ)+\log\Ntot+t\}}{n\alpha(1-\alpha)\varepsilon^2}.
\]
\end{apptheorem}

The first bound is an oracle inequality: the risk of the selected model is at most the upper curve at any model $k^\circ$ that the search visits, plus the price $\log\Ntot+t$ of the search. The probability $\Pst(\Kstop<k^\circ)$ that $k^\circ$ is not visited is controlled by Theorem~\ref{th:thm:early}. The second bound converts the risk into the relative Frobenius error of the covariance matrix, through the lower bound on the R\'enyi divergence in Lemma~\ref{th:lem:glower}.

\begin{appremark}[Choice of $t$]\label{th:rem:t}
All results in this section hold for every $t>0$. In Appendices~\ref{th:sec:exact} and \ref{th:sec:misspec}, we take $t=\log n$, which makes each failure probability of order $1/n$ at the cost of an additional $\log n$ in the curves.
\end{appremark}

\subsection{Case I: the truth is an exact 1f model}\label{th:sec:exact}

In this section, we apply the results of Appendix~\ref{th:sec:theorems} when the truth is an exact 1f model. The proofs of the results in this section and in Appendix~\ref{th:sec:misspec} are given in Appendix~\ref{th:app:proofs-cases}.

\subsubsection{Assumptions}

\begin{assumption}{(E)}{Exact 1f model}\label{th:ass:E}
$\Omega_*=\Omega_{\rho_*}(\theta_*)$ for a partition $\rho_*$ with $K_*\ge1$ communities, each with at least two variables, and $\theta_*=(\Phi_*,\tau_*)$ with $\Phi_*\succ0$.
\end{assumption}

Assumption~(E) states that the true covariance is a 1f background-block model with $K_*$ communities. The condition $\Phi_*\succ0$ rules out two communities whose factors are perfectly correlated, which would be indistinguishable from a single community. The condition that each community has at least two variables makes the factor covariance identifiable from the off-diagonal entries of $\Omega_*$. We let $d_*$ be the number of free parameters of the true model, that is, $d_{K_*}$, or $d_{K_*}-1$ if the true background is empty. The case $K_*=0$, in which all variables are independent, is not covered.

\begin{assumption}{(F)}{Fixed dimension and candidates}\label{th:ass:F}
The true partition is a candidate, $\rho_*\in\Fam_{K_*}$. The true covariance $\Omega_*$ (hence $K_*$, $\rho_*$ and $\theta_*$), the dimension $p_0$, the candidate families, the prior and $\alpha$ do not depend on $n$.
\end{assumption}

Assumption~(F) requires that the dense-subgraph search finds the true partition for at least one hyperparameter in its grid. It does not require the other candidates to be good: the lower curve accounts for them through the union bound in Lemma~\ref{th:lem:lower}. Appendix~\ref{th:sec:icons} discusses the case in which the search is run on the same data.

\begin{assumption}{(W)}{Strong decrease of the bias}\label{th:ass:W}
$m_{K-1}>\mKL_K$ for $1\le K\le K_*-1$, where $\mKL_K=\min_{\rho\in\Fam_K}\inf_\theta\KL(\rho,\theta)$ is the smallest KL divergence from the truth to model $K$.
\end{assumption}

Assumption~(W) states that the best model with $K-1$ communities, measured by the R\'enyi divergence, is still farther from the truth than the best model with $K$ communities, measured by the KL divergence. It is a substantive condition. Since $D_{1-\alpha}\le\KL$, it is stronger than the strict decrease of the bias, $\mKL_{K-1}>\mKL_K$ (Remark~\ref{th:rem:divergences}), and nesting alone does not imply it: it can fail even when every step adds a true community. Let the truth consist of two independent communities of $b$ variables each, with unit variances and within-community correlation $1/2$, let the candidates add the two true communities in turn, and let $\alpha=0.02$. Placing every variable in the background with variance $1/2$ shows that $m_0\le\{\log(1+\alpha b)-\alpha\log(1+b)\}/\alpha$, whereas $\mKL_1=\frac12\{b\log2-\log(b+1)\}$. For $b=500$ these are $113.7$ and $170.2$, so (W) fails at $K=1$: here the R\'enyi bias grows only logarithmically in $b$, whereas the KL bias grows linearly. This limits the scope of the sufficient conditions; it does not mean that the search fails. The gap between the two divergences is small when $\alpha(b-1)r$ is small for every missing community, where $b$ is the size of the community and $r$ its within-community correlation, as the following example shows.

\begin{example}[One missing community]\label{th:ex:onemissing}
Consider a model that misses one community of size $b$ with within-community correlation $r$ and unit variances. The following table reports, for several values of $b$, $r$ and $\alpha$, the KL and R\'enyi biases per observation of the best model without this community, which places its variables in a background block; each bias is minimized over the residual variance of that block and computed from Lemma~\ref{th:lem:closed}.

\begin{center}
\small
\begin{tabular}{@{}lrrrrr@{}}
\toprule
community & $\alpha$ & $\alpha(b-1)r$ & KL bias & R\'enyi bias & ratio\\
\midrule
$b=20$, $r=0.3$ & 0.02 & 0.11 & 2.44 & 2.27 & 0.93\\
$b=60$, $r=0.3$ & 0.02 & 0.35 & 9.06 & 7.65 & 0.85\\
$b=282$, $r=0.5$ & 0.0025 & 0.35 & 94.9 & 83.4 & 0.88\\
$b=282$, $r=0.5$ & 0.02 & 2.81 & 94.9 & 42.9 & 0.45\\
\bottomrule
\end{tabular}
\end{center}

The ratio of the R\'enyi bias to the KL bias is about $0.85$ or more when $\alpha(b-1)r$ is about $0.35$ or less, but drops to $0.45$ when $\alpha(b-1)r=2.81$. Now suppose that the missing communities are equally strong, that their biases add up, and that each has ratio $\vartheta$. Then (W) at step $K$ reads $\vartheta(j+1)>j$, where $j=K_*-K$ is the number of communities still missing after step $K$. Hence (W) holds when fewer than $\vartheta/(1-\vartheta)$ communities remain: about thirteen when $\vartheta=0.93$, but not even one when $\vartheta=0.45$.
\end{example}

\subsubsection{The shape of the two curves}

We first describe the shape of the two curves under Assumptions~(E) and (F).

\begin{proposition}[The curves under the exact model]\label{th:prop:curves}
Assume (E) and (F), take $t=\log n$, and use the conjugate oracles of Lemma~\ref{th:lem:oracle}. For large $n$:
\begin{enumerate}[label=(\alph*),nosep]
\item for $K<K_*$, $\Lt(K)\ge n\alpha m_K-O(\log n)$ with $m_K>0$, and $\Ut(K)\le n\alpha(\mKL_K+\delta)+O(\sqrt{n\log n})$ for any fixed $\delta>0$;
\item at $K=K_*$, $\Ut(K_*)\le\frac{d_*}2\log n+O(1)$;
\item for $K>K_*$, $\Lt(K)\ge c_0(n\alpha(1-\alpha))^{1/3}-O(\log n)$, where $c_0=\min\{c/4,\,1/(18\Vert\Omega_*\Vert_{\op}^2)\}$.
\end{enumerate}
\end{proposition}

Parts~(a) and (b) show that the upper curve drops from order $n$ to order $\log n$ at $K=K_*$, and part~(c) shows that the lower curve rises again to at least order $n^{1/3}$ beyond $K_*$. Part~(c) is where the inverse-Wishart prior matters. By Lemma~\ref{th:lem:rep}, a model with more than $K_*$ communities can reproduce $\Omega_*$ only by carving a community out of the true background or by splitting a true community, and in both cases the factor covariance must be singular. The prior assigns mass at most $e^{-c/(4\varepsilon)}$ to factor covariances with smallest eigenvalue below $\varepsilon$, while outside this set the R\'enyi risk is at least of order $n\alpha(1-\alpha)\varepsilon^2$ (Lemma~\ref{th:lem:iw}). Balancing the two exponents at $\varepsilon=(n\alpha(1-\alpha))^{-1/3}$ gives a lower bound of order $n^{1/3}$; candidates with a positive bias have a lower curve of order $n$.

\subsubsection{Selection and rate}

Combining Proposition~\ref{th:prop:curves} with Theorems~\ref{th:thm:early} and \ref{th:thm:late}, we obtain that the search selects the true model.

\begin{corollary}[The search selects the true model]\label{th:cor:exact}
Assume (E), (F) and (W). Then, for all sufficiently large $n$,
\[
\Pst\Bigl(\widehat K=K_*,\ \widehat\rho=\rho_*,\ \text{and the search stops at }K_*+1\text{ (or at }M=K_*)\Bigr)\ge1-\frac{2K_*+2+N_{K_*}}n .
\]
\end{corollary}

Corollary~\ref{th:cor:exact} shows that the search stops exactly one step after the true number of communities, so the search fits models of only one size beyond $K_*$. The failure probability grows linearly in $K_*$ and in the number $N_{K_*}$ of candidate partitions with $K_*$ communities.

Combining Corollary~\ref{th:cor:exact} with Theorem~\ref{th:thm:risk}, we obtain the convergence rate of the selected model.

\begin{corollary}[Rate]\label{th:cor:rate}
Assume (E), (F) and (W). With probability at least $1-O(1/n)$,
\[
\int R_n(\widehat\rho,\theta)\,\widehat Q(d\theta)\le\frac{d_*}2\log n+\log n+\log\Ntot+O(1),
\]
and for every sequence $A_n\to\infty$,
\[
\widehat Q\Bigl(\bigl\Vert\Omega_*^{-1/2}\Omega_{\widehat\rho}(\theta)\Omega_*^{-1/2}-I\bigr\Vert_F>A_n\sqrt{\frac{d_*\log n}{n\alpha(1-\alpha)}}\Bigr)\to0
\]
in probability.
\end{corollary}

The rate $\sqrt{d_*\log n/(n\alpha(1-\alpha))}$ is the parametric rate $\sqrt{d_*/n}$ for a model with $d_*$ parameters, up to the factor $\sqrt{\log n/\{\alpha(1-\alpha)\}}$. The parametric rate cannot be improved: by the local asymptotic minimax theorem \citep[Chapter~8]{vandervaart1998}, no estimator of $\Omega_*$ converges faster than $n^{-1/2}$ uniformly over a neighborhood of $\theta_*$, even when the true partition is known. The selected model therefore attains the optimal rate up to a logarithmic factor without knowing $K_*$ or $\rho_*$. The factor $1/\alpha$ reflects the learning rate of the fractional likelihood, which reduces the effective sample size from $n$ to $\alpha n$.

\begin{corollary}[Posterior mean]\label{th:cor:mean}
Assume (E), (F) and (W), and let $\widehat\Omega=\int\Omega_{\widehat\rho}(\theta)\,\widehat Q(d\theta)$ be the posterior mean covariance of the selected model. Then, for every sequence $A_n\to\infty$,
\[
\Pst\Bigl(\bigl\Vert\Omega_*^{-1/2}\widehat\Omega\,\Omega_*^{-1/2}-I\bigr\Vert_F>A_n\sqrt{\frac{d_*\log n}{n\alpha(1-\alpha)}}\Bigr)\to0 .
\]
\end{corollary}

Corollary~\ref{th:cor:rate} alone does not give this, because a small amount of variational posterior mass far from the truth can move the mean. The proof uses in addition that every member of the variational family concentrates around its own mean: its inverse-Wishart and inverse-gamma components have relative variances of order $1/(\alpha n)$. If the mean were far from the truth, most of the mass would be far from the truth as well, which Corollary~\ref{th:cor:rate} excludes. The constants of the argument depend on $p_0$, and we do not extend it to the growing dimension of Appendix~\ref{th:app:highdim}.

Appendix~\ref{th:app:highdim} shows that Proposition~\ref{th:prop:curves} and Corollaries~\ref{th:cor:exact} and \ref{th:cor:rate} continue to hold when $p_0$ and the community sizes grow with $n$, possibly with $p_0\gg n$, under uniform versions of Assumptions~(E), (F) and (W) and the condition $\log p_0=o(n^{1/3})$. The rate then becomes $\sqrt{(d_*\log n+(K_*+1)\log p_0)/(n\alpha(1-\alpha))}$.

\subsection{Case II: misspecified model with geometrically decaying bias}\label{th:sec:misspec}

In this section, we consider a misspecified case. If the truth is not a 1f block model and $p_0$ is fixed, every candidate keeps a bias of order one. The method then approaches the best candidate, and there is no rate. A rate requires the best number of communities to grow with $n$, so we consider a sequence of problems in which $\Omega_*$, $p_0$ and the candidate families may change with $n$. Rather than modeling $\Omega_*$ directly, we state an assumption on the two curves.

\begin{assumption}{(G)}{Geometrically decaying bias}\label{th:ass:G}
Take $t=\log n$. There exist constants $B,c_L,c_U>0$, $\vartheta\in(0,1]$ and $\gamma>\log(1/\vartheta)$ such that, for all $K\le M_n$,
\[
\Ut(K)\le n\alpha Be^{-\gamma K}+c_Ud_K\log n,\qquad\Lt(K)\ge \vartheta\,n\alpha Be^{-\gamma K}+c_Ld_K\log n .
\]
Moreover, $M_n\ge(2/\gamma)\log n$ and $\log\Ntot=O(\log n)$.
\end{assumption}

Assumption~(G) describes a bias that decays geometrically in the number of communities. The term $n\alpha Be^{-\gamma K}$ is the approximation error, and each added community removes the same fraction $1-e^{-\gamma}$ of the remaining KL bias. The factor $\vartheta\le1$ allows the R\'enyi bias in the lower curve to be smaller than the KL bias in the upper curve, and the terms $c_Ld_K\log n$ and $c_Ud_K\log n$ are the parameter costs. The condition $\gamma>\log(1/\vartheta)$, that is, $\vartheta e^\gamma>1$, plays the role of Assumption~(W): the bias term in $\Lt(K-1)$ is $\vartheta e^\gamma$ times that in $\Ut(K)$, so it is the larger of the two.

The following corollary shows that the search stops after about $(\log n)/\gamma$ communities and that the selected model attains the parametric rate up to a polylogarithmic factor.

\begin{corollary}[Geometric decay]\label{th:cor:geom}
Under (G), let $\zeta=c_U/(\vartheta e^\gamma-1)$. With probability tending to one,
\begin{gather*}
\frac{\log n}\gamma-\frac3\gamma\log\log n-C_6\ \le\ \Kstop\ \le\ C_7\frac{\log n}\gamma,\\
\int R_n(\widehat\rho,\theta)\,\widehat Q(d\theta)\le(\zeta+c_U)\frac{(\log n)^3}{2\gamma^2}\{1+o(1)\},
\end{gather*}
where $C_6=\gamma^{-1}\{\log(\zeta/(\alpha B))+2\log(2/\gamma+2)\}^++1$ and $C_7=\sqrt{(\zeta+c_U)/c_L}+1$. Consequently the relative Frobenius error is of order $\sqrt{(\log n)^3/(\gamma^2n\alpha)}$.
\end{corollary}

The stopping index grows only logarithmically in $n$, so the search visits few layers; the number of fitted models is $\sum_{K\le\Kstop}N_K$, which (G) allows to grow polynomially in $n$. The risk is polylogarithmic in $n$, so the method attains the parametric rate up to a polylogarithmic factor with a model of about $(\log n)/\gamma$ communities, without any knowledge of $\gamma$.

\subsection{Relation to ESA}\label{th:sec:esa}

In this section, we compare our analysis with that of ESA. ESA introduced the early-stopping rule used here and analyzed it for variational Bayes, for variational empirical Bayes and for frequentist aggregation. The forward stepwise search is an instance of their empirical Bayes setting, in which the partition plays the role of the hyperparameter. ESA proved that the rule does not stop too early in this setting (their Theorem~4.1) by comparing an upper bound on the deficit of model $K$ with a lower bound on the deficit of model $K-1$, and we follow this strategy. Our analysis differs from theirs in four respects.

First, our candidate families are finite. ESA allow infinitely many hyperparameters and control them through a R\'enyi-type cover of the family of priors and H\"older's inequality. This argument loses constant factors, so their condition is a ratio, $\En(K-1)>(1+c_*)\En(K)$ for an unspecified constant $c_*>0$. Since our candidates are finitely many, Lemma~\ref{th:lem:lower} uses a union bound instead, and the condition becomes the difference $\Ut(K)<\Lt(K-1)$. Under the exact model, this weakens a separation requirement on the community strengths to Assumption~(W).

Second, we compute the upper curve directly. ESA bound the deficit at the oracle through an exponential-moment assumption on the log-likelihood ratio (the second line of their Assumption~A). This assumption fails for the Gaussian covariance models considered here, because the exponential moment is infinite for parameters with small residual variances, to which our priors assign positive mass (Lemma~\ref{th:lem:A2}); it could hold on a parameter space whose residual variances are bounded away from zero. Lemma~\ref{th:lem:upper} avoids this through the decomposition in Lemma~\ref{th:lem:decomp}, in which the deficit is affine in the sample covariance.

Third, we prove a not-too-late result with empirical Bayes. ESA have a not-too-late result for variational Bayes with a fixed prior (their Theorem~3.4), but none with empirical Bayes. Theorem~\ref{th:thm:late} provides one, and Proposition~\ref{th:prop:curves}(c) verifies its condition under the exact model through the inverse-Wishart prior.

Finally, we bound the risk of a single selected model. ESA aggregate the variational posteriors of the visited models with exponential weights and prove a contraction rate for the aggregate (their Theorem~4.2). The search outputs the single best visited model, and Theorem~\ref{th:thm:risk} bounds its risk.

\subsection{Data-dependent choices, approximation, and the two-stage version}\label{th:sec:icons}

The results above concern the exploratory step with fixed candidate families, a fixed learning rate, exact maximization of the ELBO, and independent Gaussian observations. In the analyses, the families and the learning rate are computed from the same standardized data, the ELBOs are maximized numerically, and the real-data analyses use the two-stage version. We discuss these differences in turn.

\paragraph{Candidate paths.}
Suppose that the output of the dense-subgraph search equals a fixed family of paths with probability at least $1-\delta_n$. This holds, for example, when the output of the dense-subgraph search does not change under perturbations of the sample correlation matrix of max-norm smaller than some $\delta\in(0,1]$, since the sample correlation matrix is within $\delta$ of its population counterpart in max-norm with probability at least $1-C_1p_0^2e^{-c_1n\delta^2}$ for constants $C_1,c_1>0$. On this event, the search behaves exactly as with the fixed family, so every probability bound above holds after subtracting $\delta_n$.

\paragraph{The learning rate.}
The results hold for every fixed $\alpha\in(0,1)$. The exploratory step uses $\alpha_0$, which is computed from the data before any community is fitted. We treat it as fixed; a formal treatment of the plug-in would require the bounds above to hold uniformly over the values that $\alpha_0$ can take, which we do not pursue. The rate $\alpha_0$ applies the calibration principle of \citet{lyddon2019llb} to the null model and is then used for every $K$; this is our choice, and the principle does not by itself calibrate the other candidates. When communities that each contain a fixed fraction of the variables are present, $\alpha_0\approx p_0/\Vert R\Vert_F^2$ is of order $1/p_0$, so a fixed learning rate does not describe $\alpha_0$ when $p_0$ grows proportionally to $n$ with such communities.

\paragraph{Standardization.}
The analyses center the variables and divide them by their sample standard deviations before the dense-subgraph search, $\alpha_0$ and the forward stepwise search, whereas the results assume independent observations on the original scale. After standardization the rows are no longer independent, and the variables of a community are rescaled by different random factors, so the exact 1f structure holds only approximately. For fixed $p_0$ each factor is within $O_P(n^{-1/2})$ of its limit, but we do not analyze the effect on the search; when $p_0$ grows with $n$, the rescaling can dominate the covariance error measured on the correlation scale.

\paragraph{Numerical approximation.}
The results concern exact maxima of the ELBO, whereas the exploratory step maximizes Monte Carlo estimates of the ELBO with a local optimizer and evaluates the maxima with fresh draws (Appendix~\ref{app:selection-implementation}). A sufficient condition for these errors not to change the decisions is the following. Suppose that, simultaneously for every candidate that the search compares, the computed ELBO differs from the exact maximum by at most $e_n$ and the returned variational distribution has a deficit at most $e_n$ above the exact minimum, and let $\tau_n\ge0$ be the tolerance with which the implementation declares a decrease. Suppose further that every gap used in the proofs exceeds $2e_n+\tau_n$: the gaps $\Lt(K-1)-\Ut(K)$ for $1\le K\le K_*$, the gap $\Lt(K_*+1)-\Ut(K_*)$, and the gaps between the true partition and the other candidates with $K_*$ communities in the proof of Corollary~\ref{th:cor:exact}. Then, on the events of the proofs, the computed search makes the same decisions as the exact one, and the risk bound of Theorem~\ref{th:thm:risk} holds with $3e_n$ added to its right-hand side. Here the same decisions mean the same stopping layer, the same selected $K$ and the same selected partition; the winning partitions at layers below $K_*$ may differ, because the condition does not control the gaps among the candidates within those layers. If no path has more than $K_*$ communities, the gap at $K_*+1$ is not needed. The tolerance $\tau_n$ must cover every comparison that the implementation makes, including the final selection; without the gap conditions, a final selection anywhere within $\tau_n$ of the largest computed ELBO satisfies the risk bound of Theorem~\ref{th:thm:risk} with $3e_n+\tau_n$ added to its right-hand side and with $\Kstop$ replaced by the layer at which the computed search stops. The two-standard-error rule of Appendix~\ref{app:selection-implementation} guards against these errors in practice; it does not establish the condition.

\paragraph{The two-stage version.}
Along a fixed path $\pi$, stage~2 is the search above with the singleton families $\Fam_K=\{\pi_K\}$, where $\pi_K$ is the partition formed by the first $K$ communities of $\pi$. Thus $N_K=1$, and $m_K$ and $\mKL_K$ are the biases along $\pi$. The kept path $\widehat\pi$ is chosen from the data, but it is one of at most $|\mathcal H|$ paths. Let $\Pi_*$ be the set of paths whose first $K_*$ communities form the true partition, and suppose that (E) holds and that (F) and (W) hold for the singleton families of every path in $\Pi_*$. If $\widehat\pi\in\Pi_*$ and stage~2 along $\widehat\pi$ fails to select $K_*$ and $\rho_*$, then stage~2 along some fixed $\pi\in\Pi_*$ fails. Hence, by Corollary~\ref{th:cor:exact} and a union bound over $\Pi_*$, for all sufficiently large $n$,
\[
\Pst\bigl(\widehat K=K_*\text{ and }\widehat\rho=\rho_*\bigr)\ge1-\Pst(\widehat\pi\notin\Pi_*)-\frac{|\mathcal H|(2K_*+3)}n,
\]
and Corollary~\ref{th:cor:rate} carries over in the same way. The two-stage version therefore inherits the guarantees whenever, with probability tending to one, stage~1 keeps a path that contains the true partition and along which (W) holds. The first stage ensures neither condition: it compares the candidates at a single size, and the path whose first $K_{\mathrm{ref}}$ communities have the largest ELBO need not contain the true partition.

\section{Proofs for Appendix~\ref*{app:selection-theory}}
\label{th:app:proofs}

This appendix proves the results of Appendix~\ref{app:selection-theory}. Appendix~\ref{th:app:proofs-general} contains the proofs for Appendices~\ref{th:sec:curves} and~\ref{th:sec:theorems}, and Appendix~\ref{th:app:proofs-cases} those for Appendices~\ref{th:sec:exact} and~\ref{th:sec:misspec}.

\subsection{Proofs for Appendices~\ref*{th:sec:curves} and \ref*{th:sec:theorems}}\label{th:app:proofs-general}

Before giving the proofs, we give several lemmas used in the proofs. Lemmas~\ref{th:lem:dv} and \ref{th:lem:chaos} are standard probabilistic tools; Lemmas~\ref{th:lem:closed}--\ref{th:lem:A2} are facts about Gaussian distributions.

\subsubsection{Lemmas}\label{th:app:lemmas}

\begin{lemma}[Donsker and Varadhan's variational formula]\label{th:lem:dv}
Let $\Pi$ be a probability measure and $g$ a function with $\int e^{g}\,d\Pi<\infty$. Then
\[
\sup_Q\Bigl\{\int g\,dQ-\KL(Q,\Pi)\Bigr\}=\log\int e^{g}\,d\Pi ,
\]
where the supremum is over probability measures $Q$ \citep{donsker1975}. In particular, $\int g\,dQ\le\KL(Q,\Pi)+\log\int e^g\,d\Pi$ for every $Q$ with $\KL(Q,\Pi)<\infty$.
\end{lemma}

\begin{lemma}[Tail of Gaussian quadratic forms]\label{th:lem:chaos}
Let $z_1,\ldots,z_n$ be independent $\Norm_{p_0}(0,I)$ and $B$ symmetric. For $u>0$,
\[
P\Bigl(\sum_{i=1}^n\bigl(z_i^\top Bz_i-\tr B\bigr)>2\sqrt{2nu}\,\Vert B\Vert_F+2u\Vert B\Vert_{\op}\Bigr)\le2e^{-u}.
\]
\end{lemma}
\begin{proof}
Let $\mu_1,\ldots,\mu_{p_0}$ be the eigenvalues of $B$. The sum has the same law as $\sum_ja_j(Y_j^2-1)$, where the $Y_j$ are independent standard normal and $a\in\R^{np_0}$ repeats each $\mu_k$ $n$ times, so that $\Vert a\Vert_2=\sqrt n\Vert B\Vert_F$ and $\Vert a\Vert_\infty=\Vert B\Vert_{\op}$. Write $a=a^+-a^-$ with $a^\pm\ge0$. By Lemma~1 of \citet{laurent2000}, with probability at least $1-2e^{-u}$,
\[
\sum_ja_j^+(Y_j^2-1)\le2\Vert a^+\Vert_2\sqrt u+2\Vert a^+\Vert_\infty u,\qquad
-\sum_ja_j^-(Y_j^2-1)\le2\Vert a^-\Vert_2\sqrt u .
\]
Adding the two bounds,
\[
\sum_ja_j(Y_j^2-1)\le2\bigl(\Vert a^+\Vert_2+\Vert a^-\Vert_2\bigr)\sqrt u+2\Vert a\Vert_\infty u\le2\sqrt2\,\Vert a\Vert_2\sqrt u+2\Vert a\Vert_\infty u,
\]
where we use $\Vert a^+\Vert_2+\Vert a^-\Vert_2\le\sqrt2\Vert a\Vert_2$. Substituting $\Vert a\Vert_2$ and $\Vert a\Vert_\infty$ gives the claim.
\end{proof}

\begin{lemma}[Gaussian closed form]\label{th:lem:closed}
Let $\mu_1,\ldots,\mu_{p_0}$ be the eigenvalues of $L=\Omega_*^{-1/2}\Omega_\rho(\theta)\Omega_*^{-1/2}$. Then
\[
R_n(\rho,\theta)=n\alpha\,D_{1-\alpha}(\Pst\Vert P^\rho_\theta)=\frac n2\sum_{i=1}^{p_0}g_\alpha(\mu_i),\qquad g_\alpha(\mu)=\alpha\log\mu+\log\Bigl(1-\alpha+\frac\alpha\mu\Bigr)\ge0,
\]
with $D_\gamma(P\Vert Q)=(\gamma-1)^{-1}\log\int p^\gamma q^{1-\gamma}$, and $R_n(\rho,\theta)\le n\alpha\KL(\rho,\theta)$.
\end{lemma}
\begin{proof}
Write $\Omega=\Omega_\rho(\theta)$ and $p_\theta$, $p_*$ for the densities. By independence and a Gaussian integral,
\begin{align*}
R_n(\rho,\theta)&=-n\log\int p_\theta^\alpha p_*^{1-\alpha}
=\frac n2\Bigl\{\alpha\log|\Omega|+(1-\alpha)\log|\Omega_*|+\log\bigl|\alpha\Omega^{-1}+(1-\alpha)\Omega_*^{-1}\bigr|\Bigr\}\\
&=\frac n2\sum_{i=1}^{p_0}\Bigl\{\alpha\log\mu_i+\log\Bigl(1-\alpha+\frac\alpha{\mu_i}\Bigr)\Bigr\},
\end{align*}
where the last line follows by writing $\Omega=\Omega_*^{1/2}L\Omega_*^{1/2}$. By definition $-n\log\int p_\theta^\alpha p_*^{1-\alpha}=n\alpha D_{1-\alpha}(\Pst\Vert P^\rho_\theta)$. The weighted arithmetic--geometric mean inequality $(1-\alpha)+\alpha\mu^{-1}\ge\mu^{-\alpha}$ gives $g_\alpha\ge0$. Finally,
\[
g_\alpha(\mu)\le\alpha\log\mu+\Bigl(1-\alpha+\frac\alpha\mu\Bigr)-1=\alpha\Bigl(\frac1\mu-1+\log\mu\Bigr),
\]
where we use $\log x\le x-1$, and $\frac n2\sum_i(\mu_i^{-1}-1+\log\mu_i)=n\KL(\rho,\theta)$.
\end{proof}

\begin{lemma}[Lower bound on $g_\alpha$]\label{th:lem:glower}
$g_\alpha(\mu)\ge\frac29\alpha(1-\alpha)\min\{(\mu-1)^2,\frac14\}$ for all $\mu>0$. Hence, with $L$ as in Lemma~\ref{th:lem:closed},
\[
R_n(\rho,\theta)\ge\frac n9\alpha(1-\alpha)\min\Bigl\{\Vert L-I\Vert_F^2,\frac14\Bigr\}\ge\frac n9\alpha(1-\alpha)\min\Bigl\{\frac{\Vert\Omega_\rho(\theta)-\Omega_*\Vert_F^2}{\Vert\Omega_*\Vert_{\op}^2},\frac14\Bigr\}.
\]
\end{lemma}
\begin{proof}
Differentiation gives $g_\alpha'(\mu)=\alpha(1-\alpha)(\mu-1)/\{\mu((1-\alpha)\mu+\alpha)\}$, so $g_\alpha$ decreases on $(0,1)$, increases on $(1,\infty)$ and vanishes at $1$. For $\mu\in[\frac12,\frac32]$, the denominator is at most $\frac94$, so
\[
g_\alpha(\mu)=\int_1^\mu g_\alpha'(s)\,ds\ge\frac49\alpha(1-\alpha)\int_1^\mu(s-1)\,ds=\frac29\alpha(1-\alpha)(\mu-1)^2 .
\]
For $|\mu-1|>\frac12$, monotonicity gives $g_\alpha(\mu)\ge\min\{g_\alpha(\frac12),g_\alpha(\frac32)\}\ge\frac1{18}\alpha(1-\alpha)$. Summing over the eigenvalues,
\[
R_n=\frac n2\sum_ig_\alpha(\mu_i)\ge\frac n9\alpha(1-\alpha)\sum_i\min\Bigl\{(\mu_i-1)^2,\frac14\Bigr\}\ge\frac n9\alpha(1-\alpha)\min\Bigl\{\Vert L-I\Vert_F^2,\frac14\Bigr\},
\]
and the last bound of the lemma follows from $\Vert\Omega_\rho(\theta)-\Omega_*\Vert_F\le\Vert\Omega_*\Vert_{\op}\Vert L-I\Vert_F$.
\end{proof}

\begin{lemma}[Exponential moments of the log-likelihood ratio]\label{th:lem:A2}
For $s>0$, the expectation $\E_*\exp\{s\,\ld(\rho,\theta)\}$ is finite if and only if every eigenvalue of $L$ exceeds $s/(1+s)$.
\end{lemma}
\begin{proof}
For one observation $x=\Omega_*^{1/2}z$,
\[
\E_*\exp\Bigl\{s\log\frac{p_*(x)}{p_\theta(x)}\Bigr\}=\Bigl(\frac{|\Omega|}{|\Omega_*|}\Bigr)^{s/2}\E\exp\Bigl\{\frac s2z^\top(L^{-1}-I)z\Bigr\}=\Bigl(\frac{|\Omega|}{|\Omega_*|}\Bigr)^{s/2}\prod_{i=1}^{p_0}\bigl\{1-s(\mu_i^{-1}-1)\bigr\}^{-1/2}
\]
when every factor $1-s(\mu_i^{-1}-1)$ is positive, and the expectation is infinite otherwise. The condition $s(\mu_i^{-1}-1)<1$ is $\mu_i>s/(1+s)$.
\end{proof}

Every prior and every $Q\in\Qfam_K(\rho)$ gives positive mass to parameters with small residual variances, for which some $\mu_i\le s/(1+s)$. So no exponential-moment bound can hold for every $\theta$; Lemma~\ref{th:lem:upper} avoids this by working with the integrated quantity $\int\ld\,dQ$, which is linear in $S$.

\subsubsection{Proof of Lemma~\ref*{th:lem:lower}}

\begin{proof}
Fix $\rho\in\Fam_K$. By Lemma~\ref{th:lem:dv} with $g=-\alpha\ld(\rho,\cdot)$,
\[
-\Gamma(K;\rho)=\max_{Q\in\Qfam_K(\rho)}\Bigl\{-\alpha\int\ld\,dQ-\KL\bigl(Q,\Pi_K(\cdot\mid\rho)\bigr)\Bigr\}\le\log\int e^{-\alpha\ld(\rho,\theta)}\,\Pi_K(d\theta\mid\rho).
\]
Therefore, by Markov's inequality and Fubini's theorem,
\begin{align*}
\Pst\bigl(\Gamma(K;\rho)\le\En(K,\rho)-t\bigr)
&\le\Pst\Bigl(\int e^{-\alpha\ld(\rho,\theta)}\,\Pi_K(d\theta\mid\rho)\ge e^{t-\En(K,\rho)}\Bigr)\\
&\le e^{\En(K,\rho)-t}\int\E_*e^{-\alpha\ld(\rho,\theta)}\,\Pi_K(d\theta\mid\rho)\\
&=e^{\En(K,\rho)-t}\int e^{-R_n(\rho,\theta)}\,\Pi_K(d\theta\mid\rho)=e^{-t},
\end{align*}
where the first inequality follows from the last display, and the equalities use the definitions of $R_n$ and $\En(K,\rho)$. This proves the second claim. For the first claim, since $\Gamma(K)=\min_\rho\Gamma(K;\rho)$ and $\En(K,\rho)\ge\En(K)$ for every $\rho$,
\[
\Pst\bigl(\Gamma(K)\le\Lt(K)\bigr)\le\sum_{\rho\in\Fam_K}\Pst\bigl(\Gamma(K;\rho)\le\En(K,\rho)-\log N_K-t\bigr)\le N_K\,e^{-\log N_K-t}=e^{-t},
\]
where the second inequality is the second claim with $t+\log N_K$ in place of $t$.
\end{proof}

\subsubsection{Proof of Lemma~\ref*{th:lem:decomp}}

\begin{proof}
Write $\Omega_\theta=\Omega_\rho(\theta)$. Since $n\KL(\rho,\theta)=\frac n2\{\tr(\Omega_\theta^{-1}\Omega_*)-p_0+\log|\Omega_\theta|-\log|\Omega_*|\}$,
\begin{align*}
\ld(\rho,\theta)
&=\frac n2\Bigl\{\log|\Omega_\theta|-\log|\Omega_*|+\tr\bigl((\Omega_\theta^{-1}-\Omega_*^{-1})S\bigr)\Bigr\}\\
&=n\KL(\rho,\theta)+\frac n2\tr\bigl\{(\Omega_\theta^{-1}-\Omega_*^{-1})(S-\Omega_*)\bigr\}.
\end{align*}
All moments below are finite because $\Omega_\theta^{-1}\preceq D_\rho(\tau)^{-1}$. Integrating over $Q$ and using $nS=\Omega_*^{1/2}(\sum_iz_iz_i^\top)\Omega_*^{1/2}$,
\begin{align*}
\int\ld\,dQ
&=n\,\E_Q\KL(\rho,\theta)+\frac n2\tr\bigl\{(\E_Q\Omega_\theta^{-1}-\Omega_*^{-1})(S-\Omega_*)\bigr\}\\
&=n\,\E_Q\KL(\rho,\theta)+\frac12\sum_{i=1}^n\bigl(z_i^\top B_Qz_i-\tr B_Q\bigr).
\end{align*}
Multiplying by $\alpha$ and adding $\KL(Q,\Pi_K(\cdot\mid\rho))$ gives the claim.
\end{proof}

\subsubsection{Proof of Lemma~\ref*{th:lem:upper}}

\begin{proof}
Write $\rho^\circ=\rho^\circ_K$, $Q^\circ=Q^\circ_K$ and $B=B_{Q^\circ}$. By the minimality of $\Gamma(K)$ and Lemma~\ref{th:lem:decomp},
\begin{align*}
\Gamma(K)\le\Gamma(K;\rho^\circ)&\le F(K,\rho^\circ,Q^\circ)\\
&=\alpha n\,\E_{Q^\circ}\KL(\rho^\circ,\theta)+\KL\bigl(Q^\circ,\Pi_K(\cdot\mid\rho^\circ)\bigr)+\frac\alpha2\sum_{i=1}^n\bigl(z_i^\top Bz_i-\tr B\bigr).
\end{align*}
The first two terms of the last line are the first two terms of $\Ut(K)$. Therefore
\begin{align*}
\Pst\bigl(\Gamma(K;\rho^\circ)>\Ut(K)\bigr)
&\le P\Bigl(\sum_{i=1}^n\bigl(z_i^\top Bz_i-\tr B\bigr)>2\sqrt{2n(t+1)}\,\Vert B\Vert_F+2(t+1)\Vert B\Vert_{\op}\Bigr)\\
&\le2e^{-(t+1)}\le e^{-t},
\end{align*}
where we use the last display and the definition of $\Ut(K)$ for the first inequality, and Lemma~\ref{th:lem:chaos} with $u=t+1$ for the second. Since $\Gamma(K)\le\Gamma(K;\rho^\circ)$, the same bound holds for $\Gamma(K)$.
\end{proof}

\subsubsection{Proof of Theorem~\ref*{th:thm:early}}

\begin{proof}
Since $\mathrm{ELBO}_{K-1}-\mathrm{ELBO}_K=\Gamma(K)-\Gamma(K-1)$, we proceed as
\[
\Pst\bigl(\mathrm{ELBO}_K\le\mathrm{ELBO}_{K-1}\text{ for some }1\le K\le K^\dagger\bigr)\le\sum_{K=1}^{K^\dagger}\Pst\bigl(\Gamma(K)\ge\Gamma(K-1)\bigr).
\]
For each summand, we decompose it as
\begin{equation}\label{th:eq:early-decomp}
\Pst\bigl(\Gamma(K)\ge\Gamma(K-1)\bigr)\le\Pst\bigl(\Gamma(K)>\Ut(K)\bigr)+\Pst\bigl(\Gamma(K-1)\le\Ut(K)\bigr),
\end{equation}
which holds because $\Gamma(K)\le\Ut(K)<\Gamma(K-1)$ implies $\Gamma(K)<\Gamma(K-1)$. The first term of the right-hand side of \eqref{th:eq:early-decomp} is at most $e^{-t}$ by Lemma~\ref{th:lem:upper}. For the second term, the assumption $\Ut(K)<\Lt(K-1)$ and Lemma~\ref{th:lem:lower} give
\[
\Pst\bigl(\Gamma(K-1)\le\Ut(K)\bigr)\le\Pst\bigl(\Gamma(K-1)\le\Lt(K-1)\bigr)\le e^{-t}.
\]
Combining the two bounds in \eqref{th:eq:early-decomp} and summing over $K$, the probability that $\mathrm{ELBO}_K\le\mathrm{ELBO}_{K-1}$ for some $K\le K^\dagger$ is at most $2K^\dagger e^{-t}$. On the complementary event the ELBO does not decrease before $K^\dagger+1$, so $\Kstop\ge\min\{K^\dagger+1,M\}$.
\end{proof}

\subsubsection{Proof of Theorem~\ref*{th:thm:late}}

\begin{proof}
The event $\{\Kstop\ge K+1\}$ implies that the ELBO does not decrease on $\{k^\circ,\ldots,K\}$, so that $\mathrm{ELBO}_K\ge\mathrm{ELBO}_{k^\circ}$. Therefore
\[
\Pst(\Kstop\ge K+1)\le\Pst(\mathrm{ELBO}_K\ge\mathrm{ELBO}_{k^\circ})=\Pst\bigl(\Gamma(K)\le\Gamma(k^\circ)\bigr).
\]
We decompose the right-hand side as
\begin{equation}\label{th:eq:late-decomp}
\Pst\bigl(\Gamma(K)\le\Gamma(k^\circ)\bigr)\le\Pst\bigl(\Gamma(k^\circ)>\Ut(k^\circ)\bigr)+\Pst\bigl(\Gamma(K)\le\Ut(k^\circ)\bigr),
\end{equation}
which holds because $\Gamma(k^\circ)\le\Ut(k^\circ)<\Gamma(K)$ implies $\Gamma(K)>\Gamma(k^\circ)$. The first term of \eqref{th:eq:late-decomp} is at most $e^{-t}$ by Lemma~\ref{th:lem:upper}. For the second term, the assumption $\Lt(K)\ge\Ut(k^\circ)$ and Lemma~\ref{th:lem:lower} give
\[
\Pst\bigl(\Gamma(K)\le\Ut(k^\circ)\bigr)\le\Pst\bigl(\Gamma(K)\le\Lt(K)\bigr)\le e^{-t},
\]
which completes the proof.
\end{proof}

\subsubsection{Proof of Lemma~\ref*{th:lem:pacbayes}}

\begin{proof}
Fix $K$ and $\rho\in\Fam_K$, and let $g=R_n(\rho,\cdot)-\alpha\ld(\rho,\cdot)$. Since $R_n$ is non-random, Fubini's theorem and the definition of $R_n$ give
\begin{align*}
\E_*\int e^{g(\theta)}\,\Pi_K(d\theta\mid\rho)&=\int e^{R_n(\rho,\theta)}\,\E_*e^{-\alpha\ld(\rho,\theta)}\,\Pi_K(d\theta\mid\rho)\\
&=\int e^{R_n(\rho,\theta)}e^{-R_n(\rho,\theta)}\,\Pi_K(d\theta\mid\rho)=1 .
\end{align*}
Hence, by Markov's inequality and a union bound over the $\Ntot$ pairs $(K,\rho)$,
\[
\Pst\Bigl(\int e^{g}\,d\Pi_K(\cdot\mid\rho)\ge\Ntot e^t\text{ for some }(K,\rho)\Bigr)\le\sum_{(K,\rho)}\frac{e^{-t}}{\Ntot}=e^{-t}.
\]
On the complementary event, Lemma~\ref{th:lem:dv} gives, for every $(K,\rho)$ and every $Q\in\Qfam_K(\rho)$,
\begin{align*}
\int R_n\,dQ-\alpha\int\ld\,dQ&\le\KL\bigl(Q,\Pi_K(\cdot\mid\rho)\bigr)+\log\int e^g\,d\Pi_K(\cdot\mid\rho)\\
&<\KL\bigl(Q,\Pi_K(\cdot\mid\rho)\bigr)+\log\Ntot+t,
\end{align*}
which is the claim.
\end{proof}

\subsubsection{Proof of Theorem~\ref*{th:thm:risk}}

\begin{proof}
Let $\mathfrak A$ be the event of Lemma~\ref{th:lem:pacbayes} and $\mathfrak B=\{\Gamma(k^\circ)\le\Ut(k^\circ)\}$. By Lemmas~\ref{th:lem:pacbayes} and \ref{th:lem:upper}, $\Pst(\mathfrak A^c)\le e^{-t}$ and $\Pst(\mathfrak B^c)\le e^{-t}$. On $\mathfrak A\cap\mathfrak B\cap\{k^\circ\le\Kstop\}$,
\begin{align*}
\int R_n(\widehat\rho,\theta)\,\widehat Q(d\theta)
&\le F(\widehat K,\widehat\rho,\widehat Q)+\log\Ntot+t
=\Gamma(\widehat K)+\log\Ntot+t\\
&\le\Gamma(k^\circ)+\log\Ntot+t
\le\Ut(k^\circ)+\log\Ntot+t,
\end{align*}
where we use the event $\mathfrak A$ for the first inequality, the definition of $(\widehat K,\widehat\rho,\widehat Q)$ for the equality, $k^\circ\le\Kstop$ and the maximality of $\mathrm{ELBO}_{\widehat K}$ over $K\le\Kstop$ for the second inequality, and the event $\mathfrak B$ for the last inequality. This proves the first claim.

For the second claim, let $L_\theta=\Omega_*^{-1/2}\Omega_{\widehat\rho}(\theta)\Omega_*^{-1/2}$ and $A_\varepsilon=\{\theta:\Vert L_\theta-I\Vert_F>\varepsilon\}$. For $\varepsilon\le\frac12$, Lemma~\ref{th:lem:glower} gives $R_n(\widehat\rho,\theta)\ge\frac n9\alpha(1-\alpha)\varepsilon^2$ on $A_\varepsilon$. Hence, by Markov's inequality under $\widehat Q$,
\[
\widehat Q(A_\varepsilon)\le\frac{9}{n\alpha(1-\alpha)\varepsilon^2}\int R_n(\widehat\rho,\theta)\,\widehat Q(d\theta)\le\frac{9\{\Ut(k^\circ)+\log\Ntot+t\}}{n\alpha(1-\alpha)\varepsilon^2}
\]
on the same event.
\end{proof}

\subsection{Proofs for Appendices~\ref*{th:sec:exact} and \ref*{th:sec:misspec}}\label{th:app:proofs-cases}

Before giving the proofs, we give four lemmas used in the proofs: two on the structure of the 1f model (Lemmas~\ref{th:lem:rep} and \ref{th:lem:zero}), one on the inverse-Wishart prior near singular matrices (Lemma~\ref{th:lem:iw}), and one on conjugate oracles (Lemma~\ref{th:lem:oracle}).

\subsubsection{Lemmas on the 1f model}

\begin{lemma}[Representation]\label{th:lem:rep}
Assume (E). Let $\rho$ have $K$ communities and suppose $\Omega_*=A_\rho\Phi A_\rho^\top+D$ with $\Phi\succeq0$ and $D\succeq0$ diagonal and constant on the blocks of $\rho$. Then (i) $K\ge K_*$; (ii) if $K=K_*$, then $\rho=\rho_*$ up to labels; (iii) if $K>K_*$, then either some community $\mathcal V_a$ of $\rho$ lies in the true background and row $a$ of $\Phi$ vanishes, or two communities $\mathcal V_a,\mathcal V_b$ of $\rho$ lie in the same true community and rows $a,b$ of $\Phi$ coincide.
\end{lemma}
\begin{proof}
Let $\mathcal V^*_1,\ldots,\mathcal V^*_{K_*}$ be the true communities and $\mathcal V^*_0$ the true background. We first show that no variable of a true community lies in the background of $\rho$. If $j\in \mathcal V^*_h$, then $(\Omega_*)_{jk}=(\Phi_*)_{hh}>0$ for any other $k\in \mathcal V^*_h$, whereas a background variable of $\rho$ is uncorrelated with all other variables under the representation.

Next we show that each community of $\rho$ lies inside a true community or inside $\mathcal V^*_0$. Let $j\ne j'$ belong to one community of $\rho$. The representation gives $\Omega_{jl}=\Omega_{j'l}$ for every $l\notin\{j,j'\}$. If $j\in \mathcal V^*_h$ and $j'\in\mathcal V^*_0$, taking $l\in \mathcal V^*_h\setminus\{j\}$ gives
\[
(\Phi_*)_{hh}=(\Omega_*)_{jl}=(\Omega_*)_{j'l}=0,
\]
a contradiction. If $j\in \mathcal V^*_h$ and $j'\in \mathcal V^*_g$ with $g\ne h$, taking $l\in \mathcal V^*_h\setminus\{j\}$ and then $l'\in \mathcal V^*_g\setminus\{j'\}$ gives $(\Phi_*)_{hh}=(\Phi_*)_{hg}$ and $(\Phi_*)_{gg}=(\Phi_*)_{hg}$, so the $2\times2$ submatrix of $\Phi_*$ on $\{h,g\}$ has four equal entries, contradicting $\Phi_*\succ0$.

Consequently every true community contains at least one community of $\rho$, which gives (i). If $K=K_*$, each true community contains exactly one community of $\rho$, which must be all of it, and the background of $\rho$ is $\mathcal V^*_0$; this gives (ii). If $K>K_*$, either some $\mathcal V_a\subset\mathcal V^*_0$, and then $\Phi_{aa}=(\Omega_*)_{jj'}=0$ for $j\ne j'\in \mathcal V_a$, so row $a$ vanishes because $\Phi\succeq0$; or some $\mathcal V^*_h$ contains two communities $\mathcal V_a,\mathcal V_b$, and their rows of $\Phi$ coincide because the variables of $\mathcal V^*_h$ have identical covariances with all other variables. This gives (iii).
\end{proof}

\begin{lemma}[Zero bias]\label{th:lem:zero}
Assume (E). If $\inf_\theta D_{1-\alpha}(\Pst\Vert P^\rho_\theta)=0$, then $\Omega_*$ has a representation as in Lemma~\ref{th:lem:rep}. Hence $m_K>0$ for $K<K_*$, and every $\rho\in\Fam_{K_*}$ other than $\rho_*$ has a positive R\'enyi bias.
\end{lemma}
\begin{proof}
Take $\theta_m=(\Phi_m,\tau_m)$ with $D_{1-\alpha}(\Pst\Vert P^\rho_{\theta_m})\to0$. By Lemmas~\ref{th:lem:closed} and \ref{th:lem:glower}, $\Omega_\rho(\theta_m)\to\Omega_*$. Every entry of $\Phi_m$ appears as an off-diagonal entry of $\Omega_\rho(\theta_m)$, because every community has at least two variables, so $\Phi_m$ converges to some $\Phi\succeq0$. The diagonal entries of $\Omega_\rho(\theta_m)$ also converge, so $v_{m,h}$ converges to some $v_h\ge0$ for every non-empty block $\mathcal V_h$. If the background is empty, $v_{m,0}$ does not enter $\Omega_\rho(\theta_m)$, and we set $v_0=0$. Passing to the limit in $\Omega_\rho(\theta_m)$ gives the representation. The two consequences follow from Lemma~\ref{th:lem:rep}(i) and (ii).
\end{proof}

\begin{lemma}[Inverse-Wishart boundary penalty]\label{th:lem:iw}
Assume (E), let $K>K_*$, and let $\rho\in\Fam_K$ have zero R\'enyi bias. Let $\varepsilon_n=(n\alpha(1-\alpha))^{-1/3}$. If $\varepsilon_n\le\min\{c/(16\nu_KK),\Vert\Omega_*\Vert_{\op}/\sqrt2\}$, then $\En(K,\rho)\ge c_0(n\alpha(1-\alpha))^{1/3}-\log2$.
\end{lemma}
\begin{proof}
By Lemmas~\ref{th:lem:zero} and \ref{th:lem:rep}(iii), $\rho$ either carves a community $\mathcal V_a$ out of the true background or splits a true community $\mathcal V^*_h$ into $\mathcal V_a$ and $\mathcal V_b$. For $\varepsilon>0$ we split the integral as
\begin{equation}\label{th:eq:iw-split}
\int e^{-R_n(\rho,\theta)}\,\Pi_K(d\theta\mid\rho)\le\Pi_K\bigl(\lambda_{\min}(\Phi)<\varepsilon\mid\rho\bigr)+\sup_{\lambda_{\min}(\Phi)\ge\varepsilon}e^{-R_n(\rho,\theta)},
\end{equation}
and bound the two terms separately.

For the first term of \eqref{th:eq:iw-split}, note that $\lambda_{\min}(\Phi)<\varepsilon$ implies $\tr(c\Phi^{-1})>c/\varepsilon$, and that $c\Phi^{-1}\sim\Wis_K(\nu_K,I)$ under the prior, so $\tr(c\Phi^{-1})\sim\chi^2_m$ with $m=\nu_KK$. Hence, for $\varepsilon\le c/(16m)$,
\[
\Pi_K\bigl(\lambda_{\min}(\Phi)<\varepsilon\mid\rho\bigr)\le P\bigl(\chi^2_m\ge c/\varepsilon\bigr)\le\exp\Bigl\{-\frac{c/\varepsilon-m}2+\frac m2\log\frac{c/\varepsilon}m\Bigr\}\le e^{-c/(4\varepsilon)},
\]
where the second inequality is the Chernoff bound for the chi-square distribution and the third holds because $u\ge2+2\log u$ for $u=c/(\varepsilon m)\ge16$.

For the second term of \eqref{th:eq:iw-split}, let $\lambda_{\min}(\Phi)\ge\varepsilon$. In the carve-out case, $\Omega_\rho(\theta)_{jj'}=\Phi_{aa}\ge\varepsilon$ while $(\Omega_*)_{jj'}=0$ for $j\ne j'\in \mathcal V_a$, and there are at least two such ordered pairs, so $\Vert\Omega_\rho(\theta)-\Omega_*\Vert_F^2\ge2\varepsilon^2$. In the split case, $\Phi_{aa}+\Phi_{bb}-2\Phi_{ab}=(e_a-e_b)^\top\Phi(e_a-e_b)\ge2\varepsilon$, while the same combination of the true values is zero. So one of $\Phi_{aa},\Phi_{bb},\Phi_{ab}$ differs from $(\Phi_*)_{hh}$ by at least $\varepsilon/2$, and each appears in at least two off-diagonal entries; thus $\Vert\Omega_\rho(\theta)-\Omega_*\Vert_F^2\ge\varepsilon^2/2$. In both cases Lemma~\ref{th:lem:glower} gives, for $\varepsilon\le\Vert\Omega_*\Vert_{\op}/\sqrt2$,
\[
R_n(\rho,\theta)\ge\frac n9\alpha(1-\alpha)\frac{\varepsilon^2}{2\Vert\Omega_*\Vert_{\op}^2}=\frac{n\alpha(1-\alpha)\varepsilon^2}{18\Vert\Omega_*\Vert_{\op}^2}.
\]
Now take $\varepsilon=\varepsilon_n$ in \eqref{th:eq:iw-split}, so that $1/\varepsilon_n=(n\alpha(1-\alpha))^{1/3}$ and $n\alpha(1-\alpha)\varepsilon_n^2=(n\alpha(1-\alpha))^{1/3}$. Then
\begin{align*}
\int e^{-R_n(\rho,\theta)}\,\Pi_K(d\theta\mid\rho)
&\le\exp\Bigl\{-\frac c4\bigl(n\alpha(1-\alpha)\bigr)^{1/3}\Bigr\}+\exp\Bigl\{-\frac{(n\alpha(1-\alpha))^{1/3}}{18\Vert\Omega_*\Vert_{\op}^2}\Bigr\}\\
&\le2\exp\Bigl\{-c_0\bigl(n\alpha(1-\alpha)\bigr)^{1/3}\Bigr\},
\end{align*}
where the last inequality uses the definition of $c_0$. Taking $-\log$ of both sides gives the claim.
\end{proof}

For a fixed centre $\theta_c=(\Phi_c,v_c)$ with $\Phi_c\succ0$ and $v_c>0$, the \emph{conjugate oracle} $Q_{\theta_c}\in\Qfam_K(\rho)$ is
\begin{gather*}
\Phi\sim\IW_K\bigl(\tilde\nu,(\tilde\nu-K-1)\Phi_c\bigr),\qquad\tilde\nu=\nu_K+\alpha n,\\
\tau_h\sim\Gam\bigl(\tilde a_h,(\tilde a_h-1)v_{c,h}\bigr),\qquad\tilde a_h=a_\tau+\alpha n|\mathcal V_h|/2,\qquad h=0,\ldots,K .
\end{gather*}
Then $\E\Phi=\Phi_c$ and $\E v_h=v_{c,h}$, and since $\Omega_\rho$ is affine in $(\Phi,v)$, $\E\,\Omega_\rho(\theta)=\Omega_\rho(\theta_c)$. Let $B_c=\Omega_*^{1/2}\Omega_\rho(\theta_c)^{-1}\Omega_*^{1/2}-I$.

\begin{lemma}[Conjugate oracle bounds]\label{th:lem:oracle}
For a fixed centre $\theta_c$, let $d_\rho=d_K$ if the background of $\rho$ is non-empty and $d_\rho=d_K-1$ otherwise. As $\alpha n\to\infty$:
\begin{enumerate}[label=(\alph*),nosep]
\item $\KL(Q_{\theta_c},\Pi_K(\cdot\mid\rho))=\frac{d_\rho}2\log(\alpha n)+O(1)$;
\item $n\,\E_{Q_{\theta_c}}\KL(\rho,\theta)\le n\KL(\rho,\theta_c)+O(\alpha^{-1})$;
\item $\Vert B_{Q_{\theta_c}}-B_c\Vert_F=O((\alpha n)^{-1})$.
\end{enumerate}
Consequently the upper curve with this oracle satisfies
\[
\Ut(K)\le\alpha n\KL(\rho,\theta_c)+\frac{d_\rho}2\log(\alpha n)+\alpha\sqrt{2n(t+1)}\,\Vert B_c\Vert_F+\alpha(t+1)\Vert B_c\Vert_{\op}+O\bigl(1+\sqrt{t/n}+t/n\bigr).
\]
\end{lemma}
\begin{proof}
\emph{Part (a).} The divergence splits as $\KL(Q_{\theta_c},\Pi_K)=\KL_\Phi+\sum_{h=0}^K\KL_{\tau_h}$. KL divergence is invariant under the bijection $\Phi\mapsto\Phi^{-1}$, and $\Phi\sim\IW_K(\nu,\Psi)$ exactly when $\Phi^{-1}\sim\Wis_K(\nu,\Psi^{-1})$. For $W_1\sim\Wis_K(n_1,T_1)$, using $\E\log|W_1|=\psi_K(n_1/2)+K\log2+\log|T_1|$ and $\E W_1=n_1T_1$, where $\psi_K(x)=\sum_{i=1}^K\psi(x-(i-1)/2)$, we obtain
\begin{multline*}
\KL\bigl(\Wis(n_1,T_1)\Vert\Wis(n_0,T_0)\bigr)=\frac{n_0}2\log\frac{|T_0|}{|T_1|}+\frac{n_1}2\bigl\{\tr(T_0^{-1}T_1)-K\bigr\}\\
+\log\frac{\Gamma_K(n_0/2)}{\Gamma_K(n_1/2)}+\frac{n_1-n_0}2\psi_K\Bigl(\frac{n_1}2\Bigr).
\end{multline*}
We apply this with $n_1=\tilde\nu$, $T_1=\{(\tilde\nu-K-1)\Phi_c\}^{-1}$, $n_0=\nu_K$ and $T_0=c^{-1}I$. Stirling's formula and $\psi(z)=\log z-\frac1{2z}+O(z^{-2})$ then give
\[
\KL_\Phi=\frac{K(K+1)}4\log\tilde\nu+\frac{\nu_K}2\log\Bigl|\frac{\Phi_c}c\Bigr|+\frac c2\tr(\Phi_c^{-1})+C_{K,\nu_K}+O(\tilde\nu^{-1}),
\]
with $C_{K,\nu}=\log\Gamma_K(\frac\nu2)+(\frac{\nu K}2-\frac{K(K+1)}4)\log2-\frac{K(K+1)}4-\frac K2\log2\pi-\frac{K(K-1)}4\log\pi$. For shape--rate Gamma distributions,
\[
\KL\bigl(\Gam(\tilde a,\tilde\beta)\Vert\Gam(a_\tau,b_\tau)\bigr)=(\tilde a-a_\tau)\psi(\tilde a)-\log\Gamma(\tilde a)+\log\Gamma(a_\tau)+a_\tau\log\frac{\tilde\beta}{b_\tau}+\tilde a\,\frac{b_\tau-\tilde\beta}{\tilde\beta},
\]
and with $\tilde\beta=(\tilde a-1)v_c$ the same expansions give
\[
\KL_\tau=\frac12\log\tilde a-\frac12\log(2\pi e)+\log\Gamma(a_\tau)+a_\tau\log\frac{v_c}{b_\tau}+\frac{b_\tau}{v_c}+O(\tilde a^{-1}).
\]
Since $\log\tilde\nu$, and $\log\tilde a_h$ for every non-empty $\mathcal V_h$, are $\log(\alpha n)+O(1)$, the coefficient of $\log(\alpha n)$ in $\KL_\Phi+\sum_h\KL_{\tau_h}$ is $K(K+1)/4+(K+1)/2=d_K/2$. If the background $\mathcal V_0$ is empty, then $\tilde a_0=a_\tau$ does not grow, $\KL_{\tau_0}=O(1)$, and the coefficient drops to $(d_K-1)/2$.

\emph{Parts (b) and (c).} We use three standard moment bounds, valid for large $\alpha n$: (M1) $\E\Vert\Phi-\Phi_c\Vert_F^4\le C\tilde\nu^{-2}$ for the inverse Wishart distribution with mean $\Phi_c$ (Wishart and inverse Wishart moments are treated in Chapter~3 of \citealp{gupta2000}); (M2) $\E(v_h-v_{c,h})^4\le C\tilde a_h^{-2}$, from inverse-Gamma moments; (M3) $\E\tau_h^6\le C$. Let $\Omega=\Omega_\rho(\theta)$, $\Omega_c=\Omega_\rho(\theta_c)$ and $H=\Omega-\Omega_c$. Then $\E H=0$, and (M1)--(M2) give $\E\Vert H\Vert_F^4\le C(\alpha n)^{-2}$. Since $\Omega\succeq D_\rho(v)$ and $\Omega_c\succeq D_\rho(v_c)$, every $\Omega_s=\Omega_c+sH$ with $s\in[0,1]$ satisfies $\Vert\Omega_s^{-1}\Vert_{\op}\le \bar\tau:=\max_h\max\{\tau_{c,h},\tau_h\}$, and (M3) gives $\E \bar\tau^6\le C$.

For part (b), let $f(\Omega)=\KL(\Pst\Vert\Norm_{p_0}(0,\Omega))$. Its second derivative along $\Omega_s$ satisfies
\[
\frac{d^2}{ds^2}f(\Omega_s)=\tr\bigl(\Omega_s^{-1}H\Omega_s^{-1}H\Omega_s^{-1}\Omega_*\bigr)-\frac12\tr\bigl(\Omega_s^{-1}H\Omega_s^{-1}H\bigr)\le\Vert\Omega_s^{-1}\Vert_{\op}^3\Vert\Omega_*\Vert_{\op}\Vert H\Vert_F^2 .
\]
Therefore
\begin{align*}
\E f(\Omega)-f(\Omega_c)
&=\E\bigl\{f(\Omega)-f(\Omega_c)-\langle\nabla f(\Omega_c),H\rangle\bigr\}
=\E\int_0^1(1-s)\frac{d^2}{ds^2}f(\Omega_s)\,ds\\
&\le\frac12\Vert\Omega_*\Vert_{\op}\,\E\bigl\{\bar\tau^3\Vert H\Vert_F^2\bigr\}
\le\frac12\Vert\Omega_*\Vert_{\op}\bigl(\E \bar\tau^6\bigr)^{1/2}\bigl(\E\Vert H\Vert_F^4\bigr)^{1/2}=O\bigl((\alpha n)^{-1}\bigr),
\end{align*}
where we use $\E H=0$ for the first equality, Taylor's formula with integral remainder for the second, the bound on the second derivative for the first inequality, and the Cauchy--Schwarz inequality for the second. Multiplying by $n$ gives (b).

For part (c), applying the resolvent identity $\Omega^{-1}=\Omega_c^{-1}-\Omega_c^{-1}H\Omega^{-1}$ twice gives
\[
\Omega^{-1}=\Omega_c^{-1}-\Omega_c^{-1}H\Omega_c^{-1}+\Omega_c^{-1}H\Omega_c^{-1}H\Omega^{-1}.
\]
The middle term has mean zero, so
\[
\bigl\Vert\E\Omega^{-1}-\Omega_c^{-1}\bigr\Vert_F\le\Vert\Omega_c^{-1}\Vert_{\op}^2\,\E\bigl\{\Vert H\Vert_F^2\bar\tau\bigr\}\le\Vert\Omega_c^{-1}\Vert_{\op}^2\bigl(\E\Vert H\Vert_F^4\bigr)^{1/2}\bigl(\E \bar\tau^2\bigr)^{1/2}=O\bigl((\alpha n)^{-1}\bigr),
\]
and $\Vert B_{Q_{\theta_c}}-B_c\Vert_F\le\Vert\Omega_*\Vert_{\op}\Vert\E\Omega^{-1}-\Omega_c^{-1}\Vert_F$.

Finally, inserting (a)--(c) into the definition of $\Ut(K)$ gives the bound on the upper curve.
\end{proof}

\subsubsection{Proof of Proposition~\ref*{th:prop:curves}}

\begin{proof}
\emph{Part (a).} By the remark after Definition~\ref{th:def:lower}, $\Lt(K)\ge n\alpha m_K-\log N_K-\log n$, and $m_K>0$ for $K<K_*$ by Lemma~\ref{th:lem:zero}. For the upper curve, let $\rho^\circ_K$ attain $\mKL_K$ and choose a fixed centre $\theta_c$ with $\KL(\rho^\circ_K,\theta_c)\le\mKL_K+\delta$. Lemma~\ref{th:lem:oracle} with $t=\log n$ gives
\begin{align*}
\Ut(K)&\le\alpha n(\mKL_K+\delta)+\frac{d_K}2\log(\alpha n)+\alpha\sqrt{2n(\log n+1)}\,\Vert B_c\Vert_F+O(\log n)\\
&=n\alpha(\mKL_K+\delta)+O(\sqrt{n\log n}),
\end{align*}
where we use $d_\rho\le d_K$ for the first line and absorb the term $\alpha(\log n+1)\Vert B_c\Vert_{\op}$ into $O(\log n)$, which is possible because $\theta_c$ is fixed.

\emph{Part (b).} Take $\rho^\circ_{K_*}=\rho_*$ and $\theta_c=\theta_*$. Then $\KL(\rho_*,\theta_*)=0$ and $B_c=0$, so Lemma~\ref{th:lem:oracle} with $t=\log n$ gives
\[
\Ut(K_*)\le\frac{d_{\rho_*}}2\log(\alpha n)+O(1)\le\frac{d_*}2\log n+O(1),
\]
where we use $d_{\rho_*}=d_*$ and $\alpha<1$.

\emph{Part (c).} Let $K>K_*$. Each $\rho\in\Fam_K$ either has a positive R\'enyi bias $\beta_\rho$, in which case $\En(K,\rho)\ge n\alpha\beta_\rho$, or has zero bias, in which case Lemma~\ref{th:lem:iw} gives $\En(K,\rho)\ge c_0(n\alpha(1-\alpha))^{1/3}-\log2$ for large $n$. Since $\Fam_K$ is finite and $n\alpha\beta_\rho$ grows linearly in $n$, for large $n$ every $\rho\in\Fam_K$ satisfies the second bound. Hence $\En(K)\ge c_0(n\alpha(1-\alpha))^{1/3}-\log2$, and (c) follows from $\Lt(K)=\En(K)-\log N_K-\log n$.
\end{proof}

\subsubsection{Proof of Corollary~\ref*{th:cor:exact}}

\begin{proof}
Take $t=\log n$ and the oracles of Proposition~\ref{th:prop:curves}. Consider the events
\begin{align*}
\mathfrak E_1&=\{\mathrm{ELBO}_K>\mathrm{ELBO}_{K-1}\text{ for all }1\le K\le K_*\},\\
\mathfrak E_2&=\{\mathrm{ELBO}_{K_*+1}<\mathrm{ELBO}_{K_*}\},\\
\mathfrak E_3&=\{\Gamma(K_*;\rho)>\Gamma(K_*;\rho_*)\text{ for all }\rho\in\Fam_{K_*}\setminus\{\rho_*\}\},
\end{align*}
with $\mathfrak E_2$ taken to be the whole space if $M=K_*$. On $\mathfrak E_1\cap\mathfrak E_2\cap\mathfrak E_3$, the ELBO increases up to $K_*$ and decreases at $K_*+1$, so the search stops at $K_*+1$ (or at $M=K_*$), the maximum over the visited set is attained only at $K_*$, and $\widehat\rho=\rho_*$. It therefore suffices to show that $\Pst(\mathfrak E_1^c)\le2K_*/n$, $\Pst(\mathfrak E_2^c)\le2/n$ and $\Pst(\mathfrak E_3^c)\le N_{K_*}/n$.

For $\mathfrak E_1$, we verify the condition of Theorem~\ref{th:thm:early} with $K^\dagger=K_*$. For $1\le K\le K_*-1$, Proposition~\ref{th:prop:curves}(a) gives
\[
\Lt(K-1)-\Ut(K)\ge n\alpha\bigl(m_{K-1}-\mKL_K-\delta\bigr)-O(\sqrt{n\log n}),
\]
which is positive for large $n$ once $\delta<m_{K-1}-\mKL_K$; such $\delta$ exists by (W). For $K=K_*$, Proposition~\ref{th:prop:curves}(a) and (b) give $\Lt(K_*-1)-\Ut(K_*)\ge n\alpha m_{K_*-1}-O(\log n)>0$ for large $n$. Theorem~\ref{th:thm:early} then gives $\Pst(\mathfrak E_1^c)\le2K_*/n$.

For $\mathfrak E_2$, Proposition~\ref{th:prop:curves}(b) and (c) give
\[
\Lt(K_*+1)-\Ut(K_*)\ge c_0\bigl(n\alpha(1-\alpha)\bigr)^{1/3}-O(\log n)>0
\]
for large $n$, so Theorem~\ref{th:thm:late} with $k^\circ=K_*$ gives $\Pst(\mathfrak E_2^c)\le2/n$.

For $\mathfrak E_3$, every $\rho\in\Fam_{K_*}\setminus\{\rho_*\}$ has a positive R\'enyi bias $\beta_\rho$ by Lemma~\ref{th:lem:zero}, so $\En(K_*,\rho)\ge n\alpha\beta_\rho$. Hence
\begin{align*}
\Pst(\mathfrak E_3^c)
&\le\Pst\bigl(\Gamma(K_*;\rho_*)>\Ut(K_*)\bigr)+\sum_{\rho\ne\rho_*}\Pst\bigl(\Gamma(K_*;\rho)\le\Ut(K_*)\bigr)\\
&\le\frac1n+\sum_{\rho\ne\rho_*}\Pst\bigl(\Gamma(K_*;\rho)\le\En(K_*,\rho)-\log n\bigr)\le\frac{N_{K_*}}n,
\end{align*}
where we use Lemma~\ref{th:lem:upper} for the first term, the inequality $\Ut(K_*)=O(\log n)<n\alpha\beta_\rho-\log n\le\En(K_*,\rho)-\log n$ for large $n$ in the sum, and the second claim of Lemma~\ref{th:lem:lower} for the last inequality.
\end{proof}

\subsubsection{Proof of Corollary~\ref*{th:cor:rate}}

\begin{proof}
Apply Theorem~\ref{th:thm:risk} with $k^\circ=K_*$ and $t=\log n$. The proof of Corollary~\ref{th:cor:exact} gives $\Pst(\Kstop<K_*)\le\Pst(\mathfrak E_1^c)=O(1/n)$, and Proposition~\ref{th:prop:curves}(b) gives $\Ut(K_*)\le\frac{d_*}2\log n+O(1)$. This proves the first claim. For the second, note that the left-hand side is non-increasing in $A_n$, so we may assume that $A_n$ grows slowly enough that $\varepsilon=A_n\sqrt{d_*\log n/\{n\alpha(1-\alpha)\}}\le\frac12$. Taking this $\varepsilon$ in Theorem~\ref{th:thm:risk}, the right-hand side is, on the same event, at most $9\{(\frac{d_*}2+1)\log n+O(1)\}/(A_n^2d_*\log n)\to0$, since $\log\Ntot$ is bounded. As this event has probability $1-O(1/n)$, the left-hand side tends to $0$ in probability.
\end{proof}

\subsubsection{Proof of Corollary~\ref*{th:cor:mean}}

\begin{proof}
Write $r_n=\{d_*\log n/(n\alpha(1-\alpha))\}^{1/2}$, $L=\Omega_*^{-1/2}\Omega_{\widehat\rho}(\theta)\Omega_*^{-1/2}$, $\bar L=\int L\,\widehat Q(d\theta)=\Omega_*^{-1/2}\widehat\Omega\,\Omega_*^{-1/2}$ and $\Delta=\Vert\bar L-I\Vert_F$.

We first bound the spread of a member of the variational family around its mean. Let $Q\in\Qfam_K(\rho)$, let $H_\rho=\{h\in\{0,\ldots,K\}:|\mathcal V_h|>0\}$ be the blocks that contain variables, write $\bar\Phi$ and $\bar v_h$ ($h\in H_\rho$) for the means of $\Phi$ and $v_h=1/\tau_h$ under $Q$, and $\bar\Omega_Q=A_\rho\bar\Phi A_\rho^\top+D_\rho(\bar v)$ for the mean of $\Omega_\rho(\theta)$. When the background is empty, its precision does not enter $\Omega_\rho(\theta)$ and is left out below; when $K=0$, the terms involving $\Phi$ vanish. For the inverse Wishart distribution with $\tilde\nu>K+3$ degrees of freedom and mean $\bar\Phi$,
\[
\E(\Phi_{ij}-\bar\Phi_{ij})^2=\frac{(\tilde\nu-K+1)\bar\Phi_{ij}^2+(\tilde\nu-K-1)\bar\Phi_{ii}\bar\Phi_{jj}}{(\tilde\nu-K)(\tilde\nu-K-3)}
\]
\citep[Chapter~3]{gupta2000}, so that $\E\Vert\Phi-\bar\Phi\Vert_F^2\le2(K+1)\Vert\bar\Phi\Vert_F^2/(\tilde\nu-K-3)$, where we use $(\tr\bar\Phi)^2\le K\Vert\bar\Phi\Vert_F^2$. For $h\in H_\rho$, the inverse gamma distribution of $v_h$ has shape $\tilde a_h=a_\tau+\alpha n|\mathcal V_h|/2>2$ and $\E(v_h-\bar v_h)^2=\bar v_h^2/(\tilde a_h-2)$. Since $\Phi$ and the $v_h$ are independent under $Q$ and $\Vert A_\rho\Vert_{\op}^2=\max_h|\mathcal V_h|\le p_0$,
\[
\int\Vert\Omega_\rho(\theta)-\bar\Omega_Q\Vert_F^2\,Q(d\theta)\le\frac{2(K+1)p_0^2}{\tilde\nu-K-3}\Vert\bar\Phi\Vert_F^2+\sum_{h\in H_\rho}\frac{|\mathcal V_h|}{\tilde a_h-2}\,\bar v_h^2 .
\]
With $\tilde\nu=\nu_K+\alpha n$, all these coefficients are at most $C_1/(\alpha n)$ for all sufficiently large $n$, where $C_1=8(M+1)p_0^2$. Moreover, $\Vert\bar\Phi\Vert_F^2+\sum_{h\in H_\rho}\bar v_h^2\le2\Vert\bar\Omega_Q\Vert_F^2$, because each entry of $\bar\Phi$ is an off-diagonal entry of $\bar\Omega_Q$, every community having at least two variables, and each $\bar v_h$ is at most a diagonal entry of $\bar\Omega_Q$. Hence $\int\Vert\Omega_\rho(\theta)-\bar\Omega_Q\Vert_F^2\,Q(d\theta)\le2C_1\Vert\bar\Omega_Q\Vert_F^2/(\alpha n)$ for every $K\le M$, $\rho\in\Fam_K$ and $Q\in\Qfam_K(\rho)$. Applied to $\widehat Q$, with $\kappa=\Vert\Omega_*\Vert_{\op}\Vert\Omega_*^{-1}\Vert_{\op}$, this gives
\[
\int\Vert L-\bar L\Vert_F^2\,\widehat Q(d\theta)\le\Vert\Omega_*^{-1}\Vert_{\op}^2\int\Vert\Omega_{\widehat\rho}(\theta)-\widehat\Omega\Vert_F^2\,\widehat Q(d\theta)\le\frac{2C_1\kappa^2}{\alpha n}\Vert\bar L\Vert_F^2\le\frac{2C_1\kappa^2(\sqrt{p_0}+\Delta)^2}{\alpha n},
\]
where we use $\Vert\widehat\Omega\Vert_F\le\Vert\Omega_*\Vert_{\op}\Vert\bar L\Vert_F$ and $\Vert\bar L\Vert_F\le\Vert I\Vert_F+\Delta$.

Now let $A_n\to\infty$ and $B_n=A_n^{1/2}$. On the event $\{\Delta>2B_nr_n\}$, every $\theta$ with $\Vert L-\bar L\Vert_F\le\Delta/2$ satisfies $\Vert L-I\Vert_F\ge\Delta/2>B_nr_n$, so Chebyshev's inequality gives
\[
\widehat Q\bigl(\Vert L-I\Vert_F>B_nr_n\bigr)\ge1-\frac4{\Delta^2}\int\Vert L-\bar L\Vert_F^2\,\widehat Q(d\theta)\ge1-\frac{8C_1\kappa^2}{\alpha n}\Bigl(1+\frac{\sqrt{p_0}}{2B_nr_n}\Bigr)^2 .
\]
Since $\alpha nr_n^2=d_*\log n/(1-\alpha)\to\infty$, the right-hand side tends to one and exceeds $\frac12$ for all large $n$. Therefore
\[
\Pst(\Delta>2B_nr_n)\le\Pst\Bigl(\widehat Q\bigl(\Vert L-I\Vert_F>B_nr_n\bigr)\ge\tfrac12\Bigr)\to0
\]
by Corollary~\ref{th:cor:rate} applied with $B_n$ in place of $A_n$. Since $2B_n\le A_n$ for all large $n$, the claim follows.
\end{proof}

\subsubsection{Proof of Corollary~\ref*{th:cor:geom}}

\begin{proof}
Write $A(K)=n\alpha Be^{-\gamma K}$ and $V(K)=d_K\log n$, and let $K_1$ be the first $K\ge1$ with $A(K)\le\zeta V(K)$. We first show that the search reaches $K_1$. By (G) and $A(K-1)=e^\gamma A(K)$,
\[
\Lt(K-1)-\Ut(K)\ge \vartheta e^\gamma A(K)-A(K)-c_UV(K)=(\vartheta e^\gamma-1)\bigl\{A(K)-\zeta V(K)\bigr\},
\]
which is positive for $K\le K_1-1$. Theorem~\ref{th:thm:early} with $K^\dagger=K_1-1$ gives $\Kstop\ge K_1$ with probability at least $1-2K_1/n$.

Next we bound $K_1$. The condition $A(K)\le\zeta V(K)$ reads
\[
\gamma K+\log d_K\ge\log\{n\alpha B/(\zeta\log n)\}.
\]
Since $d_K\ge1$,
\[
K_1\le\frac1\gamma\log\frac{n\alpha B}{\zeta\log n}+1\le\frac{\log n}\gamma+1
\]
for large $n$. Since $d_K\le(K+2)^2$ and $K_1\le(2/\gamma)\log n$, we have $\log d_{K_1}\le2\log\log n+2\log(2/\gamma+2)$, and hence $K_1\ge(\log n)/\gamma-(3/\gamma)\log\log n-C_6+1$. This gives the lower bound on $\Kstop$.

We then bound the upper curve at $K_1$. By (G) and the definition of $K_1$,
\begin{align*}
\Ut(K_1)&\le A(K_1)+c_UV(K_1)\le(\zeta+c_U)d_{K_1}\log n\\
&\le(\zeta+c_U)\frac{(K_1+2)^2}2\log n\le(\zeta+c_U)\frac{(\log n)^3}{2\gamma^2}\{1+o(1)\},
\end{align*}
where we use (G) for the first inequality, $A(K_1)\le\zeta V(K_1)$ for the second, $d_K\le(K+2)^2/2$ for the third, and $K_1\le(\log n)/\gamma+1$ for the last.
For the upper bound on $\Kstop$, let $K>K_1$. By (G), $\Lt(K)\ge c_LV(K)\ge c_LK^2(\log n)/2$, which exceeds $\Ut(K_1)$ once $K\ge\sqrt{(\zeta+c_U)/c_L}\,(\log n)/\gamma\,\{1+o(1)\}$. Theorem~\ref{th:thm:late} with $k^\circ=K_1$ and $K=\lceil C_7(\log n)/\gamma\rceil-1$ gives the upper bound on $\Kstop$ with probability at least $1-2/n$ for large $n$. (If this $K$ exceeds $M_n$, the upper bound holds trivially because $\Kstop\le M_n$.)

Finally, on the event of the first step $K_1\le\Kstop$, so Theorem~\ref{th:thm:risk} with $k^\circ=K_1$ and $t=\log n$ gives
\[
\int R_n(\widehat\rho,\theta)\,\widehat Q(d\theta)\le\Ut(K_1)+\log\Ntot+\log n\le(\zeta+c_U)\frac{(\log n)^3}{2\gamma^2}\{1+o(1)\},
\]
and the contraction statement follows from the second part of Theorem~\ref{th:thm:risk}.
\end{proof}

\section{Selection theory in growing dimension}\label{th:app:highdim}

The results of Appendix~\ref{th:sec:exact} assume that the dimension $p_0$ is fixed. In this appendix, we show that they continue to hold, with minimal changes, when $p_0$ grows with $n$, possibly with $p_0\gg n$, provided that the number of communities is fixed and the candidates satisfy the uniform separation condition (W$'$) below. Example~\ref{th:ex:growing} gives a sequence with fixed communities and a growing background in which all conditions hold; when the communities grow with $p_0$ and $\alpha$ is fixed, (W$'$) can fail, as the example after Assumption~(W) shows. The guarantee for the posterior mean (Corollary~\ref{th:cor:mean}) is not extended here. We first state uniform versions of Assumptions~(E), (F) and (W). We then give versions of Lemmas~\ref{th:lem:oracle} and \ref{th:lem:iw} whose constants do not depend on $p_0$, and restate Proposition~\ref{th:prop:curves} and Corollaries~\ref{th:cor:exact} and \ref{th:cor:rate}. Their proofs follow those in Appendix~\ref{th:app:proofs-cases}, and we only indicate the changes. Appendix~\ref{th:sec:misspec} needs no change, since Assumption~(G) is stated directly on the two curves. Appendix~\ref{th:sec:highdim-sim} checks the results by simulation in the setting of Example~\ref{th:ex:growing}.

\subsection{Assumptions}

We consider a sequence of problems indexed by $n$. For each $n$, the dimension $p_0$, the true partition $\rho_*$ and the candidate families may change, but the number of communities, the factor covariance and the residual variances of the truth do not.

\begin{assumption}{(E$'$)}{Exact 1f model with fixed parameters}\label{th:ass:Ep}
For each $n$, Assumption~(E) holds, and $K_*$, $\Phi_*$ and $v_*$ do not depend on $n$.
\end{assumption}

\begin{assumption}{(F$'$)}{Candidates}\label{th:ass:Fp}
For each $n$, $\rho_*\in\Fam_{K_*}$. There are constants $\bar N$ and $\bar M$ such that $N_K\le\bar N$ and $M\le\bar M$, and the prior and $\alpha$ do not depend on $n$.
\end{assumption}

\begin{assumption}{(W$'$)}{Uniform decrease of the bias}\label{th:ass:Wp}
There are constants $\varpi>0$, $\beta_{\min}>0$ and $\bar C\ge1$ such that, for each $n$:
\begin{enumerate}[label=(\roman*),nosep]
\item for $1\le K\le K_*-1$, some $\rho_K\in\Fam_K$ and $\theta_K$ satisfy $m_{K-1}\ge(1+\varpi)\KL(\rho_K,\theta_K)$, where the eigenvalues of the factor covariance of $\theta_K$ and its residual variances lie in $[1/\bar C,\bar C]$;
\item $m_{K_*-1}\ge\beta_{\min}$, and every $\rho\in\Fam_K$ with $K\ge K_*$ and positive R\'enyi bias has R\'enyi bias at least $\beta_{\min}$.
\end{enumerate}
\end{assumption}

\begin{assumption}{(D)}{Dimension}\label{th:ass:D}
$\log p_0=o(n^{1/3})$.
\end{assumption}

Assumptions~(E$'$) and (F$'$) are (E) and (F) without the requirement that $p_0$ and the candidate partitions be fixed. Assumption~(W$'$) is (W) with a relative margin $\varpi$ and a uniform lower bound $\beta_{\min}$. For fixed $p_0$, the finiteness of the candidate families supplies such constants automatically; when $p_0$ grows, the biases may grow with the community sizes, so the margin in (W) must grow with them, and the biases of wrong candidates must stay bounded away from zero. Part~(i) also asks that the KL bias of model $K$ be nearly attained at a parameter in a fixed compact set, which keeps the constants below independent of $p_0$. When $p_0$ is fixed, Assumptions~(E), (F) and (W) imply (E$'$), (F$'$), (W$'$) and (D), so Appendix~\ref{th:sec:exact} is a special case. Like (W), Assumption~(W$'$) compares a R\'enyi bias with a KL bias, so it becomes restrictive when the missing communities are large compared with $1/\alpha$, because the ratio of the two biases is then small (Example~\ref{th:ex:onemissing}).

Assumption~(D) allows $p_0$ to be much larger than $n$. It is used only in the not-too-late step, where the parameter cost $\frac12\log|\mathcal V_h|$ of each residual variance must be smaller than the boundary penalty, which is at least of order $n^{1/3}$ (Lemma~\ref{th:lem:iw-p}).

\begin{example}[Fixed communities and a growing background]\label{th:ex:growing}
Fix $K_*\ge1$, a community size $b\ge2$, a correlation $r\in(0,1)$ and $\alpha\in(0,1)$. For each $n$, let the truth have $K_*$ communities of $b$ variables with independent factors, $\Phi_*=rI_{K_*}$ and residual variances $v^*_h=1-r$, and a background of $p_0-K_*b$ independent variables with variance $v^*_0=1$, so that every variable has unit variance and $r$ is the within-community correlation. Only the background grows with $n$, with $\log p_0=o(n^{1/3})$. For $K\le K_*$, let $\Fam_K$ contain only the first $K$ true communities in a fixed order; for $K_*<K\le\bar M$, let $\Fam_K$ contain the partition that adds to the true partition the first $K-K_*$ of $\bar M-K_*$ disjoint communities, fixed in advance, of at least two background variables each, so that every candidate is a prefix of one fixed path. The candidates form a deterministic family; the example does not assert that the dense-subgraph search returns them. Let $\kappa_{\mathrm{KL}}=\frac12\{-\log(1+(b-1)r)-(b-1)\log(1-r)\}$ be the KL divergence from $\Norm_b(0,(1-r)I+r\mathbf 1\mathbf 1^\top)$ to $\Norm_b(0,I)$, and let $\kappa_{\mathrm R}$ be the smallest R\'enyi divergence $D_{1-\alpha}$ from the same law to $\Norm_b(0,vI)$ over $v>0$. If $\vartheta_0=\kappa_{\mathrm R}/\kappa_{\mathrm{KL}}$ satisfies $K_*\vartheta_0>K_*-1$, then (E$'$), (F$'$), (W$'$) and (D) hold, with $\beta_{\min}=\kappa_{\mathrm R}$ and, when $K_*\ge2$, $\varpi=K_*\vartheta_0/(K_*-1)-1$. For example, $K_*=5$, $b=20$, $r=0.3$ and $\alpha=0.02$ give $\kappa_{\mathrm{KL}}=2.437$, $\kappa_{\mathrm R}=2.273$ and $\vartheta_0=0.932>0.8$, so the conditions hold with $\varpi=0.166$.
\end{example}

To verify (W$'$), let $1\le K\le K_*$ and let $\rho_{K-1}$ consist of the first $K-1$ true communities. Under any parameter $\theta$ of $\rho_{K-1}$, the variables of these communities are independent of all other variables, and the other variables are independent with a common variance $v$; under the truth, the communities and the background variables are mutually independent. The R\'enyi divergence between product measures is the sum of the divergences between their factors, and each of these is non-negative, so
\[
D_{1-\alpha}\bigl(\Pst\Vert P^{\rho_{K-1}}_\theta\bigr)\ge\sum_{h=K}^{K_*}D_{1-\alpha}\bigl(\Norm_b(0,(1-r)I+r\mathbf 1\mathbf 1^\top)\Vert\Norm_b(0,vI)\bigr)\ge(K_*-K+1)\kappa_{\mathrm R},
\]
and hence $m_{K-1}\ge(K_*-K+1)\kappa_{\mathrm R}$. For $K=K_*$ this gives $m_{K_*-1}\ge\kappa_{\mathrm R}$; every candidate with $K\ge K_*$ can reproduce the truth and has zero R\'enyi bias, so (W$'$)(ii) holds with $\beta_{\min}=\kappa_{\mathrm R}$. For $K\le K_*-1$, let $\theta_K$ have factor covariance $rI_K$, residual variances $1-r$ and background variance $1$. The same factorization gives $\KL(\rho_K,\theta_K)=(K_*-K)\kappa_{\mathrm{KL}}$, so (W$'$)(i) holds if $(j+1)\kappa_{\mathrm R}\ge(1+\varpi)j\kappa_{\mathrm{KL}}$ for $j=K_*-K\in\{1,\ldots,K_*-1\}$. Since $(j+1)/j$ decreases in $j$, it suffices that this holds for $j=K_*-1$, which is the stated choice of $\varpi$. The parameters $\theta_K$ have eigenvalues and residual variances in $\{r,1-r,1\}$, and (E$'$), (F$'$) and (D) hold by construction. The condition $K_*\vartheta_0>K_*-1$ requires the communities to be small relative to $1/\alpha$: $\vartheta_0\to1$ as $\alpha(b-1)r\to0$, whereas for fixed $\alpha$ and growing $b$, $\kappa_{\mathrm R}$ grows at most logarithmically in $b$ and $\kappa_{\mathrm{KL}}$ linearly.

\subsection{Lemmas without dimension}

The following decomposition of $\R^{p_0}$ reduces all computations for a 1f model to $K\times K$ matrices and scalars. For a partition $\rho$ with communities $\mathcal V_1,\ldots,\mathcal V_K$ and background $\mathcal V_0$, write $b_h=|\mathcal V_h|$, $u_h=b_h^{-1/2}1_{\mathcal V_h}$, $U_\rho=(u_1,\ldots,u_K)$ and $D_b=\operatorname{diag}(b_1,\ldots,b_K)$. For $1\le h\le K$, let $P_h$ be the orthogonal projection onto the vectors supported on $\mathcal V_h$ whose entries sum to zero, and let $P_0$ be the orthogonal projection onto the vectors supported on $\mathcal V_0$. Write $r_h$ for the rank of $P_h$, so that $r_h=b_h-1$ for $h\ge1$ and $r_0=|\mathcal V_0|$.

\begin{lemma}[Block decomposition]\label{th:lem:blocks}
$U_\rho U_\rho^\top+\sum_{h=0}^KP_h=I$, and for every $\theta$,
\[
\Omega_\rho(\theta)=U_\rho M_\rho(\theta)U_\rho^\top+\sum_{h=0}^Kv_hP_h,\qquad M_\rho(\theta)=D_b^{1/2}\Phi D_b^{1/2}+\operatorname{diag}(v_1,\ldots,v_K).
\]
Consequently,
\[
\Omega_\rho(\theta)^{-1}=U_\rho M_\rho(\theta)^{-1}U_\rho^\top+\sum_{h=0}^K\tau_hP_h,\qquad\log|\Omega_\rho(\theta)|=\log|M_\rho(\theta)|-\sum_{h=0}^Kr_h\log\tau_h .
\]
\end{lemma}

\begin{lemma}[Fluctuation and bias]\label{th:lem:Bkl}
For $\Omega\succ0$, let $k=\KL(\Norm_{p_0}(0,\Omega_*)\Vert \Norm_{p_0}(0,\Omega))$ and $B=\Omega_*^{1/2}\Omega^{-1}\Omega_*^{1/2}-I$. Then $\Vert B\Vert_{\op}\le\Vert B\Vert_F\le\sqrt{8k}+8k$.
\end{lemma}

\begin{lemma}[Conjugate oracle bounds without dimension]\label{th:lem:oracle-p}
Let $\rho\in\Fam_K$, let $\theta_c=(\Phi_c,v_c)$ be a centre whose factor covariance has eigenvalues in $[1/\bar C,\bar C]$ and whose residual variances lie in $[1/\bar C,\bar C]$, and let $Q=Q_{\theta_c}$ be the conjugate oracle defined before Lemma~\ref{th:lem:oracle}. Write $k_c=\KL(\rho,\theta_c)$ and $B_c=\Omega_*^{1/2}\Omega_\rho(\theta_c)^{-1}\Omega_*^{1/2}-I$, and let $d_\rho$ be as in Lemma~\ref{th:lem:oracle}. There are constants $C$ and $n_1$, depending only on $K$, $\bar C$, the prior and $\alpha$, such that for $n\ge n_1$:
\begin{enumerate}[label=(\alph*),nosep]
\item $\KL(Q,\Pi_K(\cdot\mid\rho))\le\frac{d_\rho}2\log(\alpha n)+\frac12\sum_{h:\mathcal V_h\ne\emptyset}\log|\mathcal V_h|+C$;
\item $n\,\E_Q\KL(\rho,\theta)\le n\,k_c+C(1+k_c)/\alpha$;
\item $\Vert B_Q-B_c\Vert_F\le C(1+k_c)/(\alpha n)$.
\end{enumerate}
Consequently,
\begin{multline*}
\Ut(K)\le\alpha n\,k_c+\frac{d_\rho}2\log(\alpha n)+\frac12\sum_{h:\mathcal V_h\ne\emptyset}\log|\mathcal V_h|\\
+\alpha\bigl(\sqrt{2n(t+1)}+t+1\bigr)\bigl(\sqrt{8k_c}+8k_c\bigr)+C(1+k_c)\bigl(1+\sqrt{t/n}+t/n\bigr).
\end{multline*}
\end{lemma}

\begin{lemma}[Inverse-Wishart boundary penalty without dimension]\label{th:lem:iw-p}
Assume (E), let $K>K_*$, and let $\rho\in\Fam_K$ have zero R\'enyi bias. Let $v^*_{\max}$ be the largest residual variance of the truth over its non-empty blocks, $c_0'=\min\{c/4,\,1/(9(v^*_{\max})^2)\}$ and $\varepsilon_n=(n\alpha(1-\alpha))^{-1/3}$. If $\varepsilon_n\le\min\{c/(16\nu_KK),\,v^*_{\max}/2\}$, then $\En(K,\rho)\ge c_0'(n\alpha(1-\alpha))^{1/3}-\log2$.
\end{lemma}

\subsection{Results}

With these lemmas, Proposition~\ref{th:prop:curves} and Corollaries~\ref{th:cor:exact} and \ref{th:cor:rate} take the following form. The only changes are that the constants do not depend on $p_0$, and that the residual variances add a parameter cost of at most $\frac{K_*+1}2\log p_0$.

\begin{proposition}[The curves in growing dimension]\label{th:prop:curves-p}
Assume (E$'$), (F$'$) and (W$'$), take $t=\log n$, and use the conjugate oracles centered at $\theta_K$ from (W$'$)(i) for $K<K_*$ and at $\theta_*$ for $K=K_*$. There are constants $C$ and $n_0$, depending only on $K_*$, $\Phi_*$, $v_*$, $\beta_{\min}$, $\bar C$, $\bar N$, $\bar M$, the prior and $\alpha$, such that for $n\ge n_0$:
\begin{enumerate}[label=(\alph*),nosep]
\item for $K<K_*$, $\Lt(K)\ge n\alpha m_K-C\log n$, and, with $k_K=\KL(\rho_K,\theta_K)$,
\[
\Ut(K)\le n\alpha k_K+C\bigl\{\sqrt{n\log n}\,(\sqrt{k_K}+k_K)+\log n+(K+1)\log p_0\bigr\};
\]
\item at $K=K_*$, $\Ut(K_*)\le\frac{d_*}2\log(\alpha n)+\frac12\sum_{h:\mathcal V^*_h\ne\emptyset}\log|\mathcal V^*_h|+C$;
\item for $K>K_*$, $\Lt(K)\ge\min\{c_0'(n\alpha(1-\alpha))^{1/3},\,n\alpha\beta_{\min}\}-C\log n$.
\end{enumerate}
\end{proposition}

\begin{corollary}[Selection in growing dimension]\label{th:cor:exact-p}
Assume (E$'$), (F$'$), (W$'$) and (D). Then, for all sufficiently large $n$,
\[
\Pst\Bigl(\widehat K=K_*,\ \widehat\rho=\rho_*,\ \text{and the search stops at }K_*+1\text{ (or at }M=K_*)\Bigr)\ge1-\frac{2K_*+2+N_{K_*}}n .
\]
\end{corollary}

\begin{corollary}[Rate in growing dimension]\label{th:cor:rate-p}
Assume (E$'$), (F$'$), (W$'$) and (D). With probability at least $1-O(1/n)$,
\[
\int R_n(\widehat\rho,\theta)\,\widehat Q(d\theta)\le\frac{d_*}2\log n+\frac12\sum_{h:\mathcal V^*_h\ne\emptyset}\log|\mathcal V^*_h|+\log n+\log\Ntot+C,
\]
and for every sequence $A_n\to\infty$,
\[
\widehat Q\Bigl(\bigl\Vert\Omega_*^{-1/2}\Omega_{\widehat\rho}(\theta)\Omega_*^{-1/2}-I\bigr\Vert_F>A_n\sqrt{\frac{d_*\log n+(K_*+1)\log p_0}{n\alpha(1-\alpha)}}\Bigr)\to0
\]
in probability.
\end{corollary}

The rate differs from that of Corollary~\ref{th:cor:rate} only by the term $(K_*+1)\log p_0$, which comes from the residual variances: each $v_h$ is estimated from $n|\mathcal V_h|$ entries, so its parameter cost is $\frac12\log(\alpha n|\mathcal V_h|)$ rather than $\frac12\log(\alpha n)$. The relative Frobenius error is measured over all $p_0$ coordinates, and it is small even when $p_0\gg n$, because the model has only $d_*$ parameters.

\subsection{A simulation check}\label{th:sec:highdim-sim}

We check Corollaries~\ref{th:cor:exact-p} and~\ref{th:cor:rate-p} by simulation in the setting of Example~\ref{th:ex:growing}, with $K_*=5$, $b=20$, $r=0.3$, $\alpha=0.02$ and $p_0=2n$ for $n\in\{500,1000,2000,5000\}$, so that the background grows from $900$ to $9900$ variables. As the theory assumes, the data are not standardized, $\alpha$ is fixed, and the candidates are fixed in advance: one path consisting of the five true communities followed by two communities of $20$ background variables, so that $\bar M=7$. With one path, the search proceeds along it; the ELBO is computed and compared as in Appendix~\ref{app:selection-implementation}, with an iteration limit of $3000$. For each $n$ we generate $50$ data sets. The contraction of $\widehat Q$ is the mean of $\Vert\Omega_*^{-1/2}\Omega\,\Omega_*^{-1/2}-I\Vert_F$ over $200$ draws of $\Omega$ from $\widehat Q$, and we compare it with $r_n'=\{(d_*\log n+(K_*+1)\log p_0)/(n\alpha(1-\alpha))\}^{1/2}$, the rate of Corollary~\ref{th:cor:rate-p}, with $d_*=21$. Both $\Omega_*$ and $\Omega$ are low rank plus diagonal, so the norm is computed from $K\times K$ matrices and vectors of length $p_0$, without forming $p_0\times p_0$ matrices.

\begin{table}[!htbp]
\centering
\caption{Simulation check of Corollaries~\ref{th:cor:exact-p} and~\ref{th:cor:rate-p} in the setting of Example~\ref{th:ex:growing}, with 50 data sets per row. Correct: proportion of data sets in which the selected partition is the true one. Stop at 6: proportion in which the search stopped at $K=6$. Contraction: mean of $\E_{\widehat Q}\Vert\Omega_*^{-1/2}\Omega\,\Omega_*^{-1/2}-I\Vert_F$. Ratio: contraction divided by $r_n'$. Posterior mean: the same norm for $\E_{\widehat Q}\Omega$, which has no guarantee in growing dimension.}
\label{tab:highdim-sim}
\small
\begin{tabular}{@{}rrrrrrrr@{}}
\toprule
\(n\) & \(p_0\) & Correct & Stop at 6 & Contraction & \(r_n'\) & Ratio & Posterior mean\\
\midrule
500 & 1{,}000 & 1.00 & 1.00 & 1.72 & 4.19 & 0.41 & 0.35\\
1000 & 2{,}000 & 1.00 & 1.00 & 1.26 & 3.12 & 0.40 & 0.25\\
2000 & 4{,}000 & 1.00 & 1.00 & 0.91 & 2.31 & 0.40 & 0.17\\
5000 & 10{,}000 & 1.00 & 1.00 & 0.59 & 1.55 & 0.38 & 0.10\\
\bottomrule
\end{tabular}

\end{table}

In every data set the search returned the true partition and stopped at $K=6$, as Corollary~\ref{th:cor:exact-p} states. As $n$ increased tenfold, and $p_0$ with it, the contraction decreased from $1.72$ to $0.59$, and its ratio to $r_n'$ stayed between $0.38$ and $0.41$; on the log scale, the contraction decreased with slope $-0.46$ in $\log n$, against $-0.43$ for $r_n'$. This agrees with Corollary~\ref{th:cor:rate-p}. The posterior mean was also accurate, with an error that decreased from $0.35$ to $0.10$. In design S1 with $p_0=2n$ (Section~\ref{sec:sim-selection}), by contrast, the data are standardized and the communities grow with $n$, and the error does not decrease.

\subsection{Proofs}

\subsubsection{Proof of Lemma~\ref*{th:lem:blocks}}

\begin{proof}
The $h$-th column of $A_\rho$ is $1_{\mathcal V_h}=b_h^{1/2}u_h$, so $A_\rho=U_\rho D_b^{1/2}$ and $A_\rho\Phi A_\rho^\top=U_\rho D_b^{1/2}\Phi D_b^{1/2}U_\rho^\top$. For $h\ge1$, the matrix $u_hu_h^\top+P_h$ is the orthogonal projection onto the vectors supported on $\mathcal V_h$, so
\[
D_\rho(\tau)=\sum_{h=1}^Kv_h\bigl(u_hu_h^\top+P_h\bigr)+v_0P_0 .
\]
Adding the two displays gives the decomposition of $\Omega_\rho(\theta)$, and summing the projections gives $U_\rho U_\rho^\top+\sum_hP_h=I$. Since the ranges of $U_\rho$ and of $P_0,\ldots,P_K$ are mutually orthogonal, $\Omega_\rho(\theta)$ acts as $M_\rho(\theta)$ on the range of $U_\rho$ and as $v_h$ on the range of $P_h$, which gives the inverse and the determinant.
\end{proof}

\subsubsection{Proof of Lemma~\ref*{th:lem:Bkl}}

\begin{proof}
Let $\beta_1,\ldots,\beta_{p_0}>-1$ be the eigenvalues of $B$. As in the proof of Lemma~\ref{th:lem:closed},
\[
2k=\tr(\Omega^{-1}\Omega_*)-p_0+\log|\Omega|-\log|\Omega_*|=\sum_{i=1}^{p_0}\bigl\{\beta_i-\log(1+\beta_i)\bigr\}.
\]
The function $h(\beta)=\beta-\log(1+\beta)=\int_0^\beta s/(1+s)\,ds$ satisfies
\[
h(\beta)\ge\beta^2/4\quad\text{for }-1<\beta\le1,\qquad h(\beta)\ge\beta/4\quad\text{for }\beta>1,
\]
because $s/(1+s)\ge s/2$ for $0\le s\le1$, $-s/(1+s)\ge-s$ for $-1<s<0$, $h(1)=1-\log2\ge1/4$, and $s/(1+s)\ge1/2$ for $s\ge1$. Therefore
\[
\Vert B\Vert_F^2=\sum_{|\beta_i|\le1}\beta_i^2+\sum_{\beta_i>1}\beta_i^2\le4\sum_{i=1}^{p_0}h(\beta_i)+\Bigl(\sum_{\beta_i>1}\beta_i\Bigr)^2\le8k+(8k)^2,
\]
where we use the first bound on $h$ for the first sum, and the second bound on $h$, which gives $\sum_{\beta_i>1}\beta_i\le4\sum_ih(\beta_i)=8k$, for the second. Taking square roots gives the claim.
\end{proof}

\subsubsection{Proof of Lemma~\ref*{th:lem:oracle-p}}

\begin{proof}
\emph{Part (a).} The expansions in the proof of Lemma~\ref{th:lem:oracle}(a) hold for every $p_0$, and their terms of order $O(1)$ depend only on $K$, the prior and the bounds on $\Phi_c$ and $v_c$. The only term that involves the block sizes is $\frac12\log\tilde a_h$, and $\tilde a_h=a_\tau+\alpha n|\mathcal V_h|/2\le\alpha n|\mathcal V_h|$ when $\alpha n\ge2a_\tau$. The factor of an empty background contributes $O(1)$. This gives (a).

\emph{Part (b).} Write $M=M_\rho(\theta)$, $M_c=M_\rho(\theta_c)$ and $M_*=U_\rho^\top\Omega_*U_\rho$, and let $\Lambda=M_c^{-1/2}M_*M_c^{-1/2}$. For each block with $r_h\ge1$, let $\omega_h=\tr(P_h\Omega_*)$ and $x_h=\tau_{c,h}\omega_h/r_h$, the ratio of the average variance of the truth on the range of $P_h$ to $v_{c,h}$. By Lemma~\ref{th:lem:blocks}, for every $\theta$,
\begin{equation}\label{th:eq:kl-blocks}
2\KL(\rho,\theta)=\tr(M^{-1}M_*)+\log|M|+\sum_{h=0}^K\bigl(\tau_h\omega_h-r_h\log\tau_h\bigr)-p_0-\log|\Omega_*| .
\end{equation}
We first bound $\Lambda$ and $x_h$ in terms of $k_c$. By Fischer's inequality, $\log|\Omega_*|$ is at most the sum of the log-determinants of the diagonal blocks of $\Omega_*$ in the decomposition of Lemma~\ref{th:lem:blocks}, and by the arithmetic--geometric mean inequality, the log-determinant of the block on the range of $P_h$ is at most $r_h\log(\omega_h/r_h)$. Inserting these bounds into \eqref{th:eq:kl-blocks} at $\theta=\theta_c$ and using $p_0=K+\sum_hr_h$, we obtain
\[
2k_c\ge\sum_{i=1}^K(\lambda_i-1-\log\lambda_i)+\sum_{h:r_h\ge1}r_h(x_h-1-\log x_h),
\]
where $\lambda_1,\ldots,\lambda_K$ are the eigenvalues of $\Lambda$. Every summand is non-negative, and $y\le2+2(y-1-\log y)$ for $y>0$ because $\log y\le y/2$. Hence
\begin{equation}\label{th:eq:Lambda-x}
\Vert\Lambda\Vert_{\op}\le2+4k_c,\qquad x_h\le2+4k_c .
\end{equation}

We next bound the residual-variance terms of \eqref{th:eq:kl-blocks}. Under $Q$, $v_{c,h}\tau_h\sim\Gam(\tilde a_h,\tilde a_h-1)$, so $\E_Q\tau_h=\tau_{c,h}\tilde a_h/(\tilde a_h-1)$ and $\E_Q\log\tau_h=\log\tau_{c,h}+\psi(\tilde a_h)-\log(\tilde a_h-1)\ge\log\tau_{c,h}$, where the inequality uses $\psi(a)>\log a-1/a\ge\log(a-1)$. Therefore
\[
\E_Q\bigl(\tau_h\omega_h-r_h\log\tau_h\bigr)-\bigl(\tau_{c,h}\omega_h-r_h\log\tau_{c,h}\bigr)\le\frac{\tau_{c,h}\omega_h}{\tilde a_h-1}=\frac{r_hx_h}{\tilde a_h-1}\le\frac{2x_h}{\alpha n},
\]
where the last inequality uses $r_h\le|\mathcal V_h|$ and $\tilde a_h-1\ge\alpha n|\mathcal V_h|/2$, which holds because $a_\tau>1$.

We then bound the community term of \eqref{th:eq:kl-blocks}. Let $\Xi=M_c^{-1/2}(M-M_c)M_c^{-1/2}$, so that $M=M_c^{1/2}(I+\Xi)M_c^{1/2}$. Since $M_\rho(\theta)$ is affine in $(\Phi,v)$ and the oracle satisfies $\E_Q\Phi=\Phi_c$ and $\E_Qv=v_c$, we have $\E_Q\Xi=0$. Using $(I+\Xi)^{-1}=I-\Xi+\Xi(I+\Xi)^{-1}\Xi$,
\begin{align*}
&\tr(M^{-1}M_*)+\log|M|-\tr(M_c^{-1}M_*)-\log|M_c|\\
&\quad=-\tr(\Xi\Lambda)+\tr\bigl(\Xi(I+\Xi)^{-1}\Xi\Lambda\bigr)+\log|I+\Xi|\\
&\quad\le-\tr(\Xi\Lambda)+\tr(\Xi)+\Vert\Lambda\Vert_{\op}\,\frac{\Vert\Xi\Vert_F^2}{\lambda_{\min}(I+\Xi)},
\end{align*}
where we use $\log|I+\Xi|\le\tr(\Xi)$ for the inequality. The first two terms of the last line have mean zero under $Q$. For the third, let $\Delta_\Phi=\Phi_c^{-1/2}(\Phi-\Phi_c)\Phi_c^{-1/2}$ and $\Delta_v=(v_h/v_{c,h}-1)_{h=1}^K$. Since $M_c\succeq D_b^{1/2}\Phi_cD_b^{1/2}$ and $M_c\succeq\operatorname{diag}(v_{c,1},\ldots,v_{c,K})$,
\[
\Vert\Xi\Vert_F\le\Vert\Delta_\Phi\Vert_F+\Vert\Delta_v\Vert_2,\qquad\lambda_{\min}(I+\Xi)\ge\lambda_Q:=\min\Bigl\{\lambda_{\min}(I+\Delta_\Phi),\ \min_{h\ge1}\frac{v_h}{v_{c,h}}\Bigr\}.
\]
Under $Q$, $\Phi_c^{-1/2}\Phi\Phi_c^{-1/2}\sim\IW_K(\tilde\nu,(\tilde\nu-K-1)I)$ and $v_{c,h}\tau_h\sim\Gam(\tilde a_h,\tilde a_h-1)$, whose laws depend only on $K$, $\tilde\nu$ and $\tilde a_h$. Standard moment bounds for these laws give $\E_Q(\Vert\Delta_\Phi\Vert_F+\Vert\Delta_v\Vert_2)^4\le C(\alpha n)^{-2}$ and $\E_Q\lambda_Q^{-2}\le C$. Hence, by the Cauchy--Schwarz inequality and \eqref{th:eq:Lambda-x},
\begin{align*}
&\E_Q\bigl\{\tr(M^{-1}M_*)+\log|M|\bigr\}-\tr(M_c^{-1}M_*)-\log|M_c|\\
&\qquad\le(2+4k_c)\bigl\{\E_Q(\Vert\Delta_\Phi\Vert_F+\Vert\Delta_v\Vert_2)^4\bigr\}^{1/2}\bigl(\E_Q\lambda_Q^{-2}\bigr)^{1/2}\le\frac{C(1+k_c)}{\alpha n}.
\end{align*}
Combining the last two bounds with \eqref{th:eq:kl-blocks} and \eqref{th:eq:Lambda-x} gives $2\{\E_Q\KL(\rho,\theta)-k_c\}\le C(1+k_c)/(\alpha n)$, which is (b).

\emph{Part (c).} By Lemma~\ref{th:lem:blocks}, $\E_Q\Omega_\rho(\theta)^{-1}-\Omega_\rho(\theta_c)^{-1}=U_\rho YU_\rho^\top+\sum_h(\E_Q\tau_h-\tau_{c,h})P_h$ with $Y=\E_QM^{-1}-M_c^{-1}$. By the triangle inequality,
\[
\Vert B_Q-B_c\Vert_F\le\bigl\Vert\Omega_*^{1/2}U_\rho YU_\rho^\top\Omega_*^{1/2}\bigr\Vert_F+\sum_h\bigl(\E_Q\tau_h-\tau_{c,h}\bigr)\bigl\Vert\Omega_*^{1/2}P_h\Omega_*^{1/2}\bigr\Vert_F .
\]
For each $h$, $\Vert\Omega_*^{1/2}P_h\Omega_*^{1/2}\Vert_F\le\tr(P_h\Omega_*)=\omega_h$ and $\E_Q\tau_h-\tau_{c,h}=\tau_{c,h}/(\tilde a_h-1)$, so the summand is at most $r_hx_h/(\tilde a_h-1)\le2(2+4k_c)/(\alpha n)$ by \eqref{th:eq:Lambda-x}. For the first term, write $\tilde Y=M_c^{1/2}YM_c^{1/2}=\E_Q\{(I+\Xi)^{-1}\}-I=\E_Q\{\Xi(I+\Xi)^{-1}\Xi\}$, where we use $\E_Q\Xi=0$. Then
\[
\bigl\Vert\Omega_*^{1/2}U_\rho YU_\rho^\top\Omega_*^{1/2}\bigr\Vert_F^2=\tr(YM_*YM_*)=\tr(\tilde Y\Lambda\tilde Y\Lambda)\le\Vert\Lambda\Vert_{\op}^2\,\Vert\tilde Y\Vert_F^2,
\]
and $\Vert\tilde Y\Vert_F\le\E_Q\{\Vert\Xi\Vert_F^2/\lambda_{\min}(I+\Xi)\}\le C/(\alpha n)$ by the moment bounds above. Together with \eqref{th:eq:Lambda-x}, this gives (c).

Finally, inserting (a)--(c) into the definition of $\Ut(K)$ and bounding $\Vert B_c\Vert_F$ by Lemma~\ref{th:lem:Bkl} gives the bound on the upper curve.
\end{proof}

\subsubsection{Proof of Lemma~\ref*{th:lem:iw-p}}

\begin{proof}
We follow the proof of Lemma~\ref{th:lem:iw} and only change the bound on the second term of \eqref{th:eq:iw-split}. Let $\lambda_{\min}(\Phi)\ge\varepsilon$ and $L=\Omega_*^{-1/2}\Omega_\rho(\theta)\Omega_*^{-1/2}$. If $y$ is a unit vector with $\Omega_*y=sy$, then $y^\top Ly=y^\top\Omega_\rho(\theta)y/s$. We apply this to vectors in two eigenspaces of $\Omega_*$.

In the carve-out case of Lemma~\ref{th:lem:rep}(iii), $\mathcal V_a$ lies in the true background, and every unit vector supported on the true background is an eigenvector of $\Omega_*$ with eigenvalue $v^*_0$. Take $y_1=|\mathcal V_a|^{-1/2}1_{\mathcal V_a}$ and a unit vector $y_2$ supported on $\mathcal V_a$ whose entries sum to zero. Then
\[
y_1^\top Ly_1=\frac{|\mathcal V_a|\Phi_{aa}+v_a}{v^*_0},\qquad y_2^\top Ly_2=\frac{v_a}{v^*_0},
\]
so $y_1^\top Ly_1-y_2^\top Ly_2=|\mathcal V_a|\Phi_{aa}/v^*_0\ge2\varepsilon/v^*_0$, and one of the two Rayleigh quotients differs from $1$ by at least $\varepsilon/v^*_0$.

In the split case, $\mathcal V_a,\mathcal V_b\subset \mathcal V^*_h$, and every unit vector supported on $\mathcal V^*_h$ whose entries sum to zero is an eigenvector of $\Omega_*$ with eigenvalue $v^*_h$. Take unit vectors $y_a$ and $y_b$ supported on $\mathcal V_a$ and $\mathcal V_b$ whose entries sum to zero, and let $y_{ab}=c_{ab}(|\mathcal V_a|^{-1}1_{\mathcal V_a}-|\mathcal V_b|^{-1}1_{\mathcal V_b})$ with $c_{ab}^{-2}=|\mathcal V_a|^{-1}+|\mathcal V_b|^{-1}\le1$. Then $y_a^\top Ly_a=v_a/v^*_h$, $y_b^\top Ly_b=v_b/v^*_h$, and
\[
y_{ab}^\top Ly_{ab}=\frac{c_{ab}^2}{v^*_h}\Bigl\{(e_a-e_b)^\top\Phi(e_a-e_b)+\frac{v_a}{|\mathcal V_a|}+\frac{v_b}{|\mathcal V_b|}\Bigr\}\ge\frac{2\varepsilon+\min\{v_a,v_b\}}{v^*_h},
\]
where we use $(e_a-e_b)^\top\Phi(e_a-e_b)\ge2\varepsilon$ and $c_{ab}\ge1$. If $\min\{v_a,v_b\}\le v^*_h-\varepsilon$, then $y_a$ or $y_b$ gives a Rayleigh quotient at most $1-\varepsilon/v^*_h$; otherwise $y_{ab}^\top Ly_{ab}\ge1+\varepsilon/v^*_h$.

In both cases $\Vert L-I\Vert_F\ge\Vert L-I\Vert_{\op}\ge\varepsilon/v^*_{\max}$, so Lemma~\ref{th:lem:glower} gives, for $\varepsilon\le v^*_{\max}/2$,
\[
R_n(\rho,\theta)\ge\frac n9\alpha(1-\alpha)\min\Bigl\{\frac{\varepsilon^2}{(v^*_{\max})^2},\frac14\Bigr\}=\frac{n\alpha(1-\alpha)\varepsilon^2}{9(v^*_{\max})^2}.
\]
This replaces the bound $n\alpha(1-\alpha)\varepsilon^2/(18\Vert\Omega_*\Vert_{\op}^2)$ in the proof of Lemma~\ref{th:lem:iw}, and the rest of that proof applies with $c_0'$ in place of $c_0$.
\end{proof}

\subsubsection{Proof of Proposition~\ref*{th:prop:curves-p}}

\begin{proof}
\emph{Part (a).} The bound on $\Lt(K)$ follows as in the proof of Proposition~\ref{th:prop:curves}(a), since $\En(K)\ge n\alpha m_K$ holds for every $p_0$ and $N_K\le\bar N$. The bound on $\Ut(K)$ follows from Lemma~\ref{th:lem:oracle-p} with $\rho=\rho_K$ and $\theta_c=\theta_K$, whose constants depend only on $K<K_*$, $\bar C$, the prior and $\alpha$, together with $\sum_h\log|\mathcal V_h|\le(K+1)\log p_0$.

\emph{Part (b).} Lemma~\ref{th:lem:oracle-p} with $\rho=\rho_*$ and $\theta_c=\theta_*$ gives $k_c=0$ and $B_c=0$, and $d_{\rho_*}=d_*$.

\emph{Part (c).} Let $K>K_*$. If $\rho\in\Fam_K$ has zero R\'enyi bias, Lemma~\ref{th:lem:iw-p} applies once $\varepsilon_n\le\min\{c/(16\nu_{\bar M}\bar M),\,v^*_{\max}/2\}$, a condition that does not involve $p_0$. Otherwise $\En(K,\rho)\ge n\alpha\beta_{\min}$ by (W$'$)(ii). Since $\Lt(K)=\En(K)-\log N_K-\log n$ and $N_K\le\bar N$, (c) follows.
\end{proof}

\subsubsection{Proof of Corollary~\ref*{th:cor:exact-p}}

\begin{proof}
We follow the proof of Corollary~\ref{th:cor:exact} and only change the verification of the conditions of Theorems~\ref{th:thm:early} and \ref{th:thm:late}, which now uses Proposition~\ref{th:prop:curves-p}.

For $\mathfrak E_1$, let $1\le K\le K_*-1$. By (W$'$)(i), $m_{K-1}\ge(1+\varpi)k_K$, and since $k_K\ge\mKL_K\ge m_K$, also $m_{K-1}\ge(1+\varpi)m_K$. Iterating this and using (W$'$)(ii), we get $m_{K-1}\ge m_{K_*-1}\ge\beta_{\min}$. Hence, by Proposition~\ref{th:prop:curves-p}(a),
\begin{align*}
\Lt(K-1)-\Ut(K)&\ge n\alpha(m_{K-1}-k_K)-C\bigl\{\sqrt{n\log n}\,(\sqrt{k_K}+k_K)+\log n+K_*\log p_0\bigr\}\\
&\ge\frac{\varpi}{1+\varpi}\,n\alpha m_{K-1}\\
&\qquad-C\bigl\{\sqrt{n\log n}\,(\sqrt{m_{K-1}}+m_{K-1})+\log n+K_*\log p_0\bigr\},
\end{align*}
where we use $k_K\le m_{K-1}/(1+\varpi)$ for the second line. Dividing the last line by $n\alpha m_{K-1}\ge n\alpha\beta_{\min}$, we see that it is positive for large $n$, because $\sqrt{\log n/(n\beta_{\min})}$, $\sqrt{\log n/n}$ and $(\log n+K_*\log p_0)/(n\beta_{\min})$ tend to zero by (D). For $K=K_*$, Proposition~\ref{th:prop:curves-p}(a) and (b) give $\Lt(K_*-1)-\Ut(K_*)\ge n\alpha\beta_{\min}-C\{\log n+(K_*+1)\log p_0\}>0$ for large $n$. Theorem~\ref{th:thm:early} then gives $\Pst(\mathfrak E_1^c)\le2K_*/n$.

For $\mathfrak E_2$, Proposition~\ref{th:prop:curves-p}(b) and (c), together with $\sum_h\log|\mathcal V^*_h|\le(K_*+1)\log p_0$, give
\[
\Lt(K_*+1)-\Ut(K_*)\ge\min\bigl\{c_0'(n\alpha(1-\alpha))^{1/3},\,n\alpha\beta_{\min}\bigr\}-C\bigl\{\log n+(K_*+1)\log p_0\bigr\},
\]
which is positive for large $n$ by (D). Theorem~\ref{th:thm:late} then gives $\Pst(\mathfrak E_2^c)\le2/n$.

For $\mathfrak E_3$, every $\rho\in\Fam_{K_*}\setminus\{\rho_*\}$ has a positive R\'enyi bias by Lemma~\ref{th:lem:zero}, which is at least $\beta_{\min}$ by (W$'$)(ii). Since $\Ut(K_*)\le C\{\log n+(K_*+1)\log p_0\}<n\alpha\beta_{\min}-\log n$ for large $n$, the argument of Corollary~\ref{th:cor:exact} gives $\Pst(\mathfrak E_3^c)\le N_{K_*}/n$.
\end{proof}

\subsubsection{Proof of Corollary~\ref*{th:cor:rate-p}}

\begin{proof}
We follow the proof of Corollary~\ref{th:cor:rate}. Theorem~\ref{th:thm:risk} with $k^\circ=K_*$ and $t=\log n$, together with $\Pst(\Kstop<K_*)\le\Pst(\mathfrak E_1^c)=O(1/n)$ and Proposition~\ref{th:prop:curves-p}(b), gives the first claim, where $\log(\alpha n)\le\log n$ and $\log\Ntot\le\log\{\bar N(\bar M+1)\}$. For the second, the left-hand side is non-increasing in $A_n$, so we may assume that
\[
\varepsilon=A_n\sqrt{\frac{d_*\log n+(K_*+1)\log p_0}{n\alpha(1-\alpha)}}\le\frac12 .
\]
Taking this $\varepsilon$ in Theorem~\ref{th:thm:risk}, the right-hand side is, on the same event, at most
\[
\frac{9\bigl\{(\frac{d_*}2+1)\log n+\frac{K_*+1}2\log p_0+C\bigr\}}{A_n^2\{d_*\log n+(K_*+1)\log p_0\}}\to0,
\]
where we use $\sum_h\log|\mathcal V^*_h|\le(K_*+1)\log p_0$. As this event has probability $1-O(1/n)$, the left-hand side tends to $0$ in probability.
\end{proof}

\section{Implementation details of the confirmatory step}
\label{app:implementation}

The following coordinate-ascent updates provide the implementation details
omitted from the main text. All loading moments are computed in the signed
row-sum parameterization~\eqref{eq:unit-row-sum}; no separate prior is placed on the primary loadings.

\subsection{Shared factor-score and factor-covariance updates}

The inverse-Wishart prior for \(\Phi\) is conjugate to the Gaussian factor
scores, so its variational factor remains inverse-Wishart. Let
\[
  S_f:=\sum_{i=1}^n\E_q(f_if_i^\top)
      =nV_f+\sum_{i=1}^n m_im_i^\top
\]
denote the sum of the posterior second moments of the subject-specific factor
scores. Then
\begin{equation}
  q(\Phi)
  =\operatorname{IW}_K\!\left(
  \nu_\Phi,\Psi_\Phi\right),
  \qquad
  \nu_\Phi=\nu_\Phi^{(0)}+n,
  \qquad
  \Psi_\Phi=\Psi_\Phi^{(0)}+S_f .
  \label{eq:full-phi-posterior}
\end{equation}
Consequently,
\begin{align}
  \E_q(\Phi)
  &=\frac{\Psi_\Phi}{\nu_\Phi-K-1}
    =\frac{\nu_0\tau_fI_K+S_f}{n+\nu_0},\\
  \E_q(\Phi^{-1})
  &=\nu_\Phi\Psi_\Phi^{-1}.
  \label{eq:full-phi-moments}
\end{align}
The first expression is the posterior mean factor covariance, whereas the
second is the expected factor precision. They are not interchangeable: the
factor-score update uses \(\E_q(\Phi^{-1})\), rather than
\(\E_q(\Phi)^{-1}\).

Writing \(\bar\lambda_j=\E_q(\lambda_j)\),
\(S_{\lambda j}=\E_q(\lambda_j\lambda_j^\top)\), and
\(\tau_j=\E_q(\psi_j^{-1})=a_j/b_j\), the subject-specific factor-score update
is
\begin{align}
  V_f
  &=
  \left(
  \E_q(\Phi^{-1})
  +
  \sum_{j=1}^p
  \E_q(\psi_j^{-1})S_{\lambda j}
  \right)^{-1},\\
  m_i
  &=
  V_f
  \sum_{j=1}^p
  \E_q(\psi_j^{-1})x_{ij}\bar\lambda_j .
  \label{eq:gb-score-update}
\end{align}
The covariance \(V_f\) is shared across subjects, while \(m_i\) depends on
the measurements for subject \(i\). For gg-BFA (1f), the same score and
factor-covariance formulas apply with
\(\bar\lambda_j=e_{h_j}\),
\(S_{\lambda j}=e_{h_j}e_{h_j}^\top\), and
\(\tau_j=\E_q(v_{h_j}^{-1})\).

\subsection{Option-specific loading updates}

\paragraph{mf loading update.}
For gg-BFA (mf), the loading update is most transparent after
substituting the row-sum constraint into the likelihood:
\begin{equation}
  x_{ij}
  =f_{i,h_j}+\sum_{k\ne h_j}\eta_{jk}
  \bigl(f_{ik}-f_{i,h_j}\bigr)+\epsilon_{ij}.
  \label{eq:difference-coordinate-regression}
\end{equation}
For \(d_{ijk}=f_{ik}-f_{i,h_j}\), define
\begin{align*}
  T_{jk}&=\sum_i\E_q(d_{ijk}^2),\\
  r_{jk}&=\sum_i\E_q\!\left[
  d_{ijk}\left\{x_{ij}-f_{i,h_j}
  -\sum_{\ell\ne h_j,k}\bar\eta_{j\ell}d_{ij\ell}
  \right\}\right].
\end{align*}
For a zero-mean Gaussian component with variance \(s_c^2\), its scalar update
is
\begin{equation}
  v_{jkc}=\left\{\tau_jT_{jk}+s_c^{-2}\right\}^{-1},
  \qquad
  u_{jkc}=v_{jkc}\tau_j r_{jk}.
  \label{eq:gb-loading-component-update}
\end{equation}
The two components use \(s_0^2=\sigma_{\mathrm{spike}}^2\) and
\(s_1^2=\sigma_{\mathrm{slab}}^2\).  With \(w_0=1-\pi\) and \(w_1=\pi\),
the cross-loading PIP is updated from the unnormalized posterior log weights of the two
components,
\begin{align}
  \log\tilde w_{jkc}
  &=\log w_c-\tfrac12\log s_c^2+\tfrac12\log v_{jkc}
    +\tfrac12(\tau_jr_{jk})^2v_{jkc},\\
  \rho_{jk}&=\operatorname{logit}^{-1}(\log\tilde w_{jk1}-\log\tilde w_{jk0}).
  \label{eq:gb-cross-pip-update}
\end{align}
The free-coordinate posterior moments used elsewhere in the algorithm are
\[
  \bar\eta_{jk}=(1-\rho_{jk})u_{jk0}+\rho_{jk}u_{jk1},
  \qquad
  \E_q(\eta_{jk}^2)
  =(1-\rho_{jk})(v_{jk0}+u_{jk0}^2)
   +\rho_{jk}(v_{jk1}+u_{jk1}^2).
\]
Under this option, the induced primary moments are
\begin{align}
  \E_q(\lambda_{j,h_j})
  &=1-\sum_{k\ne h_j}\bar\eta_{jk},\\
  \operatorname{Var}_q(\lambda_{j,h_j})
  &=\sum_{k\ne h_j}\operatorname{Var}_q(\eta_{jk}),
  &
  \operatorname{Cov}_q(\lambda_{j,h_j},\eta_{jk})
  &=-\operatorname{Var}_q(\eta_{jk}).
  \label{eq:induced-primary-moments}
\end{align}
Consequently, \(q(\lambda_j)\) has a singular full covariance whose null
direction is the vector of ones, exactly enforcing the signed row sum.

\paragraph{2f and 2f+s loading updates.}
For a candidate target \(k\ne h_j\), both structured options use the same
single-direction regression
\[
  x_{ij}=f_{i,h_j}+\eta_j(f_{ik}-f_{i,h_j})+\epsilon_{ij}
\]
and the Gaussian active-state moments in
Eq.~\eqref{eq:gb-loading-component-update} with variance \(\sigma_a^2\).
For gg-BFA (2f), the normalized row-state log probabilities combine this
local Gaussian evidence with
\(\E_q\log(1-\theta_h)\) for the no-overlap state and
\(\E_q\log\theta_h+\E_q\log\omega_{hk}\) for target \(k\).  The conjugate
global updates are
\begin{align}
  q(\theta_h)
  &=\operatorname{Beta}\!\left(
      a_\theta+\sum_{j\in\mathcal B_h}\Pr_q(s_j\ne0),
      b_\theta+\sum_{j\in\mathcal B_h}\Pr_q(s_j=0)
    \right),\\
  q(\omega_{h,-h})
  &=\operatorname{Dirichlet}\!\left(
      \alpha_\omega\mathbf 1+
      \left\{\sum_{j\in\mathcal B_h}\Pr_q(s_j=k)\right\}_{k\ne h}
    \right).
  \label{eq:two-global-updates}
\end{align}
For gg-BFA (2f+s), the target evidence first produces the gate log-odds
\(\ell_j\) in Eq.~\eqref{eq:truncated-bernoulli-q}; the dynamic program then
returns the exact marginal \(\Pr_q(g_j=1)\) under the block budget, after which
the Dirichlet target update uses the budgeted target responsibilities.  Loading
means and second moments are obtained by averaging
\(e_{h_j}+\eta_j(e_k-e_{h_j})\) over the mutually exclusive row states.
Accordingly, both structured options retain the negative primary--target
dependence required by Eq.~\eqref{eq:unit-row-sum}; unlike the mf option,
they also retain posterior dependence among competing target columns.

\subsection{Residual variance and convergence}

For gg-BFA (1f), the shared residual variance of community \(h\) has the update
\begin{equation}
  q(v_h)=\operatorname{IG}\!\left(
    a_0+\frac{n|\mathcal B_h|}{2},\;
    b_0+\frac12\sum_i\sum_{j\in\mathcal B_h}
    \{(x_{ij}-m_{ih})^2+(V_f)_{hh}\}\right).
  \label{eq:fixed-residual-update}
\end{equation}

For each learned-loading member, define
\(R_j=\sum_i\E_q[(x_{ij}-\lambda_j^\top f_i)^2]\). The residual update is
\begin{equation}
  q(\psi_j)=\operatorname{IG}(a_j,b_j),
  \qquad
  a_j=a_0+\frac n2,
  \qquad
  b_j=b_0+\frac{R_j}{2}.
  \label{eq:gb-residual-update}
\end{equation}

One iteration updates scores, \(q(\Phi)\), the selected option's loading states
and amplitudes, induced loading moments, and residual variances, then evaluates
the corresponding evidence lower bound.
Iteration stops after the prescribed minimum number of iterations when the
relative ELBO change is below tolerance; an ELBO decrease beyond numerical
tolerance is treated as an error.

The implementation vectorizes the loading-moment and residual operations over
variables and subjects; the shared score precision is only \(K\times K\).
The 2f option averages over \(K\) row states, while 2f+s additionally uses the
blockwise dynamic program. Runtime for model fitting is recorded separately
from posterior draws and masked-entry prediction.

\subsection{Posterior covariance calculation}

Write
\[
  \bar\Lambda_o=\E_{q_o}(\Lambda),\qquad
  \bar\psi_o=\E_{q_o}(\psi),\qquad
  \bar\Phi_o=\E_{q_o}(\Phi).
\]
Because \(q(\Phi)\) and \(q_o(\Lambda)\) are separate variational blocks, the
entrywise form of Eq.~\eqref{eq:gb-posterior-mean-covariance} is
\[
  (\widehat\Omega_o)_{jl}
  =\tr\!\left\{\bar\Phi_o
    \E_{q_o}(\lambda_l\lambda_j^\top)\right\}
    +\Ind{j=l}\bar\psi_{o,j}.
\]
Thus, the factor covariance enters through \(\bar\Phi_o\), while the loadings
enter through their joint second moments. When a variable has posterior mass
on several loading states, this expression averages the covariance implied by
each admissible state. It is generally different from the plug-in estimator
\(
\bar\Lambda_o\bar\Phi_o\bar\Lambda_o^\top+
\diagop(\bar\psi_o)
\), which combines alternative loading states in the posterior mean loading
matrix before constructing the covariance.

\subsection{Local variational prediction}
\label{app:local-vb-prediction}

For a new subject, hold the fitted global variational factors fixed and use
the observed entries \(x_O\) to obtain
\[
  q(f_{\mathrm{new}}\mid x_O)
  =\Norm(m_{\mathrm{new}},V_{\mathrm{new}}).
\]
Let \(D_H\) be the diagonal matrix of training-data scales for the masked
variables, let \(\bar\Lambda_H=\E_q(\Lambda_H)\), and define
\[
  M_{\mathrm{new}}
  =V_{\mathrm{new}}+m_{\mathrm{new}}m_{\mathrm{new}}^\top.
\]
On the original measurement scale, the point prediction is
\begin{equation}
  \widehat x_H^{\mathrm{VB}}
  =\mu_H+D_H\bar\Lambda_Hm_{\mathrm{new}},
  \label{eq:vb-predictive-mean}
\end{equation}
and its predictive covariance is
\begin{equation}
  \widehat V_H^{\mathrm{VB}}
  =D_H\left[
    \E_q\!\left\{\Lambda_HM_{\mathrm{new}}\Lambda_H^\top\right\}
    -(\bar\Lambda_Hm_{\mathrm{new}})
     (\bar\Lambda_Hm_{\mathrm{new}})^\top
    +\diagop\!\{\E_q(\psi_H)\}
  \right]D_H.
  \label{eq:vb-predictive-covariance}
\end{equation}
To obtain predictive intervals, draw
\[
  f_{\mathrm{new}}^{(b)}
  \sim q(f_{\mathrm{new}}\mid x_O),
  \qquad
  (\Lambda^{(b)},\psi^{(b)})\sim q(\Lambda,\psi),
\]
followed by
\[
  z_H^{(b)}
  \sim
  \Norm\!\left(
    \Lambda_H^{(b)}f_{\mathrm{new}}^{(b)},
    \diagop\{\psi_H^{(b)}\}
  \right),
  \qquad
  x_H^{(b)}=\mu_H+D_Hz_H^{(b)}.
\]
The interval endpoints are the 2.5th and 97.5th percentiles of
\(\{x_H^{(b)}\}\).  The masked measurements do not enter the local update.

\subsection{Defaults and interval conventions}
\label{app:defaults}

Package defaults are \(a_0=b_0=0.01\), \(\nu_0=\tau_f=1\),
\(\pi=0.1\), \(\sigma_{\mathrm{spike}}^2=10^{-4}\),
\(\sigma_{\mathrm{slab}}^2=\sigma_a^2=0.25\),
\(a_\theta=1,b_\theta=9,\alpha_\omega=1,\pi_g=0.05,\kappa=0.1\).
The cross-loading simulation uses the residual prior \(\operatorname{IG}(5,2)\) instead.
The variational fits allow at most 1,000 iterations, with relative-ELBO
tolerance \(10^{-6}\) and at least five iterations. EFA uses its own
likelihood criterion and at most 5,000 iterations in the real-data
comparison. In the real-data analyses, the variables are standardized by their mean and their
standard deviation with divisor \(n\), computed from the training subjects in the held-out
analyses and from all subjects in the full-data analyses, whose empirical covariance is
\(S=X^\top X/n\).

The package returns analytic marginal summaries where they are available and seeded posterior
draws of the five posterior objects of Section~\ref{sec:posterior-outputs}.  By default it
summarizes the mixture distribution of a loading by a normal approximation; the two main
simulations use the quantiles of the mixture itself.  Quantiles of the factor covariance use
inverse-Wishart draws.  Real-data predictive intervals use the empirical quantiles of 5,000 draws
per test subject, which we checked against 10,000 draws.  All intervals come from the variational
approximation, not from the exact posterior.

\clearpage
\section{Evaluation metric definitions}
\label{app:evaluation-metrics}

This section gives the full definitions behind the evaluation map in
Table~\ref{tab:evaluation-map}. Metrics are first computed within a simulation
replicate or held-out fold and are then averaged across replicates or across
the five folds. Monte Carlo standard errors use the replicate-level values;
real-data tables report the fold mean together with the sample standard deviation or the range
over folds.  Throughout, \(p\) is the number of evaluated variables.

\subsection{Structure recovery}
\label{app:structure-metrics}

In the structure simulations of Comparison~1 (Section~\ref{sec:sim-selection}), each procedure
returns a number \(\widehat K\) of communities or factors and a partition of the variables, and
\(\widehat K\) is averaged over the data sets.  Accuracy is the proportion of data sets in which the
selected partition equals the true one; a run that did not complete counts as incorrect.  The
adjusted Rand index \citep{hubert1985ari} compares the selected and the true partitions over all
variables, with the background treated as one block.  The covariance difference
\[
  \mathrm{Cov}_F^d=\bigl\Vert\Omega_*^{-1/2}\widehat\Omega\,\Omega_*^{-1/2}-I\bigr\Vert_F
\]
compares the estimate \(\widehat\Omega\), the posterior mean for gg-BFA and the covariance of the
fitted confirmatory model for the pipelines, with the population correlation matrix
\(\Omega_*\); it is averaged over the completed runs.

\subsection{Covariance and correlation recovery}

In the simulations of Comparison~2, the population covariance \(\Omega_*\) is known. Let
\[
  R_*=D_*^{-1/2}\Omega_*D_*^{-1/2},
  \qquad
  \widehat R=\widehat D^{-1/2}\widehat\Omega\widehat D^{-1/2},
\]
where \(D_*=\diagop(\Omega_*)\) and
\(\widehat D=\diagop(\widehat\Omega)\). Off-diagonal correlation recovery is
measured by
\[
 \operatorname{RMSE}_{\mathrm{cor}}
 =\left\{\frac{2}{p(p-1)}
   \sum_{j<l}(\widehat R_{jl}-R_{*,jl})^2\right\}^{1/2},
\]
and covariance recovery is measured by the relative Frobenius error
\[
  d_F(\widehat\Omega,\Omega_*)
  =\frac{\|\widehat\Omega-\Omega_*\|_F}{\|\Omega_*\|_F}.
\]
The former removes marginal scale; the latter retains it. Independent test
observations provide the Gaussian negative log likelihood described below.

In real data, the population covariance is unknown. If
\(S_{\mathrm{eval}}\) is the empirical covariance of the evaluation subjects,
we use
\[
  d_F(\widehat\Omega,S_{\mathrm{eval}})
  =\frac{\|\widehat\Omega-S_{\mathrm{eval}}\|_F}
  {\|S_{\mathrm{eval}}\|_F}.
\]
For the complete-data analysis, the fitted covariance and
\(S_{\mathrm{eval}}\) use the same subjects, so this is a descriptive measure
of reconstruction. For covariance generalization, the model is fitted to the
training subjects and \(S_{\mathrm{eval}}\) uses only held-out subjects.

Let \(\widehat R\) and \(R_{\mathrm{eval}}\) be the correlation matrices
corresponding to \(\widehat\Omega\) and \(S_{\mathrm{eval}}\). Their
off-diagonal standardized root mean square residual is
\[
  \operatorname{SRMR}
  =\left\{\frac{2}{p(p-1)}
    \sum_{j<l}(\widehat R_{jl}-R_{\mathrm{eval},jl})^2
    \right\}^{1/2}.
\]
For held-out observations \(x_1,\ldots,x_m\), the Gaussian negative log
likelihood per variable is
\[
  \operatorname{NLL}_{\mathrm{var}}
  =\frac{1}{2mp}\sum_{i=1}^m
  \left\{p\log(2\pi)+\log|\widehat\Omega|
  +(x_i-\widehat\mu)^\top\widehat\Omega^{-1}
  (x_i-\widehat\mu)\right\},
\]
where \(\widehat\mu\) and every preprocessing quantity are estimated from the
training subjects. A non-positive-definite covariance receives no NLL score;
we do not repair it for scoring.

All methods use the same variables and subjects within a comparison. The main
CFA comparison uses the variables assigned to the training-data community
partition. In the complete retained-panel analysis, unassigned variables enter
community-based estimates as residual-only coordinates and have zero fitted
covariance with the selected variables.

\subsection{Parameter recovery and interval calibration}

Let \(\mathcal I_\theta\) denote the evaluated coordinates of a parameter
\(\theta\). Within simulation replicate \(b\), we compute
\begin{align}
  \operatorname{RMSE}_b(\theta)
  &=\left\{
    \frac{1}{|\mathcal I_\theta|}
    \sum_{r\in\mathcal I_\theta}
    (\widehat\theta_{br}-\theta_{br})^2
  \right\}^{1/2},\\
  \operatorname{Coverage}_b(\theta)
  &=\frac{1}{|\mathcal I_\theta|}
    \sum_{r\in\mathcal I_\theta}
    \Ind{\theta_{br}\in[L_{br},U_{br}]},\\
  \operatorname{Width}_b(\theta)
  &=\frac{1}{|\mathcal I_\theta|}
    \sum_{r\in\mathcal I_\theta}(U_{br}-L_{br}).
  \label{eq:simulation-parameter-metrics}
\end{align}
We apply these definitions separately to loadings, factor covariance, factor
scores, and residual variances.

For a binary cross-loading indicator, let
\(\rho_{jk}=\Pr_q(\gamma_{jk}=1)\). Its 95\% equal-tail credible set is
\(\{0\}\) when \(\rho_{jk}\le 0.025\), \(\{1\}\) when
\(\rho_{jk}\ge 0.975\), and \(\{0,1\}\) otherwise. Coverage is reported
separately for active and inactive cross-loadings. Support recovery uses a
0.5 posterior-inclusion threshold to compute the true-positive and
false-discovery rates.

\subsection{Masked-measurement prediction}

Masked values and their predictions are on the scale of the training-standardized data.  Let
\(\mathcal M\) be the set of masked subject--variable cells,
\(y_c\) their observed values, \(\widehat y_c\) their predictive means, and
\([L_c,U_c]\) their 95\% predictive intervals. The pooled prediction summaries
are
\begin{align*}
  \operatorname{RMSE}
  &=\left\{|\mathcal M|^{-1}\sum_{c\in\mathcal M}
    (y_c-\widehat y_c)^2\right\}^{1/2},\\
  \operatorname{MAE}
  &=|\mathcal M|^{-1}\sum_{c\in\mathcal M}|y_c-\widehat y_c|,\\
  \operatorname{Coverage}_{0.95}
  &=|\mathcal M|^{-1}\sum_{c\in\mathcal M}
    \Ind{y_c\in[L_c,U_c]},\\
  \operatorname{Width}_{0.95}
  &=|\mathcal M|^{-1}\sum_{c\in\mathcal M}(U_c-L_c).
\end{align*}
Negative log predictive density is averaged over the same masked cells. The
common Gaussian predictor uses its conditional Gaussian density. For the
separate variational-posterior prediction, the reported NLPD uses a Gaussian density matched to
the predictive mean and variance, while the interval endpoints are empirical quantiles of the
joint predictive draws. Predictive coverage must be read together with interval
width and NLPD. It concerns future observations and is distinct from the
parameter coverage measured in simulation.

\subsection{Loading complexity and computation}

Loading complexity is summarized by the expected number of active
cross-loadings per selected row,
\(\sum_{k\ne h_j}\E_q\gamma_{jk}\), and by the density of the displayed
posterior-mean loading matrix,
\[
 \frac{1}{pK}\sum_{j,k}\Ind{|\E_q\lambda_{jk}|>0.05}.
\]
The displayed posterior mean may contain small entries in several target
columns even when a posterior draw permits only one cross-loading. The 2f+s
budget instead constrains the number of cross-loading rows within each
community in every draw.

Runtime is recorded separately for community discovery, model fitting,
covariance construction, posterior sampling, prediction, scoring, and I/O.
Convergence status, iteration count, and peak memory accompany the timing
results. The main GPU comparisons use estimator time---fitting plus covariance
construction---because community discovery and evaluation are shared across
methods.

\clearpage
\section{Supplementary simulation evidence}
\label{app:simulation-details}

This section gives the simulation specifications and results behind the summary in
Section~\ref{sec:sim-summary}. It first gives the designs of the structure simulations of
Comparison~1. For Comparison~2, it then gives the generating mechanisms, the results of the two
main simulations (Sections~\ref{sec:experiment1-current-results}
and~\ref{sec:experiment2-current-results}), calibration and computation, variational bagging, and the
stress test.

\subsection{Additional details of the structure simulations}
\label{app:sim-selection-design}

The structure simulations of Comparison~1 draw \(n\) independent observations
from \(N_{p_0}(0,\Omega_*)\) through the factor representation, and every
procedure uses the same data sets.  In S1, the residual variance of community
\(h\) is \((1-r_h)/r_h\), where \(r_h\) is its within-community correlation,
and with \(p_0=2n\) the communities contain 15, 13, 11, 9, and 7\% of the
variables, rounded to integers.

The following numbers support the summary in Section~\ref{sec:sim-selection}.
In S1, the exploratory step missed the true partition only in data sets in which it was not
among the candidates, and it then returned five communities with adjusted
Rand index at least \(0.98\).  The mean error of the exploratory step was within \(0.03\) of
the oracle error.  With \(p_0=200\), the error multiplied by \(n^{1/2}\)
stayed between \(8.1\) and \(9.2\); with \(\alpha\) set to the average
\(\alpha_0\), the error was between \(0.28\,r_n\) and \(0.34\,r_n\), where
\(r_n\) has an extra factor \((\log n)^{1/2}\) compared with the parametric
rate.  The standardization adds about \((2/n)\sum_h|\mathcal V_h|r_h^2\) to
the squared error; with \(p_0=2n\) this term equals \(0.53\), an error of
\(0.73\) on its own, close to the observed \(0.73\)--\(0.78\).  In S2, the
number of selected communities grew by about \(2.05\) per unit of \(\log n\).
The proof of Corollary~\ref{th:cor:geom} shows that the search reaches the
first \(K\) with \(n\alpha Be^{-\gamma K}\le\zeta d_K\log n\).  This \(K\)
grows by \((1-1/\log n)/(\gamma+\mathrm d\log d_K/\mathrm dK)\) per unit of
\(\log n\), which lies between \(1.9\) and \(2.4\) over the observed range when
\(\gamma\approx0.27\), the decay rate of the Kullback--Leibler bias along the
true path.  The error decreased with a log--log slope of \(-0.39\), faster than
the rate \(\{(\log n)^3/n\}^{1/2}\) of the corollary, whose slope over this
range is about \(-0.28\).  The two-stage version returned the true partition
in 56--86\% of the S1 data sets, and when it did not, its error was between
\(1.6\) and \(25\).  Excluding the shared community extraction, the exploratory step fitted 78--390
candidate models per data set and took 1--15 minutes on one GPU, whereas the
two-stage version fitted 19--39 models in 6--37 seconds.

\subsection{Simulation designs and interval construction}
\label{app:sim-designs}

The fixed-community simulation uses nine settings with \((n,p)\) equal to
\((40,20)\), \((40,30)\), \((40,40)\), \((80,40)\), \((80,80)\),
\((80,120)\), \((120,40)\), \((120,120)\), and \((120,200)\).
Each has 100 paired replicates, three oracle communities with sizes in the
ratio \(3:3:4\), primary-mask loadings, and block residual variances
\((0.1,0.2,0.5)\). The fitted methods are SCFA and the four gg-BFA members, without bagging.
The factor covariance is
\[
\Phi_*=
\begin{pmatrix}
2.020878&0.730920&1.153572\\
0.730920&3.130999&1.633474\\
1.153572&1.633474&3.693128
\end{pmatrix}.
\]

The cross-loading simulation fixes \(n=300,p=40,K=4\), with four oracle blocks of size 10.
It generates data from three loading structures and fits SCFA, the four gg-BFA members, and
gg-BFA (mf), (2f), and (2f+s) with bagging to each.
The factor covariance has diagonal one and off-diagonal 0.3; variable
residual variances follow \(\operatorname{IG}(5,2)\). The
multiple-cross-loading truth
uses independent Bernoulli(0.1) slab indicators, with Gaussian slab and spike
variances 0.25 and \(10^{-4}\). In the single-cross-loading truth, each row overlaps with
probability 0.1, chooses a uniformly random nonprimary target, and receives a
\(\Norm(0,0.25)\) amplitude. The sparse truth applies the same construction
but randomly retains at most one overlapping row per block. Its support is
compatible with gg-BFA (2f+s), but it is not drawn from the exact truncated prior of that member. Primary loadings are always derived from the signed row sum.
There are 100 paired seeds: factor scores, residual variances, and noise
are shared across the three truths within a seed, and all methods receive the
same matrix within a truth.

Both simulations report parameter error, empirical 95\% coverage, interval
width, runtime, and loading complexity. SCFA intervals are frequentist Wald intervals, with factor-score intervals
conditional on the fitted block residual variances; Bayesian intervals use the fitted
variational families. Intervals for the learned loadings use the quantiles of their mixture
distributions, not the moment-matched normal approximation that the package uses by default. Factor-covariance intervals use 1,000 seeded draws. In the cross-loading simulation, bagged gg-BFA
uses \(B=20\) bootstrap fits of size \(M=2n\), with 200 posterior draws per
component for marginal summaries. Its primary partition remains fixed.
Appendix~\ref{app:bagging} defines the bagged mixture.

The supplementary stress test uses \(n=60,p_0=128,K=8\), 1,000 independent test observations,
and 30 paired seeds for each of four truths. Ninety-six variables carry
factor signal and 32 are residual-only. Three truths have eight blocks of
12 signal variables, factor covariance with diagonal one and off-diagonal
0.3, signal residual variance \(0.75^2\), and noise variance one.
The first has primary loading one and no cross-loadings; the second replaces
each primary loading by \(\max\{0.05,\Norm(1,0.3^2)\}\); the third additionally
gives each signal variable probability 0.3 of one random cross-loading,
with equiprobable sign and magnitude
\(\max\{0,\Norm(0.25,0.0625^2)\}\).
Unlike the main cross-loading simulation, these continuous-loading truths need not satisfy the
unit-row-sum constraint.

The fourth truth is neutral and FactorVerse-compatible, with \(\Phi=I\).
Its 96 signal rows have Bernoulli(0.15) support conditioned on at least one
active factor per row and at least eight active rows per factor. Active
loadings have equiprobable signs and Uniform(0.5,1) magnitudes. There is no
generating primary partition; for the community-based fits only, the largest
absolute loading supplies a post-generation oracle primary label.
Unrestricted EFA and FactorVerse receive all 128 variables but no labels.
All covariance scores use the full 128-variable target.

The 14 methods are the sample covariance as a reference, SCFA, the four gg-BFA members, EFA, two
BCME variants, and five FactorVerse priors.
FactorVerse uses fixed \(K=8\), one chain, 2,000 iterations with 1,000 burn-in
for MGPS, DL, and BASS, and 8,000 with 4,000 burn-in for SSL and HS.
The primary endpoint is off-diagonal correlation RMSE; covariance Frobenius
error and test Gaussian negative log likelihood are secondary. No fitted
covariance is rescaled using the truth.

\subsection{When the community mask is correct}
\label{sec:experiment1-current-results}

In the fixed-community simulation (Appendix~\ref{app:sim-designs}), the true model uses the
binary community mask as its loading matrix, shares residual variances within
three communities, and allows an unrestricted factor covariance. gg-BFA (1f)
is therefore correctly specified for both the loading matrix and the
residual-variance structure. The three learned-loading members receive the
same correct primary labels but estimate a separate residual variance for each
variable. Comparisons with 1f consequently reflect both loading and residual
parameterization, whereas comparisons among mf, 2f, and 2f+s isolate the
loading-support restriction more closely.

The primary comparison is covariance recovery, measured by
\(d_F(\Omega)\). Parameter recovery and interval coverage help explain
differences between methods, while loading complexity and runtime describe
the cost of that recovery.

When the binary community mask is correct, additional loading flexibility
provides little improvement in covariance recovery
(Table~\ref{tab:exp1-point-current}). SCFA and gg-BFA (1f) behave almost
identically, while the three learned-loading members achieve similar
covariance and factor-score recovery. Their residual-variance estimates are
less accurate because they estimate a separate residual variance for every
variable under a truth that shares residual variation within each community.
In this setting, learning cross-loadings therefore adds flexibility without
materially improving the recovered covariance.

Because the generating model contains no cross-loadings, a well-matched
loading model should suppress unnecessary overlap. The mf member retains more
posterior mass on nonprimary loadings, whereas 2f and 2f+s remain closer to the
true binary mask. This reduction in loading complexity leaves covariance
recovery essentially unchanged and shortens fitting time. When the community
mask is correct, restricting the cross-loading support can therefore improve
parsimony and computation without sacrificing covariance accuracy.

\begin{table}[!htbp]
\centering
\caption{Fixed-mask simulation: point recovery, loading complexity, and mean
fit time per data set. All gg-BFA members use the same computing
configuration; SCFA is an analytic reference. \(d_F(\Omega)\) is relative
Frobenius covariance error. Because the truth has no cross-loadings, expected
cross-loadings per row measure false overlap.}
\label{tab:exp1-point-current}
\scriptsize
\setlength{\tabcolsep}{3.5pt}
\resizebox{\textwidth}{!}{%
\begin{tabular}{@{}lrrrrrrr@{}}
\toprule
Method & RMSE\((F)\) & RMSE\((\psi)\) & RMSE\((\Phi)\) & RMSE\((\Lambda)\) &
\(d_F(\Omega)\) & E[cross/row] & Time (s) \\
\midrule
Closed-form SCFA        & .1188 & .0116 & .4076 & fixed & .1966 & .0000 & .0005 \\
gg-BFA (1f)            & .1183 & .0127 & .4001 & fixed & .1932 & .0000 & .0198 \\
gg-BFA (mf)            & .1208 & .0625 & .4002 & .0063 & .1935 & .0298 & .3181 \\
gg-BFA (2f)            & .1208 & .0626 & .4003 & .0009 & .1933 & .0034 & .0373 \\
gg-BFA (2f+s)          & .1208 & .0626 & .4002 & .0011 & .1933 & .0052 & .0610 \\
\bottomrule
\end{tabular}}
\end{table}

\begin{table}[!htbp]
\centering
\caption{Fixed-mask simulation: empirical coverage of nominal 95\% intervals,
with mean width in parentheses.  Results pool nine settings and 100 paired
replicates per setting.  SCFA reports confidence intervals; gg-BFA reports
variational credible intervals.}
\label{tab:exp1-coverage-current}
\small
\setlength{\tabcolsep}{3.5pt}
\begin{tabular}{@{}lrrrr@{}}
\toprule
Method & \(F\) & \(\psi\) & \(\diag(\Phi)\) & \(\operatorname{offdiag}(\Phi)\) \\
\midrule
Closed-form SCFA        & 95.15 (.410) & 94.74 (.046) & 92.81 (1.993) & 94.00 (1.488) \\
gg-BFA (1f)            & 94.99 (.405) & 91.57 (.032) & 92.52 (1.904) & 93.26 (1.428) \\
gg-BFA (mf)            & 94.28 (.398) & 93.76 (.178) & 92.59 (1.901) & 93.26 (1.426) \\
gg-BFA (2f)            & 94.33 (.399) & 93.85 (.179) & 92.37 (1.902) & 93.19 (1.426) \\
gg-BFA (2f+s)          & 94.33 (.399) & 93.84 (.179) & 92.52 (1.904) & 92.93 (1.426) \\
\bottomrule
\end{tabular}
\end{table}

Point recovery and interval calibration give a more nuanced picture.
Factor-score intervals remain close to their nominal coverage, whereas
residual-variance and factor-covariance intervals show some undercoverage
(Table~\ref{tab:exp1-coverage-current}).
Figure~\ref{fig:exp1-coverage-current} shows how calibration changes across
the nine settings. Because this simulation contains no active cross-loadings,
the high coverage of inactive loading indicators assesses false-positive
control only; uncertainty for genuine cross-loadings is examined in the next
simulation.

\FloatBarrier
\subsection{When variables carry cross-community signal}
\label{sec:experiment2-current-results}

The cross-loading simulation (Appendix~\ref{app:sim-designs}) isolates the loading support more
directly: it generates truths with multiple, single, and sparse single cross-loadings per row and
fits each of the three learned-loading members to every truth.  The fitted members share the same
oracle communities and variable-specific residual parameterization.  The
crossed design therefore asks what is gained by a simpler loading explanation
when it is compatible with the truth, and what is lost when it rules out real
overlap.

The primary comparison is recovery of the loading structure, measured by
loading RMSE. Covariance error shows whether structural misspecification
affects the observation-level fit, while TPR and FDR distinguish missed
cross-loadings from false discoveries.

The relative performance of the three loading models changes with the
generating structure (Table~\ref{tab:exp2-option-tradeoff}). The mf member performs best when
variables can have multiple cross-loadings, while restricting the loading
support loses signal, especially under the additional sparsity budget. When
each variable has at most one cross-loading, 2f recovers the loading structure
more accurately and reduces false discoveries without materially changing
covariance recovery. Under sparse single cross-loadings, 2f+s provides the
most parsimonious loading representation while retaining the covariance
accuracy of 2f. Loading restrictions are therefore useful when they reflect
the form of overlap present in the data, but become harmful when they exclude
genuine cross-community signal.

\begin{table}[!htbp]
\centering
\caption{Cross-loading simulation: point recovery, loading complexity, and support
selection.  \(d_F(\Omega)\) is relative Frobenius error; TPR and FDR use a
0.5 posterior-inclusion threshold.  Truths with multiple, single, and sparse single
cross-loadings per row are fitted by gg-BFA (mf), (2f), and (2f+s).}
\label{tab:exp2-option-tradeoff}
\scriptsize
\setlength{\tabcolsep}{3.3pt}
\resizebox{\textwidth}{!}{%
\begin{tabular}{@{}llrrrrrr@{}}
\toprule
Truth & Fit & RMSE\((F)\) & RMSE\((\Lambda)\) & \(d_F(\Omega)\) &
Expected cross/row & TPR & FDR \\
\midrule
Multiple & mf & .1935 & .0305 & .1248 & .3140 & .792 & .103 \\
Multiple & 2f & .1945 & .0423 & .1353 & .2626 & .733 & .070 \\
Multiple & 2f+s & .1996 & .1141 & .2201 & .0926 & .295 & .005 \\
Single & mf & .1948 & .0125 & .1191 & .1312 & .823 & .068 \\
Single & 2f & .1946 & .0106 & .1189 & .0958 & .811 & .017 \\
Single & 2f+s & .1962 & .0353 & .1379 & .0615 & .568 & .000 \\
Sparse single & mf & .1958 & .0103 & .1180 & .0957 & .809 & .070 \\
Sparse single & 2f & .1958 & .0085 & .1176 & .0592 & .782 & .028 \\
Sparse single & 2f+s & .1958 & .0081 & .1176 & .0546 & .786 & .010 \\
\bottomrule
\end{tabular}}
\end{table}

Point recovery and interval calibration again lead to different conclusions.
Factor-score intervals remain close to nominal coverage, but the mean-field
posterior substantially undercovers active cross-loadings
(Table~\ref{tab:exp2-matched-coverage}). Variational bagging improves coverage
for these loadings, although it does not fully restore nominal coverage and
does not uniformly improve residual-variance or factor-covariance intervals.
It therefore provides a partial correction to mean-field uncertainty rather
than a complete calibration solution.

\begin{table}[!htbp]
\centering
\caption{Cross-loading simulation: pooled empirical coverage (\%) of nominal 95\%
intervals when each truth is fitted by its matching member: gg-BFA (mf), (2f), and (2f+s) for
multiple, single, and sparse single cross-loadings.  Base denotes the member without bagging and
Bagged the member with bagging. ``Active'' refers to generating slab indicators equal to one;
under the truth with multiple cross-loadings, inactive entries come from the narrow Gaussian spike.  The \(\gamma\)
column is dominated by inactive coordinates because the active base rate is
low.}
\label{tab:exp2-matched-coverage}
\scriptsize
\setlength{\tabcolsep}{3.2pt}
\resizebox{\textwidth}{!}{%
\begin{tabular}{@{}llrrrrrrr@{}}
\toprule
Truth & Fit & \(F\) & \(\psi\) & \(\diag(\Phi)\) &
\(\operatorname{offdiag}(\Phi)\) & Primary \(\Lambda\) & Active cross \(\Lambda\) & \(\gamma\) \\
\midrule
Multiple cross-loadings & Base   & 94.05 & 92.18 & 90.25 & 89.50 & 91.43 & 75.46 & 98.08 \\
Multiple cross-loadings & Bagged & 94.43 & 91.55 & 90.50 & 90.33 & 94.20 & 87.41 & 99.13 \\
Single cross-loading & Base     & 94.87 & 92.30 & 93.50 & 90.83 & 97.75 & 76.85 & 99.49 \\
Single cross-loading & Bagged   & 95.16 & 91.50 & 90.25 & 90.83 & 99.03 & 88.54 & 99.80 \\
Sparse single cross-loading & Base   & 94.91 & 91.98 & 92.75 & 91.00 & 98.35 & 74.43 & 99.61 \\
Sparse single cross-loading & Bagged & 95.22 & 91.93 & 92.75 & 90.67 & 98.90 & 81.68 & 99.78 \\
\bottomrule
\end{tabular}}
\end{table}

The high coverage of the binary inclusion indicators is dominated by inactive
coordinates and should not be interpreted as accurate support recovery.
Bagging increases sensitivity to active cross-loadings, but it also increases
false discoveries and computation time. Improved marginal coverage therefore
comes with a less selective loading summary and a substantially larger
computational cost; complete support and timing results are given in
Appendix~\ref{app:sim-recovery}.

SCFA and gg-BFA (1f), which fix the loadings at the community mask, provide a separate
misspecification check: their
residual-variance coverage deteriorates as omitted overlap grows.  Because 1f
also ties residual variances within blocks, this comparison cannot attribute
the full difference to cross-loadings alone.  The full truth-by-fit comparison
among mf, 2f, and 2f+s is the cleaner evidence for the loading-support
trade-off.

\FloatBarrier
The stress test of Appendix~\ref{sec:experiment3-design} compares the gg-BFA members with
exploratory-factor and block-covariance estimators under truths with and without a generating
community structure.

\subsection{Further results on recovery, calibration, and computation}
\label{app:sim-recovery}

Pooled coverage is the number of intervals that contain the true value divided by the number of
intervals, over all evaluated coordinates of all replicates; pooled width averages over the same
intervals. In the fixed-community simulation, expected cross-loadings per row are
0.0298, 0.0034, and 0.0052 for mf, 2f, and 2f+s.
The three members cover primary loadings, inactive cross-loadings, and \(\gamma\) in at least
99.9\% of cases; in this truth without cross-loadings, this shows that inactive support is
excluded, not that intervals for genuine cross-loadings are calibrated.  Their observation-covariance error is 0.1933--0.1935,
compared with 0.1932 for 1f.

When each truth is fitted by its matching member, the replicate-level average coverage of
active cross-loadings is 76.31\%, 76.46\%, and 75.09\% for gg-BFA (mf), (2f), and (2f+s),
compared with 87.61\%, 88.65\%, and 81.38\% after bagging. These give the same conclusion as pooled coordinate
coverage. At the 0.5 inclusion threshold, bagging the matching member raises pooled
sensitivity from 0.792 to 0.838, 0.812 to 0.859, and 0.786 to 0.813.
The corresponding false-discovery rates increase from 0.103 to 0.279,
0.017 to 0.316, and 0.010 to 0.086. Factor-score coverage of the matching members is 94.05--94.91\% without bagging and
94.43--95.22\% with it, and that of SCFA and gg-BFA (1f) is 94.13--95.10\%.

Mean fit time per cross-loading data set is 0.0006 seconds for SCFA,
0.0046 for gg-BFA (1f), and 0.1407, 0.0212, and 0.0284 for gg-BFA (mf), (2f), and (2f+s); with
bagging over 20 bootstrap fits, the last three take 3.0936, 0.7712, and 0.7834 seconds. Figure~\ref{fig:exp1-coverage-current} below shows the
setting-level coverage variation for the fixed-community simulation; the complete cross-loading
truth-by-fit results appear in Table~\ref{tab:exp2-option-tradeoff}.

\begin{figure}[!htbp]
\centering
\includegraphics[width=\textwidth]{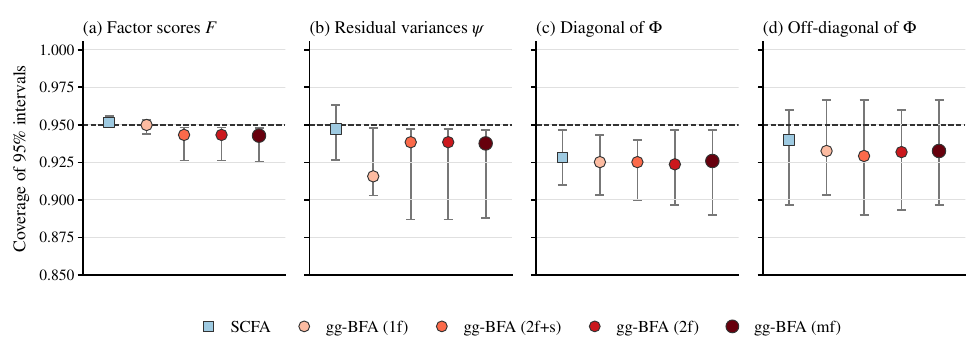}
\caption{Fixed-community simulation: empirical coverage of nominal 95\% intervals for (a) factor
scores, (b) residual variances, and the (c) diagonal and (d) off-diagonal entries of the factor
covariance.  Each point pools the nine settings, each vertical bar spans the nine setting-level
coverages, and the dashed line marks 95\%.  gg-BFA in red, mf darkest; SCFA reports confidence
intervals and gg-BFA variational credible intervals.}
\label{fig:exp1-coverage-current}
\end{figure}

\FloatBarrier
\subsection{Variational bagging as an uncertainty supplement}
\label{app:bagging}

For a fixed loading option \(o\) and primary partition, draw \(B=20\)
bootstrap samples \(X^{*(b)}\), each containing \(M=2n\) rows sampled with
replacement. Fit \(q_o^{(b)}\) to each sample. The reported distribution is
the equally weighted mixture of the \(B\) fits of the same member
\[
 q_{o,\mathrm{bag}}(\vartheta)=\frac1B\sum_{b=1}^Bq_o^{(b)}(\vartheta),
\]
where \(\vartheta\) denotes parameters expressed in a common coordinate
system. Subject-score distributions are re-evaluated on the original rows
under each fitted component. Joint draws choose one bootstrap component for
all requested parameters and retain its row support and block budget.

For component means \(\mu_b\) and covariances \(V_b\), the mixture covariance is
\[
 V_{\mathrm{mix}}=
 \frac1B\sum_b\{V_b+(\mu_b-\bar\mu)(\mu_b-\bar\mu)^\top\}.
\]
It includes within-component and between-component variation. At \(M=2n\),
the marginal-coverage experiments use this raw mixture without an additional
variance multiplier. Observation covariance is likewise
\(B^{-1}\sum_b\E_{q_o^{(b)}}(\Lambda\Phi\Lambda^\top+\Psi)\), not a product
of pooled parameter means.

The package also implements the block-wise adjustment of joint moments suggested by
variational-bagging theory \citep{fan2025variationalbagging}; it was not used in the reported
results, and choosing \(M=2n\) does not by itself guarantee nominal coverage.

\subsection{Stress test without a generating community structure}
\label{sec:experiment3-design}

The stress test asks where the advantage of community guidance ends. It
contrasts the three block-oriented truths defined above with a neutral
sparse-factor truth that has no generating primary partition. All fitted
factor models use \(K=8\). Community-oriented methods receive the same oracle
primary labels for comparison, whereas EFA and FactorVerse do not receive
those labels.

The primary comparison is off-diagonal correlation RMSE. Covariance
\(d_F\) and test NLL are supporting summaries of covariance scale and
out-of-sample density.

The three endpoints reveal different parts of the boundary.  Under the three
block-oriented truths, 1f has the lowest off-diagonal correlation RMSE:
0.0554, 0.0655, and 0.0770, compared with 0.1091, 0.1084, and 0.1100 for the
best unrestricted alternative, EFA (Table~\ref{tab:exp3-primary}).  The
covariance-scale results are similar but show a role for controlled loading
flexibility.  The lowest relative Frobenius errors among the gg-BFA members are
0.2893 for 1f under fixed loadings and 0.4373 and 0.4604 for 2f under the two
continuous-loading truths.  EFA gives 0.4740, 0.4781, and 0.4815 in the same
three settings.

Test likelihood is less aligned with direct covariance recovery.  Closed-form
SCFA has the lowest Gaussian NLL under the fixed truth, 167.51, but FactorVerse
DL has the lowest NLL under the continuous and cross-loading truths, 170.38 and
170.96, even though its Frobenius errors are larger.  Under the neutral truth,
the result is unambiguous: DL is best on correlation RMSE (0.0403),
covariance error (0.2761), and test NLL (170.35).  The strongest gg-BFA member there is mf, with 0.0895, 0.5400, and 178.81.  Assigning a primary label
after generation cannot recreate a community backbone that is absent from the
truth.

\begin{table}[!htbp]
\centering
\caption{Covariance recovery beyond the loading family: off-diagonal correlation
RMSE, mean (Monte Carlo standard error) over 30 paired seeds per truth. Fixed,
continuous, and cross-loading are block-oriented truths; neutral has no
generating community partition. Lower is better.}
\label{tab:exp3-primary}
\footnotesize
\setlength{\tabcolsep}{3pt}
\resizebox{\textwidth}{!}{
\begin{tabular}{@{}lrrrr@{}}
\toprule
Method & Fixed & Continuous & Cross-loading & Neutral \\
\midrule
Sample covariance & 0.1245 (0.0010) & 0.1254 (0.0009) & 0.1263 (0.0011) & 0.1266 (0.0007) \\
SCFA & 0.0559 (0.0017) & 0.0659 (0.0013) & 0.0775 (0.0017) & 0.1539 (0.0008) \\
gg-BFA (1f) & 0.0554 (0.0016) & 0.0655 (0.0013) & 0.0770 (0.0017) & 0.1545 (0.0008) \\
\GBmulti & 0.0568 (0.0016) & 0.0757 (0.0012) & 0.0836 (0.0015) & 0.0895 (0.0013) \\
\GBtwo & 0.0565 (0.0016) & 0.0734 (0.0013) & 0.0825 (0.0016) & 0.1082 (0.0013) \\
\GBsparse & 0.0565 (0.0016) & 0.0699 (0.0012) & 0.0796 (0.0017) & 0.1528 (0.0009) \\
EFA & 0.1091 (0.0011) & 0.1084 (0.0011) & 0.1100 (0.0012) & 0.1088 (0.0009) \\
BCME (fixed $K$) & 0.1317 (0.0017) & 0.1318 (0.0014) & 0.1343 (0.0015) & 0.1417 (0.0010) \\
BCME (ridge-ratio $K$) & 0.1629 (0.0003) & 0.1563 (0.0008) & 0.1606 (0.0008) & 0.1508 (0.0013) \\
FactorVerse MGPS & 0.1280 (0.0008) & 0.1242 (0.0011) & 0.1232 (0.0013) & 0.0774 (0.0007) \\
FactorVerse SSL & 0.2002 (0.0007) & 0.2018 (0.0008) & 0.2008 (0.0008) & 0.1749 (0.0007) \\
FactorVerse DL & 0.1342 (0.0002) & 0.1282 (0.0008) & 0.1285 (0.0009) & 0.0403 (0.0008) \\
FactorVerse HS & 0.1610 (0.0006) & 0.1596 (0.0009) & 0.1575 (0.0010) & 0.1170 (0.0006) \\
FactorVerse BASS & 0.1321 (0.0003) & 0.1265 (0.0009) & 0.1261 (0.0010) & 0.0525 (0.0007) \\
\bottomrule
\end{tabular}
}
\end{table}

Appendix~\ref{app:exp3-secondary} gives the complete covariance and likelihood tables for all
methods, including the large errors of FactorVerse SSL and HS on the generating scale and the
extreme neutral-truth likelihood of ridge-ratio BCME.  The MNL prior of FactorVerse is not
included: in two of the four truths, the mean covariances from the two halves of its chains
differed by more than 10\% in relative Frobenius norm, the stability threshold fixed before the
comparison.

\subsection{Stress-test covariance and likelihood endpoints}
\label{app:exp3-secondary}

Tables~\ref{tab:exp3-covariance} and~\ref{tab:exp3-nll} report every method.  The covariance
errors of SSL and HS are on the generating scale, with no correction that uses the truth.  The
automatic rule of BCME returns one block in every replicate of the block-oriented truths and can
select a large \(K\) under the neutral truth, where its test likelihood is unstable.

\begin{table}[!htbp]
\centering
\caption{Supplementary stress-test relative Frobenius covariance error, mean (Monte Carlo
standard error), with 30 paired seeds per truth. Lower is better.}
\label{tab:exp3-covariance}
\footnotesize
\setlength{\tabcolsep}{3pt}
\resizebox{\textwidth}{!}{
\begin{tabular}{@{}lrrrr@{}}
\toprule
Method & Fixed & Continuous & Cross-loading & Neutral \\
\midrule
Sample covariance & 0.5182 (0.0065) & 0.5164 (0.0088) & 0.5173 (0.0088) & 0.6343 (0.0075) \\
SCFA & 0.2955 (0.0096) & 0.4656 (0.0079) & 0.4950 (0.0089) & 0.8944 (0.0011) \\
gg-BFA (1f) & 0.2893 (0.0087) & 0.4610 (0.0067) & 0.4887 (0.0082) & 0.8979 (0.0011) \\
\GBmulti & 0.2934 (0.0087) & 0.4420 (0.0081) & 0.4619 (0.0082) & 0.5400 (0.0082) \\
\GBtwo & 0.2916 (0.0087) & 0.4373 (0.0085) & 0.4604 (0.0078) & 0.6590 (0.0074) \\
\GBsparse & 0.2916 (0.0087) & 0.4552 (0.0075) & 0.4794 (0.0080) & 0.9123 (0.0036) \\
EFA & 0.4740 (0.0070) & 0.4781 (0.0092) & 0.4815 (0.0095) & 0.5770 (0.0080) \\
BCME (fixed $K$) & 0.5446 (0.0052) & 0.5608 (0.0066) & 0.5649 (0.0071) & 0.8075 (0.0033) \\
BCME (ridge-ratio $K$) & 0.6852 (0.0035) & 0.6923 (0.0048) & 0.6985 (0.0054) & 0.8403 (0.0041) \\
FactorVerse MGPS & 0.6087 (0.0053) & 0.6274 (0.0084) & 0.6161 (0.0080) & 0.4927 (0.0101) \\
FactorVerse SSL & 130.9447 (0.6905) & 130.9171 (1.5323) & 125.0482 (1.0296) & 145.5509 (1.5162) \\
FactorVerse DL & 0.6105 (0.0020) & 0.6123 (0.0023) & 0.6117 (0.0025) & 0.2761 (0.0065) \\
FactorVerse HS & 3.1545 (0.0443) & 3.2983 (0.0761) & 3.1242 (0.0473) & 3.0117 (0.0422) \\
FactorVerse BASS & 0.6009 (0.0021) & 0.6027 (0.0023) & 0.6007 (0.0033) & 0.3323 (0.0065) \\
\bottomrule
\end{tabular}
}
\end{table}

\begin{table}[!htbp]
\centering
\caption{Supplementary stress-test Gaussian negative log likelihood per row, mean
(Monte Carlo standard error). Lower is better. The raw sample covariance
is singular at \(p_0>n\), so it has no prediction score; it is not regularized.}
\label{tab:exp3-nll}
\footnotesize
\setlength{\tabcolsep}{3pt}
\resizebox{\textwidth}{!}{
\begin{tabular}{@{}lrrrr@{}}
\toprule
Method & Fixed & Continuous & Cross-loading & Neutral \\
\midrule
Sample covariance & N/A & N/A & N/A & N/A \\
SCFA & 167.5082 (0.0473) & 173.9101 (0.1695) & 175.3784 (0.1696) & 202.3473 (0.3296) \\
gg-BFA (1f) & 167.5357 (0.0482) & 173.9371 (0.1694) & 175.4067 (0.1699) & 202.9832 (0.3380) \\
\GBmulti & 168.5024 (0.0712) & 173.1189 (0.1213) & 174.2469 (0.1232) & 178.8098 (0.3211) \\
\GBtwo & 168.3449 (0.0631) & 173.1482 (0.1267) & 174.3552 (0.1156) & 184.7036 (0.3946) \\
\GBsparse & 168.3443 (0.0631) & 173.7941 (0.1436) & 175.1479 (0.1484) & 198.8034 (0.3536) \\
EFA & 180.4941 (0.2071) & 181.3226 (0.2745) & 181.7056 (0.1927) & 181.0287 (0.1948) \\
BCME (fixed $K$) & 181.0983 (0.4108) & 183.9825 (0.3731) & 184.7231 (0.3892) & 194.4354 (0.3619) \\
BCME (ridge-ratio $K$) & 197.2595 (0.1020) & 197.2701 (0.3457) & 198.4086 (0.2919) & 31657.0258 (22266.6278) \\
FactorVerse MGPS & 172.0354 (0.0868) & 172.5307 (0.1230) & 172.9286 (0.0886) & 173.7421 (0.1246) \\
FactorVerse SSL & 327.3033 (0.3743) & 327.0332 (0.6010) & 324.6596 (0.6291) & 315.0310 (0.7182) \\
FactorVerse DL & 169.7801 (0.0697) & 170.3766 (0.0853) & 170.9617 (0.0858) & 170.3490 (0.0962) \\
FactorVerse HS & 180.4263 (0.0651) & 180.9837 (0.0752) & 180.7558 (0.0950) & 178.6783 (0.0726) \\
FactorVerse BASS & 170.1316 (0.0762) & 170.7284 (0.0949) & 171.3173 (0.0830) & 171.3668 (0.1070) \\
\bottomrule
\end{tabular}
}
\end{table}

\FloatBarrier
\clearpage
\section{Supplementary real-data results}
\label{app:real-data}

This appendix supplements Section~\ref{sec:applications}.  It describes the implementations of the
comparison methods and gives the complete results at the common factor budget of
Table~\ref{tab:realdata-design-current}.

\subsection{Implementations of the comparison methods}
\label{app:cfa-gpu-implementation}

The four CFA references are implemented in CuPy in double precision.  SCFA evaluates the
closed-form block moments.  ML-CFA maximizes the joint Gaussian likelihood by EM with
factor-correlation rescaling; we checked it against \texttt{lavaan}.  MFVB-CFA follows
Algorithm~2 of \citet{dang2022sem}; it is our translation, not software distributed by the
authors.  PCFA-VA implements the complete-data, oblique, diagonal-residual branch of
\texttt{vbpm} 0.9.1, including its deterministic initialization, internal standardization by the
sample standard deviation, sequential updates of the factor coordinates within each loading row,
four-stage path, and stopping rule \citep{jin2025pcfa,chen2026vbpm,jin2026sparse}.  We checked
each implementation against a CPU reference on one- and three-factor examples (fitted parameters,
covariance and stopping state), and PCFA-VA also against the R package, including its final ELBO.

ML-CFA uses relative log-likelihood tolerance \(10^{-7}\), parameter tolerance \(10^{-4}\), and
at most 3,000 iterations; MFVB-CFA uses tolerance 0.01 and at most 1,000 iterations; PCFA-VA uses
\(v_0=(0.01,0.005,0.002,0.001)\), tolerance \(10^{-4}\), and at most 5,000 iterations per stage.
Times are measured with GPU synchronization; the times of the variational methods include the
construction of the covariance estimate, and those of every method include its own data transfer
and posterior summaries.  On the real data, the GPU and CPU implementations agree for SCFA,
ML-CFA, MFVB-CFA and the converged PCFA-VA fits on Kidney and miRNA.  PCFA-VA does not meet its
tolerance on Nightingale in either implementation, and its final covariance there depends on the
numerical path, as do its nonconverged full-data fits on the imaging panels.

The held-out comparison also includes three estimators that do not produce a confirmatory loading
pattern; they are fitted on Nightingale, Kidney, and miRNA.  Maximum-likelihood EFA and the
FactorVerse Bayesian factor models \citep{dunson2026tutorial} use the same number of factors but
not the community partition, and BCME \citep{lei2026bcme} estimates a blockwise correlation matrix
with the same number of blocks.  Because the latent coordinates of EFA and FactorVerse depend on
rotation, we compare their observation covariances and Gaussian predictions rather than their
loadings.  FactorVerse uses the MGPS, SSL, DL, HS, and BASS priors; the MNL prior is not used,
because its chains failed the stability check of the stress test
(Appendix~\ref{sec:experiment3-design}).  All FactorVerse fits use our CuPy implementation of the
FactorVerse update equations, whose formulas and draws were checked against the native R/Rcpp
code; the results apply to this implementation, and each comes from a single chain.

\subsection{Complete results at the common factor budget}
\label{app:current-real-data}

The tables of this section give the results of Section~\ref{sec:applications} for all five
panels.  Table~\ref{tab:cfa-r1} gives the full-data fits on the variables assigned to
communities.  The held-out tables have two targets: the variables assigned to communities on the
training subjects, as in the main text (Tables~\ref{tab:r2-fixed-selected}
and~\ref{tab:r3-fixed-selected}), and all retained variables (Tables~\ref{tab:r2-fixed-full}
and~\ref{tab:r3-fixed-full}); values for the two targets should not be compared.  Each held-out
entry gives the mean and sample standard deviation over the same five subject folds.  Covariances
are on the scale of the training-standardized data, all predictors use zero mean in those
coordinates, and the covariances of the Bayesian CFA methods include the variance of the loadings.
\(\dagger\) marks at least one nonconverged fold, with the count in
Table~\ref{tab:runtime-status}; these values are not used to rank the methods.  \(\ddagger\) marks
the indefinite factor moment of SCFA on Nightingale, whose observation covariance is nevertheless
positive definite.  PCFA-VA was not run on the held-out folds of ABCD and UKB because of its
computational cost.  Table~\ref{tab:r23-factorverse-diagnostics} gives the single-chain
diagnostics of FactorVerse, and Table~\ref{tab:r3-local-variational} gives the predictions of the
gg-BFA members that integrate the local variational factor posterior of each test subject.

The comparison methods show that correlation, covariance scale, and joint density cannot be
inferred from one another.  With the SSL prior, FactorVerse has held-out covariance errors above
100 in Kidney and miRNA (\(d_F=131.298\) and 118.474), yet the smallest prediction errors there
(RMSE 0.461 and 0.500, against 0.497 and 0.528 for gg-BFA (mf)), because conditional means depend
only on ratios of covariances.  Its coverage of 0.998 and 1.000 comes with intervals two to three
times as wide as those of mf (width 4.005 and 6.020, against 1.880 and 1.993).  The Nightingale
EFA covariance is nearly singular: its mean NLL is about 2,600 per variable although its
covariance error is close to that of mf, and four of its five folds miss the stopping tolerance.

{\scriptsize\setlength{\tabcolsep}{3pt}
\begin{longtable}{@{}llrrr@{}}
\caption{Full-data CFA representations on the same variables assigned to communities. $d_F$ is descriptive distance to the sample covariance, not population recovery. Cross/row is the expected number of active nonprimary loading components. Seconds include estimation and covariance construction on P100. $\dagger$ marks a last iterate.}\label{tab:cfa-r1}\\
\toprule
Panel & Method & $d_F$ & Cross/row & Seconds \\
\midrule
\endfirsthead
\multicolumn{5}{c}{\tablename\ \thetable\ continued}\\
\toprule
Panel & Method & $d_F$ & Cross/row & Seconds \\
\midrule
\endhead
\midrule
\multicolumn{5}{r}{Continued on next page}\\
\endfoot
\bottomrule
\endlastfoot
Nightingale & SCFA & 0.368 & 0.000 & 0.171 \\
Nightingale & ML-CFA & 0.401 & 0.000 & 6.399 \\
Nightingale & MFVB-CFA$^{\dagger}$ & 0.872 & 0.000 & 4.708 \\
Nightingale & PCFA-VA$^{\dagger}$ & 3.390 & 6.997 & 124.863 \\
Nightingale & gg-BFA (1f) & 0.368 & 0.000 & 6.297 \\
Nightingale & gg-BFA (2f+s) & 0.357 & 0.081 & 3.037 \\
Nightingale & gg-BFA (2f) & 0.258 & 1.000 & 2.904 \\
Nightingale & gg-BFA (mf) & 0.021 & 5.680 & 6.751 \\
Kidney & SCFA & 0.366 & 0.000 & 0.196 \\
Kidney & ML-CFA & 0.362 & 0.000 & 0.601 \\
Kidney & MFVB-CFA$^{\dagger}$ & 0.674 & 0.000 & 5.032 \\
Kidney & PCFA-VA & 0.066 & 2.519 & 20.742 \\
Kidney & gg-BFA (1f) & 0.366 & 0.000 & 1.018 \\
Kidney & gg-BFA (2f+s) & 0.346 & 0.098 & 2.467 \\
Kidney & gg-BFA (2f) & 0.258 & 1.000 & 1.635 \\
Kidney & gg-BFA (mf) & 0.087 & 3.065 & 2.662 \\
miRNA & SCFA & 0.425 & 0.000 & 4.971 \\
miRNA & ML-CFA & 0.408 & 0.000 & 1.478 \\
miRNA & MFVB-CFA$^{\dagger}$ & 0.505 & 0.000 & 5.234 \\
miRNA & PCFA-VA & 0.060 & 3.078 & 28.170 \\
miRNA & gg-BFA (1f) & 0.425 & 0.000 & 0.922 \\
miRNA & gg-BFA (2f+s) & 0.397 & 0.097 & 1.686 \\
miRNA & gg-BFA (2f) & 0.260 & 0.999 & 1.167 \\
miRNA & gg-BFA (mf) & 0.081 & 4.108 & 2.575 \\
ABCD & SCFA & 0.534 & 0.000 & 0.410 \\
ABCD & ML-CFA & 0.513 & 0.000 & 2.728 \\
ABCD & MFVB-CFA$^{\dagger}$ & 2.261 & 0.000 & 60.969 \\
ABCD & PCFA-VA$^{\dagger}$ & 0.204 & 5.453 & 390.944 \\
ABCD & gg-BFA (1f) & 0.534 & 0.000 & 8.026 \\
ABCD & gg-BFA (2f+s) & 0.512 & 0.098 & 16.613 \\
ABCD & gg-BFA (2f) & 0.337 & 1.000 & 40.953 \\
ABCD & gg-BFA (mf) & 0.094 & 13.106 & 137.750 \\
UKB & SCFA & 0.411 & 0.000 & 0.799 \\
UKB & ML-CFA & 0.355 & 0.000 & 2.048 \\
UKB & MFVB-CFA$^{\dagger}$ & 1.883 & 0.000 & 139.975 \\
UKB & PCFA-VA$^{\dagger}$ & 0.146 & 6.999 & 630.452 \\
UKB & gg-BFA (1f) & 0.411 & 0.000 & 16.440 \\
UKB & gg-BFA (2f+s) & 0.413 & 0.097 & 14.297 \\
UKB & gg-BFA (2f) & 0.284 & 1.000 & 116.809 \\
UKB & gg-BFA (mf) & 0.163 & 5.648 & 25.111 \\
\end{longtable}

\begin{longtable}{@{}llrrr@{}}
\caption{Selected variables; common factor budget. Held-out covariance: Gaussian NLL per variable, relative covariance error, and correlation SRMR. Five-fold means (sample SD). $\dagger$ denotes at least one fit that did not meet its tolerance; $\ddagger$ denotes the SCFA latent-factor covariance caveat.}\label{tab:r2-fixed-selected}\\
\toprule
Panel & Method & NLL/variable & $d_F$ & SRMR \\
\midrule
\endfirsthead
\multicolumn{5}{c}{\tablename\ \thetable\ continued}\\
\toprule
Panel & Method & NLL/variable & $d_F$ & SRMR \\
\midrule
\endhead
\midrule
\multicolumn{5}{r}{Continued on next page}\\
\endfoot
\bottomrule
\endlastfoot
Nightingale & SCFA$^{\ddagger}$ & 0.569 (0.047) & 0.396 (0.022) & 0.205 (0.010) \\
Nightingale & ML-CFA & 0.380 (0.046) & 0.416 (0.029) & 0.218 (0.013) \\
Nightingale & MFVB-CFA$^{\dagger}$ & 0.385 (0.045) & 0.901 (0.228) & 0.261 (0.029) \\
Nightingale & PCFA-VA$^{\dagger}$ & -0.531 (0.221) & 1.765 (0.739) & 0.203 (0.023) \\
Nightingale & gg-BFA (1f) & 0.570 (0.047) & 0.396 (0.022) & 0.205 (0.010) \\
Nightingale & gg-BFA (2f+s) & 0.351 (0.043) & 0.394 (0.034) & 0.192 (0.007) \\
Nightingale & gg-BFA (2f) & -0.009 (0.042) & 0.300 (0.013) & 0.125 (0.007) \\
Nightingale & gg-BFA (mf) & -0.864 (0.056) & 0.187 (0.029) & 0.058 (0.008) \\
Nightingale & EFA$^{\dagger}$ & 2.60e+03 (5.82e+03) & 0.186 (0.028) & 0.058 (0.008) \\
Nightingale & FV-MGPS & -0.980 (0.052) & 3.309 (5.122) & 0.284 (0.196) \\
Nightingale & FV-SSL & -0.986 (0.050) & 0.508 (0.069) & 0.154 (0.021) \\
Nightingale & FV-DL & -0.977 (0.049) & 5.568 (1.923) & 0.499 (0.051) \\
Nightingale & FV-HS & -0.985 (0.050) & 0.569 (0.049) & 0.144 (0.016) \\
Nightingale & FV-BASS & -0.944 (0.051) & 34.660 (7.538) & 0.119 (0.031) \\
Nightingale & BCME & 0.885 (0.034) & 0.701 (0.022) & 0.389 (0.009) \\
Kidney & SCFA & 0.886 (0.054) & 0.436 (0.027) & 0.190 (0.008) \\
Kidney & ML-CFA & 0.845 (0.056) & 0.424 (0.017) & 0.185 (0.006) \\
Kidney & MFVB-CFA$^{\dagger}$ & 0.845 (0.056) & 0.572 (0.035) & 0.216 (0.010) \\
Kidney & PCFA-VA & 0.660 (0.039) & 0.243 (0.035) & 0.091 (0.012) \\
Kidney & gg-BFA (1f) & 0.886 (0.054) & 0.436 (0.027) & 0.190 (0.008) \\
Kidney & gg-BFA (2f+s) & 0.835 (0.061) & 0.423 (0.033) & 0.173 (0.008) \\
Kidney & gg-BFA (2f) & 0.765 (0.048) & 0.354 (0.032) & 0.141 (0.009) \\
Kidney & gg-BFA (mf) & 0.676 (0.045) & 0.249 (0.035) & 0.095 (0.011) \\
Kidney & EFA & 0.654 (0.045) & 0.240 (0.032) & 0.089 (0.010) \\
Kidney & FV-MGPS & 0.653 (0.043) & 0.510 (0.051) & 0.198 (0.030) \\
Kidney & FV-SSL & 1.044 (0.014) & 131.298 (10.161) & 0.305 (0.031) \\
Kidney & FV-DL & 0.655 (0.043) & 0.624 (0.023) & 0.233 (0.018) \\
Kidney & FV-HS & 0.653 (0.043) & 0.612 (0.021) & 0.240 (0.019) \\
Kidney & FV-BASS & 0.683 (0.029) & 39.648 (8.550) & 0.290 (0.016) \\
Kidney & BCME & 0.897 (0.029) & 0.481 (0.015) & 0.212 (0.005) \\
miRNA & SCFA & 0.988 (0.041) & 0.497 (0.089) & 0.204 (0.022) \\
miRNA & ML-CFA & 0.929 (0.046) & 0.483 (0.073) & 0.198 (0.016) \\
miRNA & MFVB-CFA$^{\dagger}$ & 0.929 (0.046) & 0.534 (0.121) & 0.201 (0.027) \\
miRNA & PCFA-VA & 0.718 (0.047) & 0.287 (0.130) & 0.102 (0.038) \\
miRNA & gg-BFA (1f) & 0.988 (0.041) & 0.497 (0.089) & 0.204 (0.022) \\
miRNA & gg-BFA (2f+s) & 0.912 (0.046) & 0.479 (0.099) & 0.181 (0.023) \\
miRNA & gg-BFA (2f) & 0.825 (0.048) & 0.366 (0.114) & 0.132 (0.032) \\
miRNA & gg-BFA (mf) & 0.733 (0.048) & 0.292 (0.129) & 0.105 (0.037) \\
miRNA & EFA & 0.718 (0.047) & 0.286 (0.131) & 0.102 (0.039) \\
miRNA & FV-MGPS & 0.716 (0.045) & 0.406 (0.054) & 0.137 (0.009) \\
miRNA & FV-SSL & 1.405 (0.010) & 118.474 (12.189) & 0.278 (0.015) \\
miRNA & FV-DL & 0.719 (0.045) & 0.632 (0.040) & 0.227 (0.030) \\
miRNA & FV-HS & 0.717 (0.045) & 0.584 (0.021) & 0.202 (0.018) \\
miRNA & FV-BASS & 0.720 (0.042) & 18.964 (9.030) & 0.367 (0.073) \\
miRNA & BCME & 1.040 (0.032) & 0.673 (0.033) & 0.286 (0.006) \\
ABCD & SCFA & 1.107 (0.008) & 0.549 (0.005) & 0.143 (0.002) \\
ABCD & ML-CFA & 1.067 (0.008) & 0.529 (0.008) & 0.137 (0.002) \\
ABCD & MFVB-CFA$^{\dagger}$ & 1.067 (0.008) & 2.208 (0.117) & 0.368 (0.003) \\
ABCD & gg-BFA (1f) & 1.107 (0.008) & 0.549 (0.005) & 0.143 (0.002) \\
ABCD & gg-BFA (2f+s) & 1.051 (0.008) & 0.527 (0.013) & 0.129 (0.002) \\
ABCD & gg-BFA (2f) & 0.980 (0.007) & 0.370 (0.013) & 0.088 (0.001) \\
ABCD & gg-BFA (mf) & 0.796 (0.007) & 0.184 (0.014) & 0.043 (0.002) \\
UKB & SCFA & 1.155 (0.006) & 0.421 (0.014) & 0.112 (0.003) \\
UKB & ML-CFA & 1.115 (0.008) & 0.367 (0.018) & 0.098 (0.004) \\
UKB & MFVB-CFA$^{\dagger}$ & 1.116 (0.008) & 1.803 (0.186) & 0.225 (0.016) \\
UKB & gg-BFA (1f) & 1.155 (0.006) & 0.421 (0.014) & 0.112 (0.003) \\
UKB & gg-BFA (2f+s) & 1.119 (0.007) & 0.421 (0.010) & 0.100 (0.005) \\
UKB & gg-BFA (2f) & 1.082 (0.008) & 0.289 (0.005) & 0.072 (5.56e-04) \\
UKB & gg-BFA (mf) & 1.052 (0.004) & 0.170 (0.004) & 0.045 (5.39e-04) \\
\end{longtable}

\begin{longtable}{@{}llrrrr@{}}
\caption{Selected variables; common factor budget. Masked-measurement prediction uses the common Gaussian rule; coverage is a proportion, and width is in training-standardized units. Five-fold means (sample SD). $\dagger$ denotes at least one fit that did not meet its tolerance; $\ddagger$ denotes the SCFA latent-factor covariance caveat.}\label{tab:r3-fixed-selected}\\
\toprule
Panel & Method & RMSE & NLPD & Coverage & Width \\
\midrule
\endfirsthead
\multicolumn{6}{c}{\tablename\ \thetable\ continued}\\
\toprule
Panel & Method & RMSE & NLPD & Coverage & Width \\
\midrule
\endhead
\midrule
\multicolumn{6}{r}{Continued on next page}\\
\endfoot
\bottomrule
\endlastfoot
Nightingale & SCFA$^{\ddagger}$ & 0.413 (0.024) & 0.532 (0.062) & 0.941 (0.012) & 1.588 (0.044) \\
Nightingale & ML-CFA & 0.434 (0.031) & 0.312 (0.127) & 0.948 (0.006) & 1.519 (0.119) \\
Nightingale & MFVB-CFA$^{\dagger}$ & 0.433 (0.030) & 0.316 (0.121) & 0.950 (0.006) & 1.527 (0.114) \\
Nightingale & PCFA-VA$^{\dagger}$ & 0.177 (0.017) & -0.710 (0.295) & 0.956 (0.016) & 0.714 (0.419) \\
Nightingale & gg-BFA (1f) & 0.414 (0.024) & 0.534 (0.062) & 0.940 (0.012) & 1.593 (0.045) \\
Nightingale & gg-BFA (2f+s) & 0.405 (0.029) & 0.280 (0.109) & 0.948 (0.006) & 1.437 (0.122) \\
Nightingale & gg-BFA (2f) & 0.292 (0.032) & -0.110 (0.087) & 0.949 (0.006) & 1.015 (0.101) \\
Nightingale & gg-BFA (mf) & 0.172 (0.016) & -1.020 (0.160) & 0.949 (0.003) & 0.504 (0.062) \\
Nightingale & EFA$^{\dagger}$ & 0.154 (0.021) & -1.199 (0.181) & 0.945 (0.007) & 0.437 (0.070) \\
Nightingale & FV-MGPS & 0.151 (0.022) & -1.153 (0.183) & 0.963 (0.005) & 0.451 (0.072) \\
Nightingale & FV-SSL & 0.151 (0.022) & -1.154 (0.183) & 0.963 (0.005) & 0.450 (0.072) \\
Nightingale & FV-DL & 0.151 (0.022) & -1.153 (0.183) & 0.963 (0.005) & 0.452 (0.072) \\
Nightingale & FV-HS & 0.151 (0.022) & -1.154 (0.183) & 0.963 (0.005) & 0.450 (0.072) \\
Nightingale & FV-BASS & 0.151 (0.022) & -1.142 (0.181) & 0.969 (0.004) & 0.475 (0.076) \\
Nightingale & BCME & 0.553 (0.066) & 0.841 (0.078) & 0.966 (0.016) & 2.611 (0.084) \\
Kidney & SCFA & 0.585 (0.033) & 0.878 (0.054) & 0.944 (0.005) & 2.275 (0.115) \\
Kidney & ML-CFA & 0.583 (0.033) & 0.837 (0.057) & 0.945 (0.006) & 2.225 (0.118) \\
Kidney & MFVB-CFA$^{\dagger}$ & 0.583 (0.033) & 0.837 (0.056) & 0.946 (0.006) & 2.235 (0.118) \\
Kidney & PCFA-VA & 0.485 (0.020) & 0.652 (0.043) & 0.944 (0.006) & 1.812 (0.025) \\
Kidney & gg-BFA (1f) & 0.585 (0.033) & 0.878 (0.054) & 0.944 (0.005) & 2.273 (0.115) \\
Kidney & gg-BFA (2f+s) & 0.574 (0.035) & 0.827 (0.063) & 0.945 (0.006) & 2.200 (0.121) \\
Kidney & gg-BFA (2f) & 0.540 (0.032) & 0.758 (0.052) & 0.945 (0.006) & 2.056 (0.084) \\
Kidney & gg-BFA (mf) & 0.497 (0.026) & 0.672 (0.051) & 0.945 (0.006) & 1.880 (0.062) \\
Kidney & EFA & 0.479 (0.022) & 0.645 (0.047) & 0.945 (0.006) & 1.806 (0.015) \\
Kidney & FV-MGPS & 0.479 (0.021) & 0.644 (0.045) & 0.948 (0.006) & 1.834 (0.016) \\
Kidney & FV-SSL & 0.461 (0.020) & 1.016 (0.014) & 0.998 (0.001) & 4.005 (0.085) \\
Kidney & FV-DL & 0.479 (0.022) & 0.645 (0.046) & 0.947 (0.006) & 1.836 (0.017) \\
Kidney & FV-HS & 0.478 (0.021) & 0.643 (0.045) & 0.948 (0.006) & 1.837 (0.016) \\
Kidney & FV-BASS & 0.485 (0.020) & 0.668 (0.033) & 0.965 (0.007) & 2.104 (0.078) \\
Kidney & BCME & 0.581 (0.020) & 0.884 (0.027) & 0.962 (0.004) & 2.554 (0.028) \\
miRNA & SCFA & 0.641 (0.026) & 0.972 (0.043) & 0.942 (0.009) & 2.502 (0.025) \\
miRNA & ML-CFA & 0.637 (0.027) & 0.913 (0.052) & 0.944 (0.010) & 2.422 (0.060) \\
miRNA & MFVB-CFA$^{\dagger}$ & 0.637 (0.027) & 0.913 (0.051) & 0.945 (0.010) & 2.433 (0.060) \\
miRNA & PCFA-VA & 0.519 (0.020) & 0.697 (0.050) & 0.943 (0.009) & 1.956 (0.045) \\
miRNA & gg-BFA (1f) & 0.641 (0.026) & 0.973 (0.043) & 0.942 (0.009) & 2.500 (0.025) \\
miRNA & gg-BFA (2f+s) & 0.628 (0.031) & 0.896 (0.058) & 0.944 (0.010) & 2.387 (0.055) \\
miRNA & gg-BFA (2f) & 0.576 (0.024) & 0.807 (0.052) & 0.944 (0.011) & 2.192 (0.052) \\
miRNA & gg-BFA (mf) & 0.528 (0.021) & 0.712 (0.052) & 0.943 (0.010) & 1.993 (0.046) \\
miRNA & EFA & 0.518 (0.020) & 0.696 (0.051) & 0.940 (0.010) & 1.928 (0.045) \\
miRNA & FV-MGPS & 0.517 (0.020) & 0.694 (0.049) & 0.944 (0.009) & 1.964 (0.045) \\
miRNA & FV-SSL & 0.500 (0.018) & 1.364 (0.020) & 1.000 (1.45e-04) & 6.020 (0.122) \\
miRNA & FV-DL & 0.518 (0.020) & 0.695 (0.050) & 0.945 (0.010) & 1.976 (0.045) \\
miRNA & FV-HS & 0.517 (0.020) & 0.693 (0.049) & 0.945 (0.010) & 1.971 (0.045) \\
miRNA & FV-BASS & 0.517 (0.020) & 0.695 (0.046) & 0.951 (0.008) & 2.053 (0.056) \\
miRNA & BCME & 0.677 (0.029) & 1.028 (0.035) & 0.961 (0.006) & 2.959 (0.024) \\
ABCD & SCFA & 0.731 (0.010) & 1.099 (0.013) & 0.949 (0.002) & 2.854 (0.008) \\
ABCD & ML-CFA & 0.726 (0.009) & 1.056 (0.013) & 0.951 (0.002) & 2.796 (0.013) \\
ABCD & MFVB-CFA$^{\dagger}$ & 0.726 (0.009) & 1.057 (0.013) & 0.951 (0.002) & 2.798 (0.013) \\
ABCD & gg-BFA (1f) & 0.731 (0.010) & 1.099 (0.013) & 0.949 (0.002) & 2.854 (0.008) \\
ABCD & gg-BFA (2f+s) & 0.720 (0.008) & 1.041 (0.011) & 0.951 (0.002) & 2.765 (0.014) \\
ABCD & gg-BFA (2f) & 0.675 (0.006) & 0.969 (0.010) & 0.951 (0.002) & 2.587 (0.012) \\
ABCD & gg-BFA (mf) & 0.582 (0.005) & 0.781 (0.008) & 0.950 (0.001) & 2.190 (0.007) \\
UKB & SCFA & 0.768 (0.009) & 1.148 (0.010) & 0.950 (9.63e-04) & 2.997 (0.035) \\
UKB & ML-CFA & 0.761 (0.011) & 1.106 (0.020) & 0.952 (3.45e-04) & 2.931 (0.050) \\
UKB & MFVB-CFA$^{\dagger}$ & 0.762 (0.009) & 1.109 (0.016) & 0.952 (3.72e-04) & 2.935 (0.043) \\
UKB & gg-BFA (1f) & 0.768 (0.009) & 1.148 (0.010) & 0.950 (9.63e-04) & 2.997 (0.035) \\
UKB & gg-BFA (2f+s) & 0.764 (0.012) & 1.111 (0.016) & 0.951 (4.31e-04) & 2.940 (0.047) \\
UKB & gg-BFA (2f) & 0.735 (0.011) & 1.072 (0.020) & 0.952 (3.99e-04) & 2.831 (0.051) \\
UKB & gg-BFA (mf) & 0.715 (0.009) & 1.041 (0.018) & 0.952 (3.73e-04) & 2.751 (0.043) \\
\end{longtable}

\begin{longtable}{@{}llrrr@{}}
\caption{All retained variables; common factor budget. Held-out covariance: Gaussian NLL per variable, relative covariance error, and correlation SRMR. Five-fold means (sample SD). $\dagger$ denotes at least one fit that did not meet its tolerance; $\ddagger$ denotes the SCFA latent-factor covariance caveat.}\label{tab:r2-fixed-full}\\
\toprule
Panel & Method & NLL/variable & $d_F$ & SRMR \\
\midrule
\endfirsthead
\multicolumn{5}{c}{\tablename\ \thetable\ continued}\\
\toprule
Panel & Method & NLL/variable & $d_F$ & SRMR \\
\midrule
\endhead
\midrule
\multicolumn{5}{r}{Continued on next page}\\
\endfoot
\bottomrule
\endlastfoot
Nightingale & SCFA$^{\ddagger}$ & 0.844 (0.042) & 0.622 (0.021) & 0.273 (0.004) \\
Nightingale & ML-CFA & 0.714 (0.043) & 0.632 (0.014) & 0.278 (0.002) \\
Nightingale & MFVB-CFA$^{\dagger}$ & 0.717 (0.042) & 0.927 (0.160) & 0.295 (0.011) \\
Nightingale & PCFA-VA$^{\dagger}$ & 0.087 (0.149) & 1.597 (0.590) & 0.272 (0.011) \\
Nightingale & gg-BFA (1f) & 0.845 (0.042) & 0.622 (0.021) & 0.273 (0.004) \\
Nightingale & gg-BFA (2f+s) & 0.694 (0.041) & 0.622 (0.027) & 0.268 (0.005) \\
Nightingale & gg-BFA (2f) & 0.446 (0.042) & 0.583 (0.017) & 0.249 (0.006) \\
Nightingale & gg-BFA (mf) & -0.142 (0.045) & 0.548 (0.021) & 0.237 (0.008) \\
Nightingale & EFA$^{\dagger}$ & 1.83e+03 (4.09e+03) & 0.231 (0.037) & 0.070 (0.007) \\
Nightingale & FV-MGPS & -0.374 (0.040) & 3.129 (4.786) & 0.263 (0.168) \\
Nightingale & FV-SSL & -0.378 (0.039) & 0.509 (0.070) & 0.141 (0.015) \\
Nightingale & FV-DL & -0.371 (0.039) & 5.131 (1.657) & 0.435 (0.040) \\
Nightingale & FV-HS & -0.378 (0.039) & 0.569 (0.051) & 0.136 (0.012) \\
Nightingale & FV-BASS & -0.350 (0.038) & 33.848 (6.870) & 0.201 (0.019) \\
Nightingale & BCME & 1.081 (0.061) & 0.773 (0.005) & 0.346 (0.003) \\
Kidney & SCFA & 1.000 (0.032) & 0.646 (0.029) & 0.271 (0.013) \\
Kidney & ML-CFA & 0.967 (0.033) & 0.639 (0.040) & 0.268 (0.016) \\
Kidney & MFVB-CFA$^{\dagger}$ & 0.967 (0.033) & 0.716 (0.048) & 0.282 (0.018) \\
Kidney & PCFA-VA & 0.819 (0.040) & 0.567 (0.048) & 0.236 (0.021) \\
Kidney & gg-BFA (1f) & 1.000 (0.032) & 0.646 (0.029) & 0.271 (0.013) \\
Kidney & gg-BFA (2f+s) & 0.959 (0.034) & 0.640 (0.028) & 0.263 (0.015) \\
Kidney & gg-BFA (2f) & 0.904 (0.029) & 0.609 (0.036) & 0.251 (0.017) \\
Kidney & gg-BFA (mf) & 0.833 (0.037) & 0.569 (0.046) & 0.237 (0.020) \\
Kidney & EFA & 0.713 (0.043) & 0.266 (0.038) & 0.095 (0.011) \\
Kidney & FV-MGPS & 0.711 (0.042) & 0.525 (0.056) & 0.191 (0.027) \\
Kidney & FV-SSL & 1.095 (0.009) & 133.747 (12.523) & 0.303 (0.024) \\
Kidney & FV-DL & 0.712 (0.042) & 0.642 (0.027) & 0.228 (0.017) \\
Kidney & FV-HS & 0.711 (0.042) & 0.624 (0.019) & 0.231 (0.015) \\
Kidney & FV-BASS & 0.738 (0.027) & 40.676 (9.217) & 0.314 (0.025) \\
Kidney & BCME & 1.011 (0.031) & 0.596 (0.012) & 0.248 (0.006) \\
miRNA & SCFA & 1.143 (0.038) & 0.711 (0.048) & 0.241 (0.006) \\
miRNA & ML-CFA & 1.105 (0.041) & 0.704 (0.040) & 0.239 (0.005) \\
miRNA & MFVB-CFA$^{\dagger}$ & 1.106 (0.041) & 0.729 (0.067) & 0.240 (0.007) \\
miRNA & PCFA-VA & 0.969 (0.040) & 0.634 (0.053) & 0.213 (0.006) \\
miRNA & gg-BFA (1f) & 1.143 (0.038) & 0.710 (0.048) & 0.241 (0.006) \\
miRNA & gg-BFA (2f+s) & 1.094 (0.041) & 0.703 (0.053) & 0.233 (0.005) \\
miRNA & gg-BFA (2f) & 1.038 (0.042) & 0.658 (0.053) & 0.220 (0.006) \\
miRNA & gg-BFA (mf) & 0.979 (0.041) & 0.635 (0.053) & 0.214 (0.006) \\
miRNA & EFA & 0.873 (0.042) & 0.345 (0.116) & 0.105 (0.026) \\
miRNA & FV-MGPS & 0.872 (0.040) & 0.444 (0.056) & 0.129 (0.007) \\
miRNA & FV-SSL & 1.546 (0.012) & 137.955 (13.373) & 0.273 (0.011) \\
miRNA & FV-DL & 0.874 (0.040) & 0.670 (0.033) & 0.201 (0.021) \\
miRNA & FV-HS & 0.872 (0.040) & 0.611 (0.020) & 0.179 (0.014) \\
miRNA & FV-BASS & 0.875 (0.038) & 18.218 (8.214) & 0.393 (0.068) \\
miRNA & BCME & 1.174 (0.035) & 0.772 (0.051) & 0.263 (0.007) \\
ABCD & SCFA & 1.172 (0.006) & 0.655 (0.007) & 0.150 (0.003) \\
ABCD & ML-CFA & 1.140 (0.006) & 0.642 (0.009) & 0.147 (0.003) \\
ABCD & MFVB-CFA$^{\dagger}$ & 1.141 (0.006) & 2.039 (0.098) & 0.310 (0.004) \\
ABCD & gg-BFA (1f) & 1.172 (0.006) & 0.655 (0.007) & 0.150 (0.003) \\
ABCD & gg-BFA (2f+s) & 1.127 (0.005) & 0.641 (0.012) & 0.142 (0.002) \\
ABCD & gg-BFA (2f) & 1.070 (0.005) & 0.543 (0.013) & 0.121 (0.002) \\
ABCD & gg-BFA (mf) & 0.923 (0.007) & 0.459 (0.012) & 0.104 (0.003) \\
UKB & SCFA & 1.234 (0.003) & 0.606 (0.003) & 0.129 (6.72e-04) \\
UKB & ML-CFA & 1.206 (0.004) & 0.578 (0.004) & 0.123 (8.99e-04) \\
UKB & MFVB-CFA$^{\dagger}$ & 1.207 (0.004) & 1.649 (0.148) & 0.188 (0.008) \\
UKB & gg-BFA (1f) & 1.234 (0.003) & 0.606 (0.003) & 0.129 (6.72e-04) \\
UKB & gg-BFA (2f+s) & 1.209 (0.004) & 0.605 (0.006) & 0.124 (0.001) \\
UKB & gg-BFA (2f) & 1.183 (0.004) & 0.543 (0.013) & 0.114 (0.002) \\
UKB & gg-BFA (mf) & 1.162 (0.006) & 0.503 (0.015) & 0.107 (0.003) \\
\end{longtable}

\begin{longtable}{@{}llrrrr@{}}
\caption{All retained variables; common factor budget. Masked-measurement prediction uses the common Gaussian rule; coverage is a proportion, and width is in training-standardized units. Five-fold means (sample SD). $\dagger$ denotes at least one fit that did not meet its tolerance; $\ddagger$ denotes the SCFA latent-factor covariance caveat.}\label{tab:r3-fixed-full}\\
\toprule
Panel & Method & RMSE & NLPD & Coverage & Width \\
\midrule
\endfirsthead
\multicolumn{6}{c}{\tablename\ \thetable\ continued}\\
\toprule
Panel & Method & RMSE & NLPD & Coverage & Width \\
\midrule
\endhead
\midrule
\multicolumn{6}{r}{Continued on next page}\\
\endfoot
\bottomrule
\endlastfoot
Nightingale & SCFA$^{\ddagger}$ & 0.652 (0.053) & 0.807 (0.079) & 0.945 (0.008) & 2.312 (0.171) \\
Nightingale & ML-CFA & 0.662 (0.051) & 0.655 (0.122) & 0.949 (0.005) & 2.263 (0.204) \\
Nightingale & MFVB-CFA$^{\dagger}$ & 0.662 (0.050) & 0.658 (0.119) & 0.951 (0.005) & 2.269 (0.201) \\
Nightingale & PCFA-VA$^{\dagger}$ & 0.572 (0.071) & -0.053 (0.279) & 0.955 (0.012) & 1.704 (0.411) \\
Nightingale & gg-BFA (1f) & 0.653 (0.053) & 0.808 (0.080) & 0.944 (0.008) & 2.315 (0.170) \\
Nightingale & gg-BFA (2f+s) & 0.649 (0.055) & 0.632 (0.121) & 0.950 (0.005) & 2.206 (0.220) \\
Nightingale & gg-BFA (2f) & 0.605 (0.060) & 0.365 (0.108) & 0.950 (0.005) & 1.917 (0.221) \\
Nightingale & gg-BFA (mf) & 0.571 (0.071) & -0.262 (0.199) & 0.950 (0.003) & 1.563 (0.271) \\
Nightingale & EFA$^{\dagger}$ & 0.416 (0.054) & -0.529 (0.226) & 0.948 (0.004) & 1.126 (0.195) \\
Nightingale & FV-MGPS & 0.415 (0.054) & -0.497 (0.222) & 0.960 (0.003) & 1.141 (0.198) \\
Nightingale & FV-SSL & 0.415 (0.054) & -0.498 (0.222) & 0.960 (0.003) & 1.139 (0.195) \\
Nightingale & FV-DL & 0.415 (0.054) & -0.497 (0.221) & 0.961 (0.003) & 1.143 (0.196) \\
Nightingale & FV-HS & 0.414 (0.054) & -0.498 (0.222) & 0.960 (0.003) & 1.138 (0.195) \\
Nightingale & FV-BASS & 0.415 (0.054) & -0.489 (0.220) & 0.967 (0.002) & 1.200 (0.202) \\
Nightingale & BCME & 0.699 (0.060) & 1.032 (0.091) & 0.946 (0.013) & 2.699 (0.064) \\
Kidney & SCFA & 0.696 (0.021) & 0.995 (0.029) & 0.944 (0.006) & 2.623 (0.035) \\
Kidney & ML-CFA & 0.695 (0.023) & 0.962 (0.031) & 0.945 (0.006) & 2.583 (0.044) \\
Kidney & MFVB-CFA$^{\dagger}$ & 0.695 (0.023) & 0.962 (0.031) & 0.946 (0.006) & 2.591 (0.044) \\
Kidney & PCFA-VA & 0.631 (0.035) & 0.815 (0.043) & 0.944 (0.006) & 2.254 (0.075) \\
Kidney & gg-BFA (1f) & 0.696 (0.021) & 0.995 (0.029) & 0.944 (0.006) & 2.621 (0.034) \\
Kidney & gg-BFA (2f+s) & 0.690 (0.022) & 0.955 (0.033) & 0.945 (0.006) & 2.564 (0.042) \\
Kidney & gg-BFA (2f) & 0.667 (0.024) & 0.899 (0.032) & 0.945 (0.006) & 2.449 (0.028) \\
Kidney & gg-BFA (mf) & 0.639 (0.030) & 0.831 (0.038) & 0.945 (0.006) & 2.309 (0.045) \\
Kidney & EFA & 0.516 (0.019) & 0.706 (0.041) & 0.943 (0.007) & 1.922 (0.029) \\
Kidney & FV-MGPS & 0.515 (0.019) & 0.704 (0.039) & 0.946 (0.007) & 1.950 (0.029) \\
Kidney & FV-SSL & 0.495 (0.019) & 1.069 (0.017) & 0.998 (8.86e-04) & 4.250 (0.111) \\
Kidney & FV-DL & 0.516 (0.019) & 0.705 (0.039) & 0.946 (0.007) & 1.951 (0.030) \\
Kidney & FV-HS & 0.515 (0.019) & 0.703 (0.039) & 0.947 (0.007) & 1.952 (0.029) \\
Kidney & FV-BASS & 0.521 (0.017) & 0.725 (0.028) & 0.964 (0.008) & 2.224 (0.080) \\
Kidney & BCME & 0.669 (0.020) & 1.006 (0.030) & 0.943 (0.005) & 2.586 (0.014) \\
miRNA & SCFA & 0.791 (0.027) & 1.134 (0.038) & 0.945 (0.008) & 3.005 (0.033) \\
miRNA & ML-CFA & 0.789 (0.027) & 1.095 (0.043) & 0.946 (0.008) & 2.953 (0.042) \\
miRNA & MFVB-CFA$^{\dagger}$ & 0.789 (0.027) & 1.096 (0.042) & 0.947 (0.008) & 2.961 (0.042) \\
miRNA & PCFA-VA & 0.731 (0.023) & 0.956 (0.041) & 0.945 (0.008) & 2.653 (0.040) \\
miRNA & gg-BFA (1f) & 0.791 (0.027) & 1.134 (0.038) & 0.945 (0.008) & 3.004 (0.033) \\
miRNA & gg-BFA (2f+s) & 0.784 (0.029) & 1.085 (0.047) & 0.947 (0.008) & 2.931 (0.042) \\
miRNA & gg-BFA (2f) & 0.758 (0.025) & 1.027 (0.044) & 0.946 (0.009) & 2.805 (0.043) \\
miRNA & gg-BFA (mf) & 0.735 (0.024) & 0.966 (0.042) & 0.946 (0.008) & 2.677 (0.042) \\
miRNA & EFA & 0.622 (0.022) & 0.858 (0.043) & 0.941 (0.008) & 2.296 (0.023) \\
miRNA & FV-MGPS & 0.621 (0.021) & 0.856 (0.041) & 0.945 (0.008) & 2.338 (0.023) \\
miRNA & FV-SSL & 0.598 (0.020) & 1.515 (0.011) & 1.000 (9.62e-05) & 7.077 (0.091) \\
miRNA & FV-DL & 0.622 (0.022) & 0.857 (0.041) & 0.946 (0.008) & 2.352 (0.023) \\
miRNA & FV-HS & 0.620 (0.021) & 0.855 (0.041) & 0.946 (0.008) & 2.345 (0.023) \\
miRNA & FV-BASS & 0.621 (0.022) & 0.857 (0.039) & 0.952 (0.007) & 2.440 (0.029) \\
miRNA & BCME & 0.786 (0.025) & 1.165 (0.033) & 0.943 (0.006) & 3.044 (0.027) \\
ABCD & SCFA & 0.791 (0.010) & 1.162 (0.012) & 0.950 (0.002) & 3.066 (0.004) \\
ABCD & ML-CFA & 0.787 (0.010) & 1.128 (0.013) & 0.951 (0.002) & 3.019 (0.011) \\
ABCD & MFVB-CFA$^{\dagger}$ & 0.787 (0.010) & 1.128 (0.014) & 0.952 (0.002) & 3.021 (0.011) \\
ABCD & gg-BFA (1f) & 0.791 (0.010) & 1.162 (0.012) & 0.950 (0.002) & 3.066 (0.004) \\
ABCD & gg-BFA (2f+s) & 0.783 (0.009) & 1.116 (0.011) & 0.951 (0.002) & 2.995 (0.010) \\
ABCD & gg-BFA (2f) & 0.751 (0.009) & 1.058 (0.011) & 0.951 (0.002) & 2.851 (0.010) \\
ABCD & gg-BFA (mf) & 0.685 (0.008) & 0.907 (0.011) & 0.951 (0.001) & 2.534 (0.011) \\
UKB & SCFA & 0.844 (0.007) & 1.229 (0.007) & 0.951 (6.73e-04) & 3.273 (0.025) \\
UKB & ML-CFA & 0.840 (0.009) & 1.200 (0.014) & 0.952 (2.73e-04) & 3.227 (0.040) \\
UKB & MFVB-CFA$^{\dagger}$ & 0.840 (0.009) & 1.202 (0.013) & 0.952 (2.91e-04) & 3.229 (0.039) \\
UKB & gg-BFA (1f) & 0.844 (0.007) & 1.229 (0.007) & 0.951 (6.73e-04) & 3.273 (0.025) \\
UKB & gg-BFA (2f+s) & 0.842 (0.008) & 1.203 (0.012) & 0.952 (3.54e-04) & 3.233 (0.037) \\
UKB & gg-BFA (2f) & 0.824 (0.009) & 1.176 (0.014) & 0.952 (2.78e-04) & 3.156 (0.043) \\
UKB & gg-BFA (mf) & 0.811 (0.011) & 1.154 (0.015) & 0.952 (2.74e-04) & 3.100 (0.047) \\
\end{longtable}

\begin{longtable}{@{}llrl@{}}
\caption{Estimator time in seconds on a P100 GPU (five-fold mean, with the sample standard deviation in parentheses) and stopping status at the common factor budget.  For the CFA references and the gg-BFA members, the time includes fitting and covariance construction, as in Figure~\ref{fig:cfa-error-time}; for EFA, FactorVerse, and BCME, it is the fitting time in the held-out runs.  Times exclude the shared community extraction and scoring; data transfers and posterior summaries differ across implementations.  Fractions count converged folds, and MCMC denotes fixed-length sampling, not a claim of convergence.  $\dagger$: at least one fold did not converge; $\ddagger$: the SCFA latent-factor covariance caveat.}\label{tab:runtime-status}\\
\toprule
Panel & Method & Seconds & Status \\
\midrule
\endfirsthead
\multicolumn{4}{c}{\tablename\ \thetable\ continued}\\
\toprule
Panel & Method & Seconds & Status \\
\midrule
\endhead
\midrule
\multicolumn{4}{r}{Continued on next page}\\
\endfoot
\bottomrule
\endlastfoot
Nightingale & SCFA$^{\ddagger}$ & 0.172 (0.015) & closed form \\
Nightingale & ML-CFA & 4.838 (0.326) & 5/5 \\
Nightingale & MFVB-CFA$^{\dagger}$ & 4.595 (0.084) & 0/5 \\
Nightingale & PCFA-VA$^{\dagger}$ & 123.081 (2.092) & 0/5 \\
Nightingale & gg-BFA (1f) & 10.062 (8.465) & 5/5 \\
Nightingale & gg-BFA (2f+s) & 2.319 (0.653) & 5/5 \\
Nightingale & gg-BFA (2f) & 4.384 (1.553) & 5/5 \\
Nightingale & gg-BFA (mf) & 8.449 (0.736) & 5/5 \\
Nightingale & EFA$^{\dagger}$ & 112.301 (61.265) & 1/5 \\
Nightingale & FV-MGPS & 20.335 (0.440) & MCMC \\
Nightingale & FV-SSL & 69.938 (1.700) & MCMC \\
Nightingale & FV-DL & 32.965 (0.848) & MCMC \\
Nightingale & FV-HS & 67.039 (0.608) & MCMC \\
Nightingale & FV-BASS & 17.956 (0.230) & MCMC \\
Nightingale & BCME & 2.940 (0.243) & 5/5 \\
Kidney & SCFA & 1.212 (2.009) & closed form \\
Kidney & ML-CFA & 0.668 (0.469) & 5/5 \\
Kidney & MFVB-CFA$^{\dagger}$ & 5.173 (0.126) & 0/5 \\
Kidney & PCFA-VA & 23.539 (5.171) & 5/5 \\
Kidney & gg-BFA (1f) & 7.805 (3.781) & 5/5 \\
Kidney & gg-BFA (2f+s) & 2.832 (0.898) & 5/5 \\
Kidney & gg-BFA (2f) & 1.839 (0.978) & 5/5 \\
Kidney & gg-BFA (mf) & 2.324 (0.361) & 5/5 \\
Kidney & EFA & 8.239 (3.209) & 5/5 \\
Kidney & FV-MGPS & 100.058 (1.775) & MCMC \\
Kidney & FV-SSL & 381.750 (7.082) & MCMC \\
Kidney & FV-DL & 110.094 (1.772) & MCMC \\
Kidney & FV-HS & 375.033 (5.813) & MCMC \\
Kidney & FV-BASS & 94.554 (1.064) & MCMC \\
Kidney & BCME & 2.996 (0.224) & 5/5 \\
miRNA & SCFA & 2.125 (2.450) & closed form \\
miRNA & ML-CFA & 0.527 (0.160) & 5/5 \\
miRNA & MFVB-CFA$^{\dagger}$ & 5.229 (0.096) & 0/5 \\
miRNA & PCFA-VA & 24.562 (5.218) & 5/5 \\
miRNA & gg-BFA (1f) & 5.819 (3.015) & 5/5 \\
miRNA & gg-BFA (2f+s) & 1.530 (0.334) & 5/5 \\
miRNA & gg-BFA (2f) & 1.488 (0.760) & 5/5 \\
miRNA & gg-BFA (mf) & 3.418 (0.824) & 5/5 \\
miRNA & EFA & 2.363 (0.237) & 5/5 \\
miRNA & FV-MGPS & 78.757 (2.621) & MCMC \\
miRNA & FV-SSL & 294.840 (14.046) & MCMC \\
miRNA & FV-DL & 90.489 (4.764) & MCMC \\
miRNA & FV-HS & 288.136 (12.819) & MCMC \\
miRNA & FV-BASS & 72.903 (2.955) & MCMC \\
miRNA & BCME & 7.478 (0.607) & 5/5 \\
ABCD & SCFA & 2.997 (2.449) & closed form \\
ABCD & ML-CFA & 2.806 (0.253) & 5/5 \\
ABCD & MFVB-CFA$^{\dagger}$ & 54.963 (0.341) & 0/5 \\
ABCD & gg-BFA (1f) & 3.740 (0.298) & 5/5 \\
ABCD & gg-BFA (2f+s) & 14.131 (0.386) & 5/5 \\
ABCD & gg-BFA (2f) & 33.018 (8.692) & 5/5 \\
ABCD & gg-BFA (mf) & 160.744 (16.246) & 5/5 \\
UKB & SCFA & 0.694 (0.273) & closed form \\
UKB & ML-CFA & 1.621 (0.304) & 5/5 \\
UKB & MFVB-CFA$^{\dagger}$ & 132.145 (3.060) & 0/5 \\
UKB & gg-BFA (1f) & 10.646 (2.770) & 5/5 \\
UKB & gg-BFA (2f+s) & 19.030 (17.839) & 5/5 \\
UKB & gg-BFA (2f) & 93.536 (13.555) & 5/5 \\
UKB & gg-BFA (mf) & 20.005 (2.261) & 5/5 \\
\end{longtable}

\begin{longtable}{@{}llrr@{}}
\caption{FactorVerse single-chain diagnostics at the common factor budget across five folds.  Occupied is the mean number of occupied groups, and the half-chain range is the range of the relative difference between the first- and second-half posterior mean covariances.  These diagnostics are not optimizer stopping criteria.}\label{tab:r23-factorverse-diagnostics}\\
\toprule
Panel & Prior & Occupied & Half-chain range \\
\midrule
\endfirsthead
\multicolumn{4}{c}{\tablename\ \thetable\ continued}\\
\toprule
Panel & Prior & Occupied & Half-chain range \\
\midrule
\endhead
\midrule
\multicolumn{4}{r}{Continued on next page}\\
\endfoot
\bottomrule
\endlastfoot
Nightingale & BASS & 6.3 & 0.078--0.124 \\
Nightingale & DL & 8.0 & 0.122--0.204 \\
Nightingale & HS & 8.0 & 0.130--0.164 \\
Nightingale & MGPS & 8.0 & 0.076--0.115 \\
Nightingale & SSL & 7.9 & 0.145--0.172 \\
Kidney & BASS & 5.8 & 0.464--0.670 \\
Kidney & DL & 6.0 & 0.052--0.276 \\
Kidney & HS & 6.0 & 0.024--0.052 \\
Kidney & MGPS & 6.0 & 0.127--0.334 \\
Kidney & SSL & 6.0 & 0.093--0.168 \\
miRNA & BASS & 7.6 & 0.524--0.738 \\
miRNA & DL & 8.0 & 0.167--0.292 \\
miRNA & HS & 8.0 & 0.027--0.049 \\
miRNA & MGPS & 8.0 & 0.072--0.205 \\
miRNA & SSL & 8.0 & 0.064--0.118 \\
\end{longtable}

\begin{longtable}{@{}llrrrrr@{}}
\caption{Local variational prediction on all retained variables for the four gg-BFA members at the common factor budget.  Metrics pool the masked cells of the five folds; these predictions integrate the local variational factor posterior of each test subject and are separate from the common Gaussian predictions of Table~\ref{tab:r3-fixed-full}.}\label{tab:r3-local-variational}\\
\toprule
Panel & Member & RMSE & NLPD & Coverage & Width & Cells \\
\midrule
\endfirsthead
\multicolumn{7}{c}{\tablename\ \thetable\ continued}\\
\toprule
Panel & Member & RMSE & NLPD & Coverage & Width & Cells \\
\midrule
\endhead
\midrule
\multicolumn{7}{r}{Continued on next page}\\
\endfoot
\bottomrule
\endlastfoot
Nightingale & gg-BFA (mf) & 0.575 & -0.264 & 0.950 & 1.563 & 80654 \\
Nightingale & gg-BFA (2f) & 0.607 & 0.365 & 0.950 & 1.917 & 80654 \\
Nightingale & gg-BFA (2f+s) & 0.650 & 0.632 & 0.950 & 2.206 & 80654 \\
Nightingale & gg-BFA (1f) & 0.655 & 0.808 & 0.944 & 2.315 & 80654 \\
Kidney & gg-BFA (mf) & 0.640 & 0.830 & 0.945 & 2.309 & 253046 \\
Kidney & gg-BFA (2f) & 0.668 & 0.899 & 0.945 & 2.449 & 253046 \\
Kidney & gg-BFA (2f+s) & 0.690 & 0.955 & 0.945 & 2.564 & 253046 \\
Kidney & gg-BFA (1f) & 0.697 & 0.995 & 0.944 & 2.621 & 253046 \\
miRNA & gg-BFA (mf) & 0.735 & 0.966 & 0.946 & 2.677 & 182240 \\
miRNA & gg-BFA (2f) & 0.758 & 1.027 & 0.946 & 2.805 & 182240 \\
miRNA & gg-BFA (2f+s) & 0.784 & 1.085 & 0.947 & 2.931 & 182240 \\
miRNA & gg-BFA (1f) & 0.791 & 1.134 & 0.945 & 3.004 & 182240 \\
ABCD & gg-BFA (mf) & 0.685 & 0.907 & 0.951 & 2.533 & 5136637 \\
ABCD & gg-BFA (2f) & 0.751 & 1.058 & 0.951 & 2.851 & 5136637 \\
ABCD & gg-BFA (2f+s) & 0.783 & 1.116 & 0.951 & 2.995 & 5136637 \\
ABCD & gg-BFA (1f) & 0.791 & 1.162 & 0.950 & 3.066 & 5136637 \\
UKB & gg-BFA (mf) & 0.811 & 1.154 & 0.952 & 3.100 & 21196804 \\
UKB & gg-BFA (2f) & 0.824 & 1.176 & 0.952 & 3.156 & 21196804 \\
UKB & gg-BFA (2f+s) & 0.842 & 1.203 & 0.952 & 3.233 & 21196804 \\
UKB & gg-BFA (1f) & 0.844 & 1.229 & 0.951 & 3.273 & 21196804 \\
\end{longtable}

}

\clearpage

\section{Computational reproducibility}
\label{app:reproducibility}
\label{app:provenance}

Each analysis was run in its own directory under {\footnotesize\path{experiments/}}, which stores the
configuration, a snapshot of the source code, the logs, the evaluation records, and the requested
and actual computing device.  The gg-BFA fits and the four CFA references ran on NVIDIA P100
GPUs; the dense-subgraph search and the \texttt{psych}, \texttt{lavaan} and \texttt{semopy}
pipelines ran on CPUs.  Every table and figure under {\footnotesize\path{manuscript/scfa_vb/generated/}} is
written by a script that records its input files and their SHA-256 hashes.  The reported results
come from the following directories.

{\raggedright
\begin{description}[style=nextline,leftmargin=1.5em]
\item[Structure simulations (Section~\ref{sec:sim-selection}).]
One directory per design and sample size, listed in
{\footnotesize\path{manuscript/scfa_vb/generated/sim_selection_20261009/numbers.json}}.  The
exploratory-to-confirmatory pipelines ran in
{\footnotesize\path{20261009_123933_sim_efa_cfa_pipelines_full_cpu_seed2610}}.
\item[Simulation check in growing dimension (Appendix~\ref{th:sec:highdim-sim}).]
Four directories, listed in
{\footnotesize\path{manuscript/scfa_vb/generated/sim_appendix_d_20261009/numbers.json}}.
\item[Simulations of Comparison~2 (Appendix~\ref{app:simulation-details}).]
Fixed-community simulation: {\footnotesize\path{20260914_experiment1_fixed_loading_gpu_full_seed24010000}};
cross-loading simulation: {\footnotesize\path{20260904_112640_experiment2_option_matrix_full_cpu_seed24020000}};
stress test: {\footnotesize\path{20260904_205544_experiment3_known_covariance_plotfix_full_cpu_seed24030000}}.
The outputs used from the last two are recorded with their hashes in
{\footnotesize\path{manuscript/scfa_vb/generated/naming_20260908/provenance.json}}.
\item[Real data at the common factor budget (Section~\ref{sec:applications}).]
Full data: {\footnotesize\path{20260920_002338_dual_k_real_data_r1_full_gpu1_seed260917101}}.  Held-out
covariance and prediction: {\footnotesize\path{20260922_082452_dual_k_real_data_r23_full_gpu1_seed260918101}},
with ML-CFA in {\footnotesize\path{20260925_002837_classical_ml_cfa_r23_full_gpu1_seed260925201}} and SCFA,
MFVB-CFA and PCFA-VA in {\footnotesize\path{20260925_100728_cfa_r23_extension_full_gpu1_seed260925301}}.
\item[Complete pipeline on the full data (Section~\ref{sec:raw-to-fit} and Figure~\ref{fig:method}).]
The input runs are listed in
{\footnotesize\path{manuscript/scfa_vb/generated/enar_figures_20261013/raw_to_fit_mf_PROVENANCE.json}} and
{\footnotesize\path{manuscript/scfa_vb/generated/enar_figures_20261015b/kidney_figures_PROVENANCE.json}}.
\end{description}\par}

The software is released as the Python package \texttt{ggbfa} (version 0.1.0).  Its function
\texttt{efa} runs the exploratory step (Algorithm~\ref{alg:two-stage} by default and
Algorithm~\ref{alg:selection} on request), and \texttt{cfa} fits the members 1f, 2f+s, 2f and mf,
each with posterior summaries, the covariance estimate with loading second moments, prediction,
and optional bagging.  The analyses above were run with the research code from which
\texttt{ggbfa} was extracted; \texttt{ggbfa} reproduces its community selections and
confirmatory fits.

\clearpage
\bibliographystyle{plainnat}
\providecommand{\bibroot}{}
\bibliography{\bibroot references}
\end{document}